\pdfoutput=1 
\documentclass[11pt,a4paper]{article}
\usepackage{mathtools}
\usepackage{jheppub}
\usepackage{float}
\usepackage{graphicx}
\usepackage{url}
\usepackage{cancel}
\usepackage{caption}
\usepackage[dvipsnames]{xcolor}
\usepackage{placeins}
\usepackage{feynman}
\usepackage{xspace}
\usepackage{bm}
\usepackage{hyperref}
\usepackage{amsmath}
\usepackage{amssymb}
\usepackage{listings}
\usepackage{framed}
\usepackage[toc, page]{appendix}
\usepackage{enumitem} 
\usepackage{color}
\usepackage{booktabs} 
\usepackage{subcaption}
\usepackage{braket}

\newcommand{\nnpdf}{{\tt NNPDF4.0}\xspace}

\newcommand{\madgraph}{\texttt{MadGraph5\_aMC@NLO}{}}

\def\G1{{\bf \gamma^{(1)}_N}}

\title{Hide and seek: how PDFs can conceal New Physics}
\author[a]{Elie Hammou}
\author[a]{, Zahari Kassabov}
\author[b]{, Maeve Madigan}
\author[c]{, Michelangelo L. Mangano}
\author[a]{, Luca Mantani}
\author[a]{, James Moore}
\author[a]{, Manuel Morales Alvarado}
\author[a]{and Maria Ubiali}
\affiliation[a]{DAMTP, University of Cambridge, Wilberforce Road, Cambridge, CB3 0WA, United Kingdom}
\affiliation[b]{Institut für Theoretische Physik, Universität Heidelberg, Germany}
\affiliation[c]{Theoretical Physics Department, CERN, Geneva, Switzerland}
\emailAdd{eh651@cam.ac.uk, zk261@cam.ac.uk, madigan@thphys.uni-heidelberg.de, Michelangelo.Mangano@cern.ch, luca.mantani@maths.cam.ac.uk, jmm232@cam.ac.uk, mom28@cam.ac.uk, M.Ubiali@damtp.cam.ac.uk}

\abstract{The interpretation of LHC data, and the assessment of possible 
hints of new physics,
  require the precise
  knowledge of the proton structure in terms of parton distribution
  functions (PDFs). 
  We present a systematic methodology designed to determine whether and how
  global PDF fits might inadvertently ‘fit away’ signs of new physics in the high-energy tails of the distributions. 
  We showcase a scenario for the High-Luminosity LHC, in which the PDFs may completely
  absorb such signs of new physics, thus biasing theoretical
  predictions and interpretations. We discuss strategies to
  single out the effects in this scenario, and disentangle the inconsistencies that stem from them.  
  Our study brings to light the synergy between the high luminosity programme at the LHC and
  future low-energy non-LHC measurements of large-$x$ sea quark distributions. 
  The analysis code 
  used in this work is made public so that 
  any users can test the robustness of the signal associated to a given BSM model against 
  absorption by the PDFs.}

\arxivnumber{2307.10370}
\subheader{\small CERN-TH-2023-137}

\begin{document}

\maketitle
\flushbottom

\section{Introduction}
\label{sec:intro}

Theoretical predictions at hadron colliders depend on Parton Distribution Functions (PDFs), which parametrise the structure
of the proton in terms of its elementary constituents, collectively called \textit{partons}. While the PDF dependence on the fraction of
longitudinal momentum  $x$ of the proton carried by each parton cannot be computed using perturbation theory, the PDF dependence
on the energy scale of the hard process is accurately predicted by perturbative QCD. Moreover, in the region of validity of collinear factorisation,
PDFs are universal, process-independent objects. Thus, the PDF dependence on $x$ can be extracted from a fit to experimental data. 
For recent reviews of state-of-the-art PDF fits see for example Refs. ~\cite{Amoroso:2022eow,Gao:2017yyd,Kovarik:2019xvh}.

Before the Large Hadron Collider (LHC) began collecting data, PDFs were mostly extracted from data collected by fixed-target, HERA and Tevatron experiments, and used as an input for theory predictions at the LHC. 
Now, the LHC data itself has become an invaluable input in the global fits of PDFs, 
and modern PDF analyses include an increasing portion of LHC data~\cite{Ball:2021leu,Bailey:2020ooq,Hou:2019efy,Ball:2017nwa,Alekhin:2017kpj}. 
For example, in the recent \nnpdf{} analysis~\cite{Ball:2021leu}, nearly 30\% of the data points included in the fit are LHC data. 
The LHC data places significant constraints on the PDFs, with Drell-Yan high-mass distributions providing a strong handle on the 
light quark and antiquark distributions in the medium and large-$x$ regions~\cite{Jing:2023isu}, and the high-$E_T$ jet and top data constraining 
the gluon in the large-$x$ region~\cite{Nocera:2017zge}. Despite the wealth of constraints coming from LHC data, the large-$x$ 
region still displays the largest PDF uncertainty, thus limiting the discovery potential for BSM signatures 
in the multi-TeV mass region. 
It is therefore paramount to devise and include new observables that can give us a robust understanding of PDFs and their uncertainties 
at large-$x$, such as, for example, the Drell-Yan forward-backward asymmetry~\cite{Accomando:2019vqt,Fiaschi:2021okg,Fiaschi:2022wgl,Ball:2022qtp,Fu:2023rrs}.  

In this hunt for new constraints on the large-$x$ structure of the proton, it is crucial to be aware that the observables 
that constrain the PDFs in the large-$x$ region are also those that are most likely to be affected by the presence of heavy new physics in the high-energy tails of the distributions.
In the past decade there has been an increasing amount of activity aimed at assessing the interplay between the determination of the proton structure
and the presence of new physics signatures in the data, for example by devising observables that are sensitive to new physics effects but have a reduced PDF-dependence ~\cite{Mangano:2012mh,Fiaschi:2021sin}.
Conversely, more recent analyses have been determined how much room there is in the proton for weakly interacting 
particles such as dark photons~\cite{McCullough:2022hzr} or Lorentz and CPT-violating effects~\cite{ZEUS:2022msi}. 
Moreover, several studies 
tackled the interplay between the fits of PDFs and the fits of model-independent parametrisations of the effects of heavy new particle via EFT 
coefficients, first in the context of DIS data~\cite{H1:2011ije,ZEUS:2019cou,Carrazza:2019sec}, then in the context of high-energy Drell-Yan 
tails~\cite{DYpaper,Iranipour:2022iak,Liu:2022plj} and in the top sector~\cite{Kassabov:2023hbm,Gao:2022srd}, and through an experimental analysis of the 
jets data~\cite{CMS:2021yzl}. 
A new tool for the simultaneous fit of PDFs and SMEFT coefficients was presented~\cite{Iranipour:2022iak} and applied first in the context of Drell-Yan 
data and then to the simultaneous fit of PDFs and up to 25 Wilson coefficients parametrising the top sector~\cite{Kassabov:2023hbm}. 

These studies highlight that the effects of the interplay between SMEFT effects and PDF fits is observable dependent. 
For example, in the top quark sector, the large-$x$ gluon density extracted from current top quark data including SMEFT effects or ignoring them 
differ by an amount comparable to its uncertainty. In the Drell-Yan sector the PDFs 
extracted from High-Luminosity LHC (HL-LHC) Drell-Yan projections do not shift as compared to the PDFs  
determined by ignoring SMEFT effects, however they exhibit larger PDF uncertainties. 
On the other hand, at the level of Wilson coefficients, in the top sector the results are unchanged upon the variation of PDFs in the fit. 
This was not the case in the Drell-Yan sector, when instead the bounds broaden significantly once SMEFT effects are propagated in the PDF fits.  
The results so far suggest that Drell-Yan distributions play a crucial role in determining the large-$x$ quark and antiquark distributions and 
that a global PDF fit could be significantly distorted if BSM physics were to be present in the high-mass Drell-Yan tails. 
In Ref.~\cite{Madigan:2021uho} a further step was taken by investigating the definition of conservative PDFs. 

A crucial question is still partially unanswered and should be fully
addressed if we want to make sure that the LHC programme of indirect searches
remains unbiased by our theoretical assumptions:  
how can one assess whether new physics effects are inadvertently
absorbed into the flexible parametrisation of the non-perturbative proton
structure, as more and more high-energy tails of the distributions are used
to determine the PDFs of the proton? Imagine that the true law of Nature
contains some new heavy particles that have mild effects on the high energy tails
of some LHC distributions that are normally used as an input in PDF fits, and then suppose that we perform a PDF fit including such distributions,
but using the SM in our theoretical predictions. There are two options:
either the data-theory agreement for those distributions will be poor,
thus the data would be excluded from the fit because it would be
inconsistent with the bulk of the other data included in it, and consequently flagged for further investigation;
alternatively, the data-theory agreement will be acceptable because the
fitted PDFs have somehow managed to adapt to the data affected by new physics,
without deteriorating
the data-theory agreement with the other data that are unaffected
by new physics. In the latter case, PDFs would be ``contaminated''
and with our current tools we would not know that this is the case.

In this study, after identifying two showcase scenarios in which  
contamination might happen, we explore two natural follow-up
questions. First, if we were using a ``contaminated'' PDF set, would we miss
other signals of new physics that Nature would exhibit in other observables
(typically not included in a PDF fit)?
Second, are there suitable observables that constrain the large-$x$ region but
are unaffected by new physics, which would uncover the inconsistency of the
tails of the distributions that we include in a PDF fit? These datasets
would help us preventing the contamination of the PDFs from occurring. 
Alternatively, we could ask whether a targeted analysis of the data included in a 
PDF fit would help detecting an inconsistency due to the presence of new physics and distinguish it from 
an inconsistency that has a different source.  

The structure of this paper is as follows. In Sect.~\ref{sec:methodology} we describe 
the methodology underlying our analysis. 
In Sect.~\ref{sec:scenarios} we present two UV scenarios and their low-energy 
SMEFT parametrisation, determining  the region of the EFT validity in each of the scenarios that we consider. 
In Sect.~\ref{sec:dy} we present the main results of the paper. We show that in one of the two scenarios that we consider, namely in 
a flavour-universal $W'$ model, contamination does happen, and we identify the phenomenological consequences of the contamination. 
Finally in Sect.~\ref{sec:solution} we explore several ways to disentangle contamination. 
We summarise our results and highlight how users can utilise the tools developed here to 
explore other BSM scenarios in Sect.~\ref{sec:summary}.
Technical details of the analysis are collected in the appendices. App.~\ref{app:fit} contains the details about the 
quality of the various fits presented in this analysis, App.~\ref{app:pdfs} shows the effect of new physics contamination on 
individual partons, while App.~\ref{app:random} explores the dependence of the 
results on the random fluctuation of the Monte Carlo data that we produce in this analysis, proving the stability and robustness of the results.

\section{Methodology}
\label{sec:methodology}

In order to systematically study contamination effects from new physics (NP) in PDF fits and to address
the questions introduced in Sect.~\ref{sec:intro}, we work in a setting in which we pretend
we know the underlying law of Nature. In our case the law of Nature consists of the ``true''
PDFs, which are low-energy quantities that have nothing to do with new physics, and the ``true''
Lagrangian of Nature, which at low energy is well approximated by the Standard Model (SM) 
Lagrangian, but to which we add some heavy new particles.
We use these assumptions to generate the artificial data that enter our analysis.
We inject the effect of the new particles that we introduce in the Lagrangian in the artificial Monte Carlo (MC) data, and their effect 
will be visible in some high-energy distributions depending on the underlying model. 

The methodology we use throughout this study is based on the NNPDF \textit{closure test} framework,
first introduced in Ref.~\cite{NNPDF:2014otw}, and explained in more detail in Ref.~\cite{DelDebbio:2021whr}.
This method was developed
in order to assess the quality and the robustness of the NNPDF fitting methodology; in brief it follows three basic steps: (i) assume that Nature's PDFs
are given by some fixed reference set; (ii) generate artificial MC data based on this assumption, which we term \textit{pseudodata}; 
(iii) fit PDFs to the pseudodata using the NNPDF methodology. Various statistical estimators, described in Ref.~\cite{DelDebbio:2021whr}, can then be applied to 
check the quality of the fit (in broad terms, assessing its difference from the true PDFs), hence verifying the accuracy of the fitting methodology. In this study, the closure test methodology is adapted to account for the fact that the true theory of Nature may \textit{not} be the 
SM. 

In this Section, we describe this adapted closure test methodology in more detail. 
In Sect.~\ref{subsec:contamination_definition}, we carefully define the terms \textit{baseline fit} and \textit{contaminated fit}, 
which shall be used throughout this paper. We then briefly remind the reader of the NNPDF fitting methodology, in particular discussing 
Monte-Carlo error propagation. In Sect.~\ref{subsec:pseudodata_generation}, we provide more details on how the MC data are generated in this work. 
Finally, in Sect~\ref{subsec:postfit} we give an overview of the types of analysis we perform on the fits we obtain.

\subsection{Basic definitions and fitting methodology}
\label{subsec:contamination_definition}
Let us suppose that the true theory of Nature is given by the SM, plus some NP contribution. 
Under this assumption, observables $T \equiv T(\theta_{\text{SM}}, \theta_{\text{NP}})$ have a dependence on both the SM 
parameters $\theta_{\text{SM}}$ (in our work, exclusively the PDFs), and the NP parameters $\theta_{\text{NP}}$ 
(for example, masses and couplings of new particles). Let us further fix notation by writing the true values of the 
parameters $\theta_{\text{SM}}, \theta_{\text{NP}}$ as $\theta_{\text{SM}}^*, \theta_{\text{NP}}^*$ respectively; 
for convenience, we shall also write the true value of the observable $T$ as $T^* \equiv T(\theta_{\text{SM}}^*, \theta_{\text{NP}}^*)$.

Suppose that we wish to perform a fit of some of the theory parameters using experimental measurements of $N_{\rm obs}$ observables, which we package as a single vector $T = (T_1, T_2,..., T_{N_{\rm obs}})$. All measurements are subject to random observational noise. For additive Gaussian uncertainties, the observed data is distributed according to:
\begin{equation}
\label{eq:observed_data}
D_0 = T^* + \eta,
\end{equation}
where $\eta$ is drawn from the multivariate Gaussian distribution $\mathcal{N}(0,\Sigma)$, with $\Sigma$ the experimental covariance matrix describing the uncertainties of the measurements and the correlations between them.
The general procedure, which also accounts for multiplicative uncertainties and positivity effects, is implemented in the NNPDF code~\cite{NNPDF:2021uiq}.

In the context of the NNPDF closure test methodology~\cite{NNPDF:2014otw}, the true values of the observables $T^*$ are referred to as \textit{level 0 pseudodata} (L0), whilst the fluctuated values $D_0$ are referred to as \textit{level 1 pseudodata} (L1).

Once we have generated a sample $D_0$ of L1 pseudodata, we may perform a fit of some of the theory parameters to this pseudodata. In this work, we shall perform various types of fits with different choices of $\theta_{\text{SM}}^*, \theta_{\text{NP}}^*$, and different choices of which parameters we are fitting. We define the types of fits as follows:
\begin{enumerate}[label = (\arabic*)]
\item \textbf{Baseline fit.} If there is no new physics, $\theta_{\text{NP}}^* \equiv 0$, then the SM is the true theory of Nature. We generate L1 pseudodata $D_0$ according to the SM. If we subsequently fit the parameters $\theta_{\text{SM}}$, we say that we are performing a \textit{baseline fit}. This is precisely equivalent to performing a standard NNPDF closure test.
\item \textbf{Contaminated fit.} If new physics exists, $\theta_{\text{NP}}^* \neq 0$, then the SM is \textit{not} the true theory of Nature. We generate L1 pseudodata $D_0$ according to the SM plus the NP contribution. If we subsequently \textit{only} fit the parameters $\theta_{\text{SM}}$ whilst ignoring the NP parameters $\theta_{\text{NP}}$, we say that we are performing a \textit{contaminated fit}.
\item \textbf{Simultaneous fit.} If new physics exists, $\theta_{\text{NP}}^* \neq 0$, we again generate L1 pseudodata $D_0$ according to the SM plus the NP contribution. If we subsequently fit \textit{both} the parameters $\theta_{\text{SM}}$ \textit{and} $\theta_{\text{NP}}$, we say that we are performing a \textit{simultaneous fit}. A closure test of this type is performed in Ref.~\cite{Iranipour:2022iak} in order to benchmark the {\tt SIMUnet} methodology. However, we do not perform such fits in this work, with our main goal being to assess the possible deficiencies associated with performing \textit{contaminated} fits.
\end{enumerate}

\begin{table}
  \centering
  \begin{tabular}{l|l|l}
    \toprule
    {\bf Fit name} & {\bf Nature}  & {\bf Fitted parameters}\\
    \midrule
    Baseline & Standard Model: $\theta_{\text{NP}}^* \equiv 0$ & Standard Model only: $\theta_{\text{SM}}$ \\
    Contaminated & SM + new physics: $\theta_{\text{NP}}^* \neq 0$ & Standard Model only: $\theta_{\text{SM}}$\\
    \bottomrule
    \end{tabular}
   \caption{\label{table:definition} A summary of the definitions of \textit{baseline} and \textit{contaminated} fits used throughout this work.}
  \end{table}
%
\vspace*{0.2cm}
\noindent Throughout this work, we shall perform only \textit{baseline} and \textit{contaminated} fits; that is, we shall only fit SM 
parameters, but we shall fit them to pseudodata generated either assuming the law of Nature is given by the SM only, 
or that it is given by the SM plus some NP contribution. A summary of the relevant definitions is given for convenient reference in Table~\ref{table:definition}.

The NNPDF methodology makes use of the Monte-Carlo (MC) replica method to propagate errors to the PDFs. 
This involves the generation of an additional layer of pseudodata, referred to as \textit{level 2 pseudodata} (L2). Given an L1 pseudodata sample $D_0$, 
we generate L2 pseudodata by augmenting $D_0$ with further random noise $\epsilon$:
\begin{equation}
\label{eq:mc_pseudodata}
D = D_0 + \epsilon = T^* + \eta + \epsilon,
\end{equation}
where $\epsilon$ is an independent multivariate Gaussian variable, distributed according to $\epsilon \sim \mathcal{N}(0,\Sigma)$, with $\Sigma$ the experimental covariance matrix. Whilst the L1 pseudodata is sampled only once, the L2 pseudodata $D$ is sampled $N_{\text{rep}}$ times, and the best-fit PDFs are obtained for each of the L2 pseudodata samples. This provides an ensemble of PDFs from which statistical estimators, in particular uncertainty bands, can be constructed.

\subsection{Pseudodata generation}
\label{subsec:pseudodata_generation}
As described above, we assume that the true theory of Nature is the SM plus some new physics. 
More specifically, in this work we take the ``true SM'' to mean SM perturbation theory to NNLO QCD accuracy. 
The true PDF set which shall be used throughout this work is the \nnpdf{} set~\cite{Ball:2021leu} (in principle, we are of course allowed to choose any PDF set).

On the other hand, we inject two different NP signals in this work. In each case, we base our NP scenario on specific UV-complete models. 
Furthermore, we choose NP scenarios which are characterised by scales much higher than the energy scales probed by the data, which allows us to justify matching 
the UV-complete models to a SMEFT parametrisation. The advantage of this approach is 
that theory predictions become polynomial in the SMEFT Wilson coefficients, which is not necessarily the case in UV-complete models; this allows
us maximum flexibility to trial many different values for the ``true'' NP parameters. To justify the SMEFT approximation, in Sect.~\ref{sec:scenarios} 
we study the validity of the EFT in each case, checking that we only use values of the SMEFT Wilson coefficients that provide good agreement with the 
UV theory at the linear or quadratic levels, as appropriate. We also make a $K$-factor approximation (the validity of which is explicitly checked in Ref.~\cite{DYpaper} in the case of the $\hat{W},\hat{Y}$ 
parameter scenarios) to avoid expensive computation of fast interpolation grids for the PDFs. As a result, the formula for the ``true'' value of 
an observable takes the schematic form:
\begin{equation}
T \equiv \left(1 + cK_{\text{lin}} + c^2K_{\text{quad}}\right) \hat{\sigma}^{\text{SM}} \otimes \mathcal{L},
\end{equation}
where $\mathcal{L}$ denotes either the PDFs or PDF luminosities for \nnpdf{} (depending on whether the observable is a deep inelastic scattering or hadronic observable), $c$ denotes the SMEFT Wilson coefficient(s) under consideration, $\hat{\sigma}^{\text{SM}}$ is the SM partonic cross-section computed at NNLO in QCD perturbation theory, and $K_{\text{lin}}, K_{\text{quad}}$ are the SMEFT $K$-factors.


%

\subsection{Post-fit analysis}
\label{subsec:postfit}

Once we have produced a contaminated fit, where PDFs have been fitted using SM theory to data produced with the SM plus some NP contribution, several natural questions arise.

\paragraph{Detection of contamination.} Is it possible to detect contamination of the PDF fit by the NP effects? If there are many datasets entering the fit that are \textit{not} affected by NP, it might be the case that datasets that \textit{are} affected by NP could appear inconsistent, and might be poorly described by the resulting fit.

In order to address this point, we use the NNPDF dataset selection criteria, discussed in detail in Ref.~\cite{Ball:2021leu}. We consider both the $\chi^2$-statistic of the resulting contaminated PDF fit to each dataset entering the fit, and also consider the number of standard deviations 
\begin{equation}
\label{eq:nsigma}
n_{\sigma} = (\chi^2 - n_{\rm dat}) / \sqrt{2n_{\rm dat}}
\end{equation} 
of the $\chi^2$-statistic from the expected $\chi^2$ for each dataset. 
If $\chi^2/n_{\rm dat} > 1.5$ and $n_{\sigma} > 2$ for a particular dataset, the dataset would be flagged by the NNPDF selection criteria, indicating an inconsistency with the other data entering the fit.

There are two possible outcomes of performing such a dataset selection analysis on a contaminated fit. In the first instance, the datasets affected by NP are flagged by the dataset selection criterion. 
If a dataset is flagged according to this condition, then a weighted fit is performed, 
i.e. a fit in which a dataset is given a
larger weight inversely proportional to the number of data points. In more detail, if the $j$th dataset has been flagged, then the $\chi^2$-loss used in the subsequent weighted fit is modified to:
\begin{equation}
\chi^2_w=\frac{1}{n_{\rm dat}-n_{\rm
    dat}^{(j)}}\,\sum_{i=1,i\neq j}^{n_{\rm exp}} \,n_{\rm
  dat}^{(i)}\,\chi^2_i \, + \,w^{(j)}\chi^2_j,
  \end{equation}
where $\chi^2_i$ denotes the usual $\chi^2$-loss for the $i$th dataset, and where the \textit{weight} is defined by:
  \begin{equation}
    \label{eq:wgts}
    w^{(j)} = n_{\rm dat}/n_{\rm dat}^{(j)}.
    \end{equation}
If the data-theory agreement improved for the set under investigation, to the extent that it now satisfies the selection criteria, \textit{and further} the
data-theory agreement of the other datasets does not deteriorate in
any statistically significant way, then the dataset is kept; otherwise, the dataset is discarded, 
on the basis of inconsistency with the remaining datasets.

In the second instance, the ``contaminated'' datasets are \textit{not} flagged, and are consistent enough that the contaminated fit 
would pass undetected as a \textit{bona fide} SM PDF fit. We introduce the following terms to describe each of these cases: 
in the former case, we say that the PDF was \textit{unable to absorb} the NP; in the latter case, we say that the PDF has \textit{absorbed} the NP.

\paragraph{Distortion of NP bounds.} Can using a contaminated fit in a subsequent fit of NP effects lead to misleading bounds? In more detail, suppose that we construct a contaminated fit which has absorbed NP - that is, the contamination would go undetected by the NNPDF dataset selection criterion. In this case, we would trust that our contaminated fit was a perfectly consistent SM PDF fit, and might try to use it to subsequently fit the underlying parameters in the NP scenario. 

There are two possible outcomes of such a fit. The contamination of the PDFs may be weak enough for the NP bounds that we obtain to be perfectly sensible, containing the true values of the NP parameters. On the other hand, the absorption of the NP may be strong enough for the NP bounds to be distorted, no longer capturing the true underlying values of the NP parameters. The second case is particularly concerning, and if it can occur, points to a clear need to disentangle PDFs and possible NP effects.

\paragraph{Distortion of SM predictions.} Finally, we ask: can using a contaminated fit lead to poor agreement on new datasets that are not affected by NP? In particular, suppose that we are again in the case where NP has been absorbed by a contaminated fit, so that the NP signal has gone undetected. If we were to use this contaminated fit to make predictions for an observable that is \textit{not} affected by the NP, it is interesting to see whether the data is well-described or not; if the contamination is sufficiently strong, it may appear that the dataset is inconsistent with the SM. This could provide a route for disentangling PDFs and NP; we shall discuss this point in Sect.~\ref{sec:solution}.


\section{New physics scenarios}
\label{sec:scenarios}
As discussed in Sect.~\ref{sec:methodology}, throughout this work we have assumed that the theory of Nature is the SM plus some new physics, 
and generated pseudodata accordingly. 
In this Section, following the methodology presented in Sect.~\ref{subsec:pseudodata_generation}, we
extend the SM to UV-complete models by introducing heavy new fields. 
%
We choose simple extensions of the SM corresponding to ``universal theories''~\cite{Wells:2015uba},
the effect of which can be well-described with an EFT approximation using the oblique corrections $\hat{Y}$ and $\hat{W}$~\cite{Peskin:1991sw,Altarelli:1991fk,Barbieri:2004qk}. 
It is important to bear in mind that such theories can be considered "universal" only if the RGE running of the Wilson coefficients~\cite{Jenkins:2013zja,Jenkins:2013wua,Alonso:2013hga,Aoude:2022aro} is neglected or additional prescriptions are devised~\cite{Wells:2015cre}. Neglecting the RGE running of the Wilson coefficients is an approximation that is completely adequate for our case studies, given that the running of the Wilson coefficients of the four-fermion operators and their mixing are expected to be sub-dominant for the observables considered in the study~\cite{DYpaper}.
We consider the following two scenarios:

\begin{itemize}
	\item {\bf Scenario I:} A flavour universal $Z'$, charged under a $U(1)_Y$ gauge symmetry. We give a mass to the field assuming it is generated by some higher energy physics. It corresponds to a new heavy neutral bosonic particle. At the EFT level, the effect of this model on our dataset can be  
		described by the $\hat{Y}$ parameter.
	\item {\bf Scenario II:} A flavour universal $W'$ charged under $SU(2)_L$.
		Once again, we directly add a mass term to the Lagrangian. This gives rise to two new heavy charged bosonic particles ($W'^{+}$ and $W'^{-}$) as well as a new heavy neutral boson, similar to a $Z'$ that only couples to left-handed fermions. At the EFT level, the effect of this model on our dataset can 
	be described by the $\hat{W}$ parameter.
\end{itemize}

\noindent The following subsections are devoted to describing each of these NP scenarios.
In particular, in each case we use tree-level matching to obtain a parametrisation of the model in terms of dim-6 operators of the SMEFT,
making use of the matching provided in Ref.~\cite{deBlas:2017xtg} to do so.
We identify the observables in our dataset affected by each NP scenario, and in each case we compare the UV and EFT predictions.  Finally, we identify values of 
the model parameters for which the EFT description is justified at the projected energy of the HL-LHC, and for which existing constraints on the models are avoided.

\subsection{Scenario I: A flavour-universal $Z'$ model}
\label{subsec:model-zprime}

The addition to the SM of a new spin-1 boson $Z'$ associated to a gauge symmetry $U(1)_Y$, a mass $M_{Z'}$ and a coupling coefficient $g_{Z'}$ yields the following Lagrangian:
\begin{equation} 
\label{eq:Zprime}
	\begin{split}
		\mathcal{L}^{Z'}_{\text{UV}} &= \mathcal{L}_{\text{SM}} - \frac{1}{4} Z'_{\mu \nu} {Z'}^{\mu \nu} + \frac{1}{2} M_{Z'}^{2} Z'_{\mu} {Z'}^{\mu} \\
		&\qquad - g_{Z'} Z'_{\mu} \sum_{\substack{f}} Y_{f} \bar{f} \gamma^{\mu} f - Y_{\varphi} g_{Z'} ( Z'_{\mu} \varphi^{\dagger} i  D^{\mu} \varphi + \textrm{h.c.} ) \, .
	\end{split}
\end{equation}
We sum the interactions with the fermions $f \in \{Q^i_L, u^i _R, d^i _R, \ell^i _L, e^i _R \text{ for i} \in  \{1, 2, 3\}\}$ , where $Y_f$ is the corresponding hypercharge: $( Y_{Q_L}, Y_{u_R}, Y_{d_R}, Y_{l_L}, Y_{e_R}, Y_{\varphi} ) = (\frac{1}{6}, \frac{2}{3}, -\frac{1}{3}, -\frac{1}{2}, -1, \frac{1}{2} ) $. The kinetic term is given by $Z'_{\mu \nu} = \partial_{\mu} Z'_{\nu} - \partial_{\nu} Z'_{\mu}$. The covariant derivative is given by 
\begin{equation}
    D_{\mu} = \partial_{\mu} + \frac{1}{2} ig \sigma^{a} W_{\mu}^{a} + i g^{'} Y_{\varphi} B_{\mu} + i g_{Z'} Y_{\varphi} Z'_{\mu},
\end{equation}
where $\sigma ^a$ are the Pauli matrices for $ a \in  \{1, 2, 3\}$. We neglect the mixing between the $Z'$ and the SM gauge bosons.  Note that quark and lepton flavour indices are suppressed, and that the couplings to quark and leptons are flavour diagonal.
The new gauge interaction is anomaly free, as it has the same hypercharge-dependent couplings to fermions as the SM fields~\cite{Allanach:2018vjg}.
Models of $Z'$ bosons and their impact on LHC data have been widely studied; see for example Refs.~\cite{Salvioni:2009mt,Salvioni:2009jp,Langacker:2008yv,Panico:2021vav}.

\begin{table}[H]
\begin{center}
\begin{tabular}{ |r|l| }
 \hline
	Bosonic & $\mathcal{O}_{\varphi D}$, $\mathcal{O}_{\varphi \Box}$, $\mathcal{O}_{\varphi l}^{(1)}$, $\mathcal{O}_{\varphi q}^{(1)}$, $\mathcal{O}_{\varphi e}$, $\mathcal{O}_{\varphi u}$, $\mathcal{O}_{\varphi d}$ \\
	4-fermion $(\bar{L} L)(\bar{L} L)$ & $\mathcal{O}_{ll}$, $\mathcal{O}_{qq}^{(1)}$, $\mathcal{O}_{lq}^{(1)}$ \\
	4-fermion $(\bar{R} R)(\bar{R} R)$ & $\mathcal{O}_{ee}$, $\mathcal{O}_{uu}$, $\mathcal{O}_{dd}$, $\mathcal{O}_{ed}$, $\mathcal{O}_{eu}$, $\mathcal{O}_{ud}^{(1)}$\\
	4-fermion $(\bar{L} L)(\bar{R} R)$ & $\mathcal{O}_{l e}$, $\mathcal{O}_{l d}$, $\mathcal{O}_{l u}$, $\mathcal{O}_{q u}^{(1)}$, $\mathcal{O}_{q d}^{(1)}$\\
 \hline
\end{tabular}
\end{center}
	\caption{\label{tab:Zpops} Warsaw basis operators generated by the $Z'$ model of Eq.~\eqref{eq:Zprime}.}
\end{table}
Tree-level matching of $\mathcal{L}^{Z'}_{\text{UV}}$ to the dim-6 SMEFT produces the Warsaw basis~\cite{deBlas:2017xtg} operators in Table~\ref{tab:Zpops}. The complete SMEFT Lagrangian is given by:
\begin{equation} 
\label{eq:ZprimeEFT}
	\begin{split}
		\mathcal{L}^{Z'}_{\text{SMEFT}} &= \mathcal{L}_{\text{SM}} -\frac{g_{Z'}^{2}}{ M_{Z'}^{2}} \Big( { 2 Y_{\varphi}^{2} \mathcal{O}_{\varphi D} + \frac{1}{2} Y_{\varphi}^{2} \mathcal{O}_{\varphi \Box}}\\
		&\qquad+ Y_{\varphi} Y_{l} \mathcal{O}_{\varphi l}^{(1)} + Y_{\varphi} Y_{q}  \mathcal{O}_{\varphi q}^{(1)}  + Y_{\varphi} Y_{e}  \mathcal{O}_{\varphi e} +  Y_{\varphi} Y_{d}  \mathcal{O}_{\varphi d} + Y_{\varphi} Y_{u}  \mathcal{O}_{\varphi u} \\
		&\qquad+ \frac{1}{2} Y_{l}^{2} \mathcal{O}_{ll} + \frac{1}{2} Y_{q}^{2} \mathcal{O}_{qq}^{(1)} + Y_{q} Y_{l} \mathcal{O}_{lq}^{(1)}\\
		&\qquad+  \frac{1}{2} Y_{e}^{2} \mathcal{O}_{ee} + \frac{1}{2} Y_{u}^{2} \mathcal{O}_{uu} + \frac{1}{2} Y_{d}^{2} \mathcal{O}_{dd} + Y_{e} Y_{d} \mathcal{O}_{ed} + Y_{e} Y_{u} \mathcal{O}_{eu}  + Y_{u} Y_{d} \mathcal{O}_{ud}^{(1)}\\
		&\qquad+ Y_{e} Y_{l} \mathcal{O}_{l e} + Y_{u} Y_{l} \mathcal{O}_{l u} + Y_{d} Y_{l} \mathcal{O}_{l d} + Y_{e} Y_{q} \mathcal{O}_{q e} + Y_{u} Y_{q} \mathcal{O}_{q u}^{(1)} + Y_{d} Y_{q} \mathcal{O}_{q d}^{(1)} \Big) \, .
	\end{split}
\end{equation}
The leading effect of this model on the data entering our analysis is to modify the Drell-Yan and Deep Inelastic Scattering datasets; 
in particular the high-mass neutral current Drell-Yan 
tails~\cite{DYpaper} will be affected.  
The $Z'$ may have an additional impact on top quark and dijet data through four-quark operators such as $\mathcal{O}_{qq}^{(1)}$, however the 
effect is negligible and we do not consider it here.

The effect of the $Z'$ on high-mass Drell-Yan is dominated by the energy-growing four-fermion operators~\cite{Farina:2016rws,Torre:2020aiz,Panico:2021vav}.
By neglecting the subdominant operators involving the Higgs doublet $\varphi$,
we can add the operators of the last three lines of Eq.~\eqref{eq:ZprimeEFT} in the following way:
\begin{equation}
\label{eq:zprimeL}
	\mathcal{L}^{Z'}_{\text{SMEFT}} = \mathcal{L}_{\text{SM}} -\frac{g_{Z'}^{2}}{2 M_{Z'}^{2}} J^{\mu}_Y J_{Y, \mu}, \qquad J^{\mu}_Y = \sum_{\substack{f}} Y_{f} \bar{f} \gamma^{\mu} f \, .
\end{equation}
We can describe the new physics introduced in this type of scenario with the $\hat{Y}$ parameter:
\begin{equation}
	\mathcal{L}^{Z'}_{\text{SMEFT}} = \mathcal{L}_{\text{SM}} -\frac{{g'}^2 \hat{Y}}{2 m^2_W} J^{\mu}_Y J_{Y, \mu}, \qquad \hat{Y} = \frac{g_{Z'}^{2}}{M_{Z'}^{2}} \frac{m^2_W}{{g'}^2} \, .
\end{equation}
The $\hat{Y}$ parameter allows us to write the Lagrangian using SM parameters. We can write the relation between $\hat{Y}$ and the $Z'$ parameters $g_{Z'}$, $M_{Z'}$ as follows,
\begin{equation}
	\frac{g_{Z'}^{2}}{M_{Z'}^{2}} = 4 \sqrt{2} G_{F} \hat{Y} \left( \frac{m_{Z}^{2} - m_{W}^{2}}{m_{W}^{2}}  \right) \, ,
\end{equation}
where we make use of the \{$m_W$, $m_Z$, $G_F$\} electroweak input scheme, and take the following as input parameters:
%
%
%
%
\begin{equation*}
	G_{F} = 1.16639 \times 10^{-5} \text{ GeV}, \qquad m_{W} = 80.352 \text{ GeV}, \qquad m_{Z} = 91.1876 \text{ GeV} \, .
\end{equation*}
%

\begin{figure}[htb]
    \centering
	\includegraphics[width=0.49\linewidth]{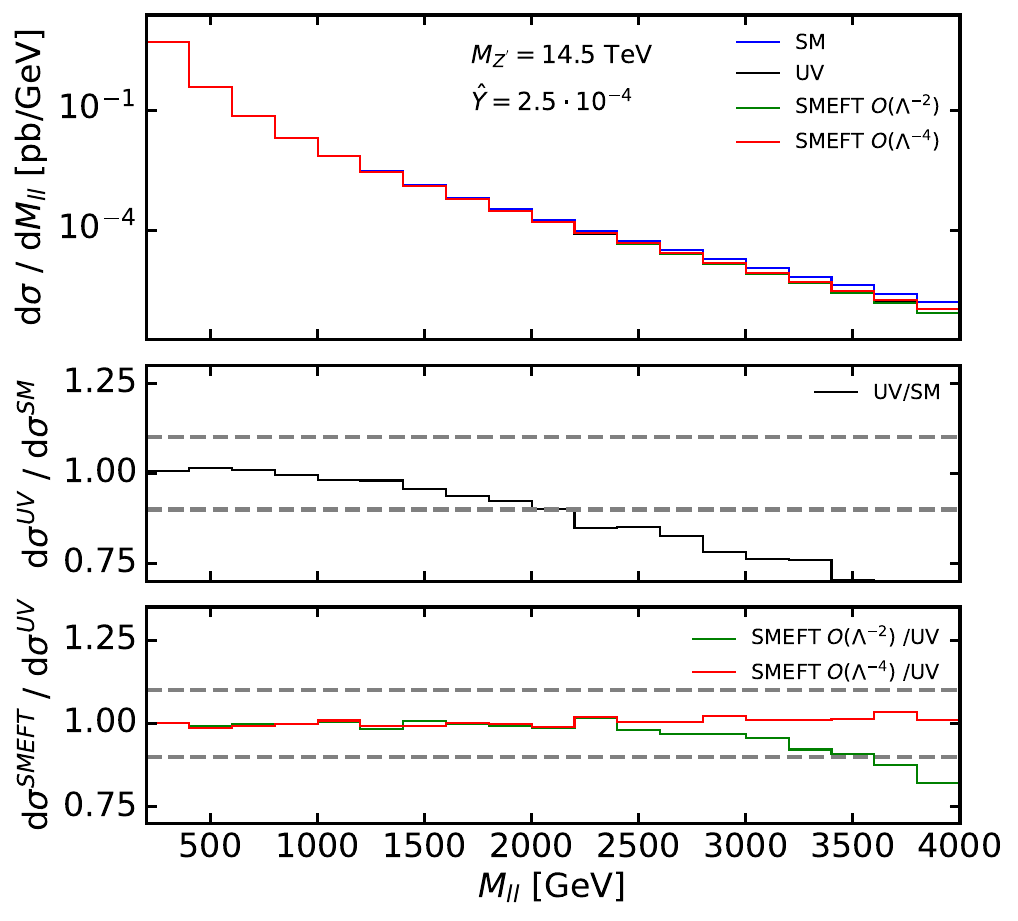}
	\includegraphics[width=0.49\linewidth]{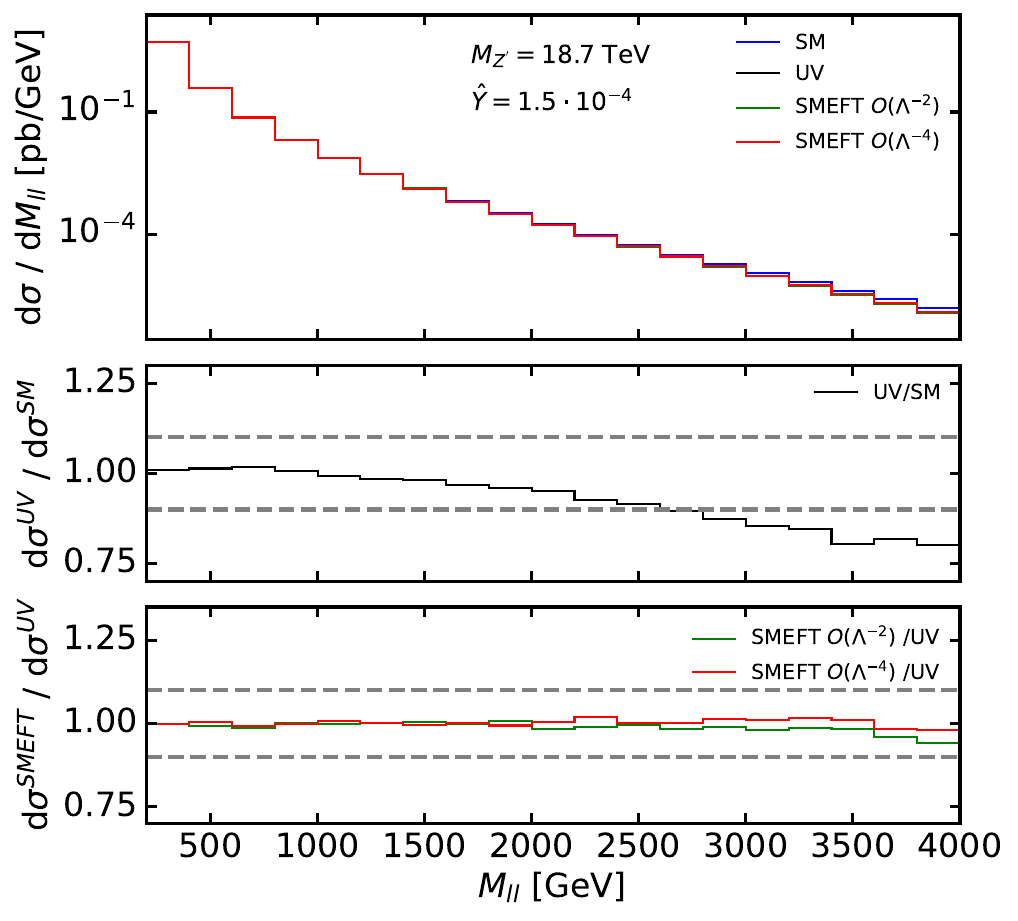}\\
	\includegraphics[width=0.49\linewidth]{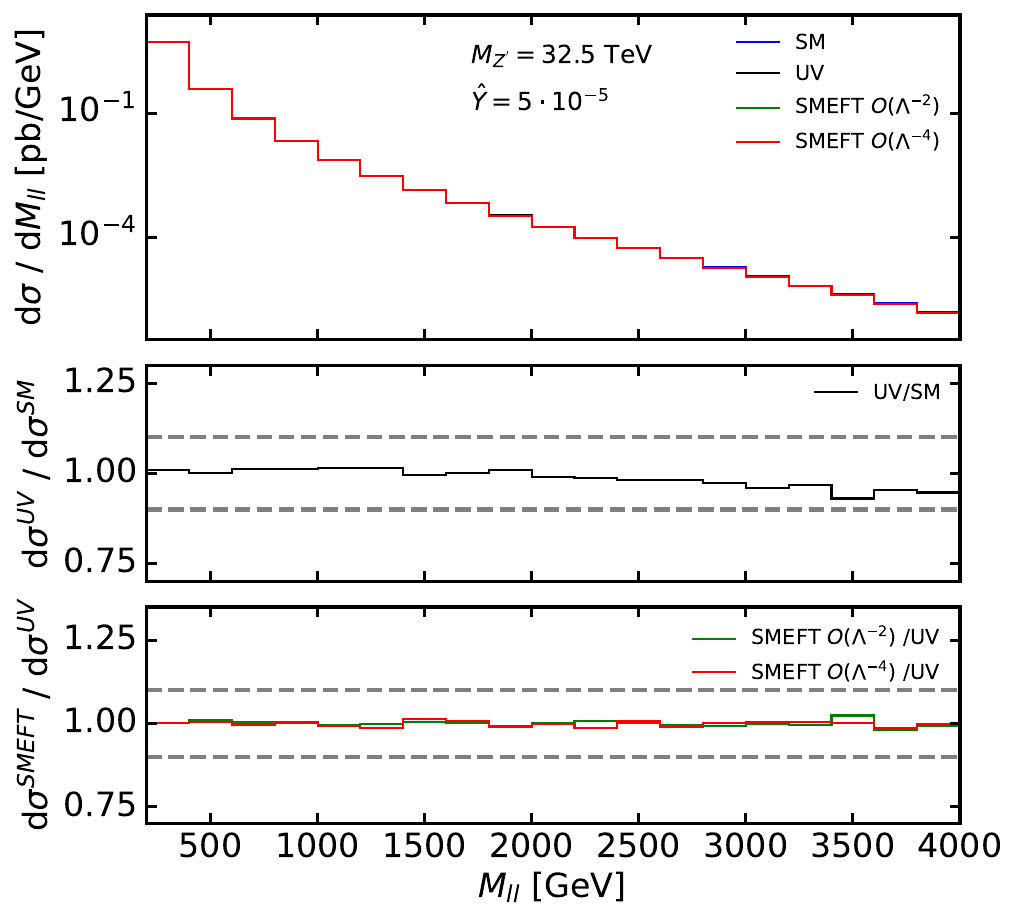}
	\caption{Predictions for neutral current Drell-Yan inclusive cross sections differential in the dilepton invariant mass $m_{\ell\ell}$, with $\ell=(e,\mu$).  
 Factorisation and renormalisation scales are set to $\mu_F=\mu_R=m_{\ell\ell}$, where $m_{\ell\ell}$ is the centre of each invariant mass bin. 
    We show the SM predictions compared to the predictions for a $Z'$ of different masses corresponding to different $\hat{Y}$ values, assuming $g_{Z'}=1$. 
    Top left: mass of $14.5$ TeV, corresponding to $\hat{Y}=25 \cdot 10^{-5}$. Top right: mass of $18.7$ TeV, corresponding to $\hat{Y}=15 \cdot 10^{-5}$. 
    Bottom: mass of $32.5$ TeV, corresponding to $\hat{Y}=5 \cdot 10^{-5}$.}
	\label{fig:Zp_DY_Comp}
\end{figure}

In Fig.~\ref{fig:Zp_DY_Comp} we compare the predictions of the UV-complete $Z'$ model and the corresponding EFT parametrisation
for the differential cross section of the Drell-Yan process $p p \rightarrow \ell^+ \ell^-$, where $\ell=e,\,\mu$.
The predictions are computed assuming $\sqrt{s}=14$ TeV, using \madgraph{} and setting $\mu_F=\mu_R=m_{\ell\ell}$, where $m_{\ell\ell}$ is the centre of each invariant mass bin.  
We compare the SM, the full UV-complete model, the linear-only $\mathcal{O}(\Lambda^{-2})$ EFT and linear-plus-quadratic $\mathcal{O}(\Lambda^{-4})$ EFT predictions 
assuming $g_{Z'} = 1$, for 
three benchmark values of the $Z'$ mass: $M_{Z'} = 14.5$ TeV, $M_{Z'} = 18.7$ TeV and $M_{Z'} = 32.5$ TeV.
Such large values of $M_{Z'}$ are clearly well beyond the possible direct reach of direct $Z'$ searches at ATLAS and CMS~\cite{Workman:2022ynf}. In the top panel we plot the differential cross section with respect to the dilepton invariant mass, 
in the middle panel we plot the ratio of the full $Z'$ model to the SM, and in the
lower panel we plot the ratio of the EFT to the full $Z'$ model predictions. 

First, we observe that the UV model predictions differ from the SM predictions by more than $20\%$ for the smaller masses of the $Z'$. 
In the lower panels, we observe the point at which the linear EFT corrections fail to describe the UV physics, and the quadratic EFT contributions begin to become non-negligible; in the same way, the quadratic dim-6 EFT description starts failing when the dim-8 SMEFT operators become important~\cite{Boughezal:2021tih}.

As displayed in the top right panel of Fig.~\ref{fig:Zp_DY_Comp}, for $M_{Z'}=14.5$ TeV the linear EFT corrections start failing to describe the UV model at higher energies. For $M_{Z'}=18.7$ TeV and heavier the linear EFT describes the UV physics faithfully for dilepton invariant masses up to 4 TeV. The deviations from new physics are over $20\%$ for $M_{Z'}=18.7$ and $M_{Z'}=14.5$ TeV.
We will implement our PDF ``contamination'' by working in the area of the parameter space where the linear EFT describes the UV physics faithfully and where the UV extension to the SM impacts significantly the observables.

Finally, it is worth noting that we have compared those SMEFT predictions involving only four-fermion operators with 
those obtained 
additionally including the SMEFT operators containing the Higgs doublet, such as $\mathcal{O}_{\varphi l}^{(1)}$ and $\mathcal{O}_{\varphi e}$. We find that these operators have no visible impact on the observables. 
They are only competitive with the four-fermion corrections at lower energies (around 500 GeV),
and at this scale the new physics has very little impact on the SM predictions. 
When the influence of the heavy new physics starts to be noticeable at higher invariant mass, the four-fermion operators, whose impact grows faster with energy, completely dominate the SMEFT corrections. Thus, we have verified that our parametrisation
in terms of the $\hat{Y}$ parameter reflects the UV physics to a very good degree.

\subsection{Scenario II: A flavour universal $W'$ model}
\label{subsec:model-wprime}

We now consider a new $SU(2)_L$ triplet field ${W'}^{a, \mu}$, where $a \in \{1,2,3\}$ denotes an $SU(2)_L$ index. We add a mass $M_{W'}$, and denote the $W'$ coupling coefficient by $g_{W'}$. Similarly to what happens with the SM $W$ field, the $W'^1$ and $W'^2$ components mix to form the $W'^+$ and $W'^-$ particles, while the $W'^3$ component gives a neutral boson similar to the $Z'$, but which only couples to left-handed fields. The 
model is described by the following Lagrangian:
\begin{equation} \label{eq:Wprime}
	\begin{split}
		\mathcal{L}^{W'}_{\text{UV}} &= \mathcal{L}_{\text{SM}} - \frac{1}{4} {W'}^{a}_{\mu \nu} {W'}^{a, \mu \nu} + \frac{1}{2} M_{W'}^{2} {W'}_{\mu}^{a} {W'}^{a, \mu}\\
		&\qquad - g_{W'} {W'}^{a,\mu} \sum_{\substack{f_L}} \bar{f}_L T^{a} \gamma^{\mu} f_L
		 - g_{W'} ({W'}^{a, \mu} \varphi^{\dagger} T^{a} i D_{\mu} \varphi + \textrm{h.c.}) \, ,
	\end{split}
\end{equation}
where we sum over the left-handed fermions: $f_L \in \{Q_L^i, \ell_L^i, \,{\rm for}\,\,\,i\in\{1,2,3\}\}$. The $SU(2)_L$ generators are given by $T^a = \frac{1}{2} \sigma ^a$ where $\sigma ^a$ are the Pauli matrices. The kinetic term is given by $W^{'a}_{\mu \nu} = \partial_{\mu} {W'}^{a}_{\nu} - \partial_{\nu} {W'}^{a}_{\mu}- ig_{W'} [{W'}^{a}_{\mu}, {W'}^{a}_{\nu} ]$. The covariant derivative is given by
\begin{equation}
    D_{\mu} = \partial_{\mu} + \frac{1}{2} ig \sigma^{a} W_{\mu}^{a} + i g^{'} Y_{\varphi} B_{\mu} + \frac{1}{2} ig_{W'} \sigma^{a} {W'}_{\mu}^{a}.
\end{equation}
As above, we neglect the mixing with the SM gauge fields.  

Tree-level matching of $\mathcal{L}^{W'}_{\text{UV}}$ to the dim-6 SMEFT produces the Warsaw basis operators in Table~\ref{tab:Wpops}, where we
have distinguished the operators
$(O_{ll})_{ij} = (l_{i} \gamma^{\mu} l_{i})(l_{j} \gamma^{\mu} l_{j})$ and $(O_{ll}')_{ij} = (l_{i} \gamma^{\mu} l_{j})(l_{j} \gamma^{\mu} l_{i})$~\cite{Brivio:2017vri}. 
\begin{table}[h!]
\begin{center}
\begin{tabular}{ |r|l| }
 \hline
	Bosonic &  $\mathcal{O}_{\varphi \Box}$, $\mathcal{O}_{\varphi}$, $\mathcal{O}_{\varphi l}^{(3)}$, $\mathcal{O}_{\varphi q}^{(3)}$\\
	Yukawa & $\mathcal{O}_{e \varphi}$, $\mathcal{O}_{d \varphi}$, $\mathcal{O}_{u \varphi}$ \\
	4-fermion $(\bar{L} L)(\bar{L} L)$ & $\mathcal{O}_{ll}$,$\mathcal{O}_{ll}^{'}$, $\mathcal{O}_{qq}^{(3)}$, $\mathcal{O}_{lq}^{(3)}$ \\
 \hline
\end{tabular}
\end{center}
        \caption{\label{tab:Wpops} Warsaw basis operators generated by the $W'$ model of Eq.~\eqref{eq:Wprime}.}
\end{table}
The complete SMEFT Lagrangian is given by~\cite{deBlas:2017xtg}:
\begin{equation} 
\label{eq:WprimeEFT}
        \begin{split}
		\mathcal{L}^{W'}_{\text{SMEFT}} &= \mathcal{L}_{\text{SM}} - \frac{g_{W}^{2}}{M_{W}^{2}} \Big(- \frac{1}{8} \mathcal{O}_{ll} + \frac{1}{4} \mathcal{O}_{ll}^{'} + \frac{1}{8} \mathcal{O}_{qq}^{(3)} + \frac{1}{4} \mathcal{O}_{lq}^{(3)}\\
		&\qquad+ \lambda_{\varphi} \mathcal{O}_{\varphi} + \frac{3}{8} \mathcal{O}_{\varphi \Box}+ \frac{1}{4} \mathcal{O}_{\varphi q}^{(3)}  + \frac{1}{4} \mathcal{O}_{\varphi l}^{(3)}     \\
		&\qquad + \frac{1}{4} (y_{e})_{ij} (\mathcal{O}_{e \varphi})_{ij} + \frac{1}{4} (y_{u})_{ij} (\mathcal{O}_{u \varphi})_{ij} + \frac{1}{4} (y_{d})_{ij} (\mathcal{O}_{d \varphi})_{ij} \Big) \, . 
	\end{split}
\end{equation}
As in the case of the $Z'$, 
the leading effect of this model on our dataset 
is to modify the Drell-Yan and Deep Inelastic Scattering datasets; however, this time both charged current and neutral current Drell-Yan will
be affected.  
This impact is dominated by the four-fermion
interactions in the first line of Eq.~\eqref{eq:WprimeEFT}, which sum to:
\begin{equation}
\label{eq:wprimeL}
	\mathcal{L}^{W'}_{\text{SMEFT}} = \mathcal{L}_{\text{SM}} -\frac{g_{W'}^{2}}{2 M^2_{W'}} J^{a, \mu}_L J^a_{L, \mu}, \qquad J^{a, \mu}_L = \sum_{\substack{f_L}} \bar{f_L} T^a \gamma^{\mu} f_L \, .
\end{equation}
We can describe the new physics introduced in this type of scenario with the $\hat{W}$ parameter:
\begin{equation}
	\mathcal{L}^{W'}_{\text{SMEFT}} = \mathcal{L}_{\text{SM}} -\frac{g^{2} \hat{W}}{2 m^2_W} J^{a, \mu}_L J^a_{L, \mu}, \qquad \hat{W} = \frac{g_{W'}^{2}}{g^{2}} \frac{m_{W}^{2}}{M_{W'}^{2}} \, .
\end{equation}
Using Fermi's constant, we can write the relation between the UV parameters and $\hat{W}$ in the following way:
\begin{equation}
	\frac{g_{W'}^{2}}{M_{W'}^{2}} = 4 \sqrt{2} G_{F} \hat{W} \, .
\end{equation}
Again, by fixing $g_{W'}=1$, each $M_{W'}$ can be associated to a value of $\hat{W}$.

\begin{figure}[htb]
    \centering
	\includegraphics[width=0.49\linewidth]{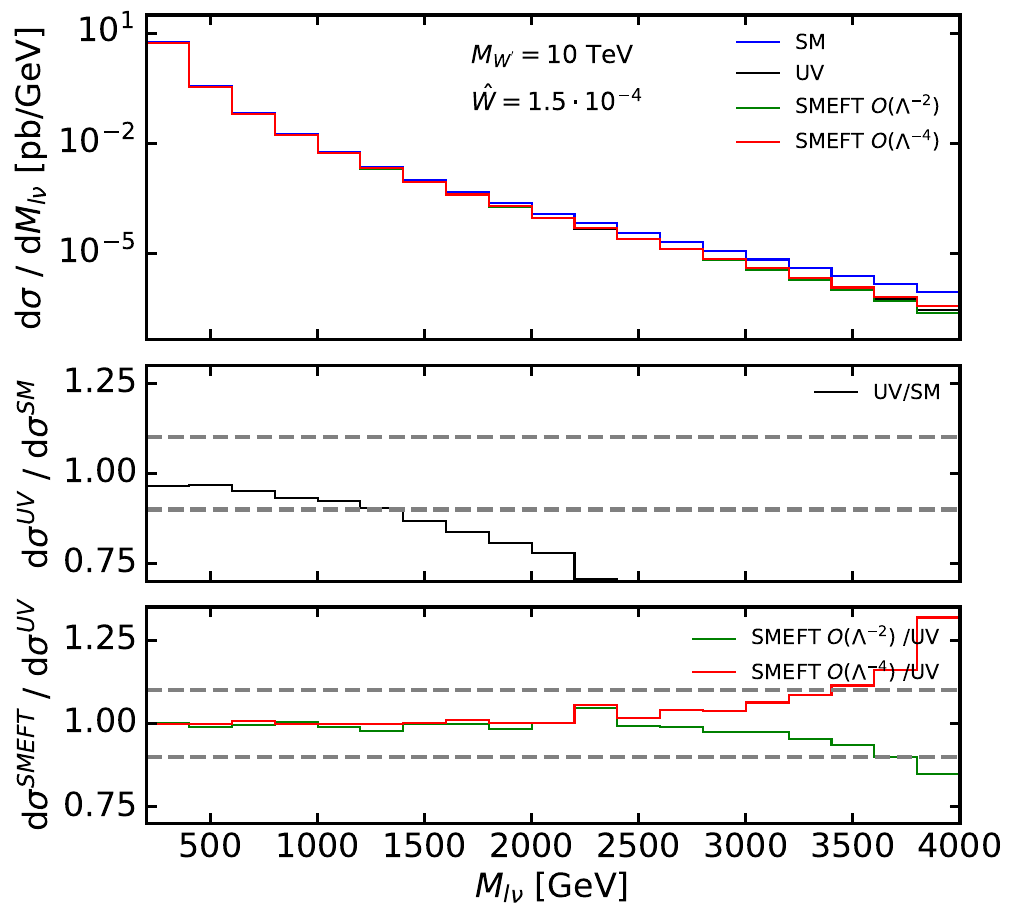}
	\includegraphics[width=0.49\linewidth]{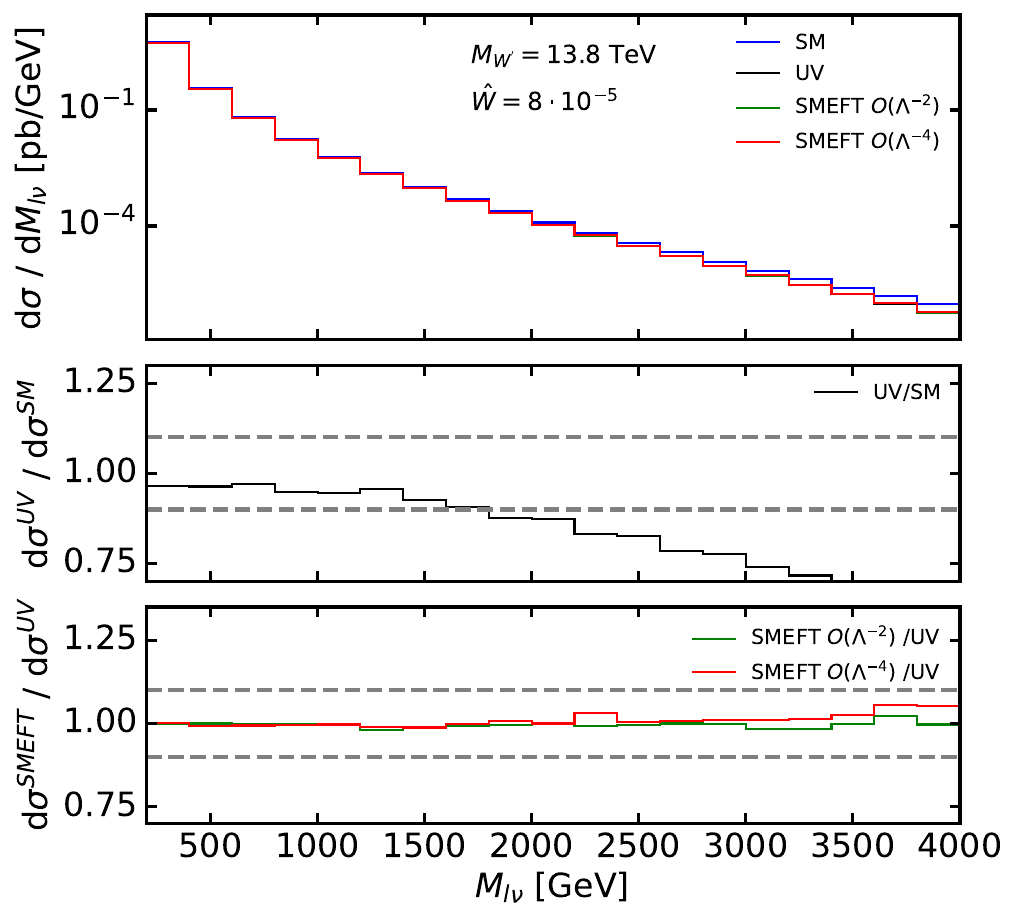}\\
	\includegraphics[width=0.49\linewidth]{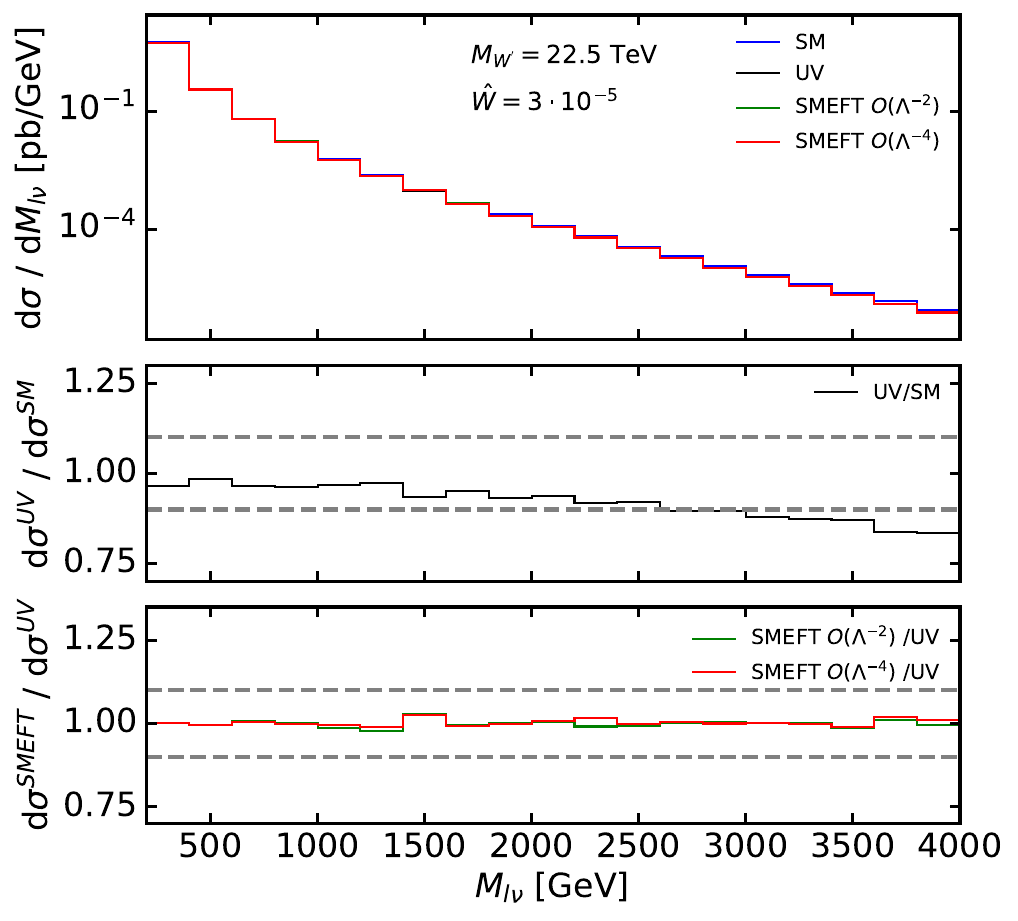}
	\caption{Predictions for charged current Drell-Yan ($pp \rightarrow l^- \bar{\nu}$ inclusive cross sections differential in $M_{\ell\nu}$, with $\ell=(e,\mu$). 
 Factorisation and renormalisation scales are set to $\mu_F=\mu_R=M_T$, where $M_T$ is the transverse mass of each bin. 
 We show the SM predictions compared to the predictions for a $W'$ of different masses corresponding to different $\hat{W}$ values, assuming $g_{W'}=1$. 
 Top left: mass of $10$ TeV, corresponding to $\hat{W}=15 \cdot 10^{-5}$. Top right: mass of $13.8$ TeV, corresponding to $\hat{W}=8 \cdot 10^{-5}$. Bottom: mass of $22.5$ TeV, corresponding to 
    $\hat{W}=3 \cdot 10^{-5}$.}
 	\label{fig:Wp_DY_Comp}
\end{figure}

In Fig.~\ref{fig:Wp_DY_Comp}
we perform a comparison of the UV-complete $W'$ model and the EFT predictions.
We assess the differences between the EFT parametrisation and the UV model description by studying the charged current Drell-Yan process, $pp \rightarrow \ell^- \bar{\nu}$,
assuming $g_{W'} = 1$, at
three benchmark values of the $W'$ mass: $M_{W'} = 10$ TeV, $M_{W'} = 13.8$ TeV and $M_{W'} = 22.5$ TeV.
A similar comparison could be made in neutral current Drell-Yan; however, we expect the dominant effect of the $W'$ to occur in charged current Drell-Yan,
and therefore this process will provide the leading sensitivity to differences between the UV model and the EFT parametrisation.
As displayed in the top right panel of Fig.~\ref{fig:Wp_DY_Comp}, for $M_{W'}=10$ TeV the linear EFT corrections start failing to describe the UV model at higher energies. For $M_{W'}=13.8$ TeV and heavier the linear EFT describes the UV physics faithfully for dilepton invariant masses up to 4 TeV. The deviations from new physics are over $20\%$ for $M_{W'}=13.8$ and $M_{W'}=10$ TeV.
We will again implement our PDF ``contamination'' by working in the area of the parameter space where the linear EFT describes the UV physics faithfully and where the UV extension to the SM impacts the observables noticeably.
Finally, our analysis also reveals that the SMEFT operators involving a Higgs doublet $\varphi$ have no impact on the predictions, for the same reason we presented in the $Z'$ case. This means that this model built with the $\hat{W}$ parameters also describes the UV physics faithfully.


\section{Contamination from Drell-Yan large invariant-mass
  distributions}
  \label{sec:dy}

In this Section, after presenting the analysis settings in terms of
theory predictions and data, we explore in detail the effects of new
heavy vector  bosons in the high-mass Drell-Yan distribution tails and how fitting
this data assuming the SM would modify the data-theory agreement and
the PDFs. We will see that in some scenarios the PDFs manage to mimic
the effects of new physics in the high tails without deteriorating the
data-theory agreement in any visible way. In this cases PDFs can
actually ``fit away'' the effects of new physics. In the following
sections we will explore the phenomenological consequences of using
such ``contaminated'' PDF sets and we will see that they might
significantly distort the interpretation of HL-LHC measurements. Subsequently, in Sect.~\ref{sec:solution},
we conclude by devising strategies to spot the contamination by
including in a PDF fit complementary observables that highlight the
incompatibility of the high-mass Drell-Yan tails with the bulk of the
data.

\subsection{Analysis settings}

For this analysis we generate a set of artificial  Monte Carlo data, which comprises 4771 data points,
spanning a broad range of processes.
The Monte Carlo data that we generate are either taken from current Run I and
Run II LHC data or from HL-LHC projections. The uncertainties in the
former category are more realistic, as they are taken from the
experimental papers (we remind the reader that the central
measurement is generated by the underlying law of Nature according to Eq.~\eqref{eq:observed_data}), 
while the uncertainties on projected HL-LHC data are generated according to specific projections. 

As far as the current data is concerned, we generate MC data that
cover all the observables included in the \nnpdf analysis~\cite{Ball:2021leu}. 
In particular, in the category of Drell-Yan, we
include the NC Drell-Yan that follows the kinematic
distributions and the errors analysed by ATLAS at \(7\) and \(8\) TeV \cite{ATLAS:2013xny, ATLAS:2016gic}, and
CMS at \(7\), \(8\), and \(13\) TeV \cite{CMS:2013zfg, CMS:2014jea,
  CMS:2018mdl}. These LHC measurements are not only used to constrain the PDFs, but are also
sufficiently sensitive to the BSM scenarios considered in Sect.~\ref{sec:scenarios}.

For the HL-LHC pseudo-data, we include the high-mass Drell-Yan projections that we produced in
Ref.~\cite{DYpaper}, inspired by the HL-LHC projections studied
in Ref.~\cite{AbdulKhalek:2018rok}. The invariant mass distribution
projections are generated at \(\sqrt{s} = 14\) TeV, assuming an
integrated luminosity of \(\mathcal{L} = 6 \text{ ab}^{-1}\) ($3 \text{ ab}^{-1}$ collected by ATLAS and  \(3
\text{ ab}^{-1}\) by CMS). 
Both in the case of NC and CC Drell-Yan cross sections, the MC data
were generated using the {\tt MadGraph5\_aMCatNLO} NLO Monte Carlo event
generator~\cite{Frederix:2018nkq} with additional $K$-factors to include the NNLO QCD and
NLO EW corrections. The MC data consist of four datasets
(associated with NC/CC distributions with muons/electrons in the final
state), each comprising 16 bins in the $m_{ll}$
invariant mass distribution or transverse mass $m_T$ distributions
with both $m_{ll}$ and $m_T$ greater than 500 GeV , with the highest energy bins reaching $m_{ll}=4$
TeV ($m_T=3.5$ TeV) for NC (CC) data.
The rationale behind the choice of number of bins and the width of each bin was outlined 
in Ref.~\cite{DYpaper}, and stemmed from the requirement that the expected number of events
per bin was big enough to ensure the applicability of Gaussian
statistics. The choice of binning for the $m_{ll}$ ($m_T$) distribution at the
HL-LHC is displayed in Fig.~5.1 of Ref.~\cite{DYpaper}.

\begin{figure}[t!]
  \centering
  \includegraphics[width=\textwidth]{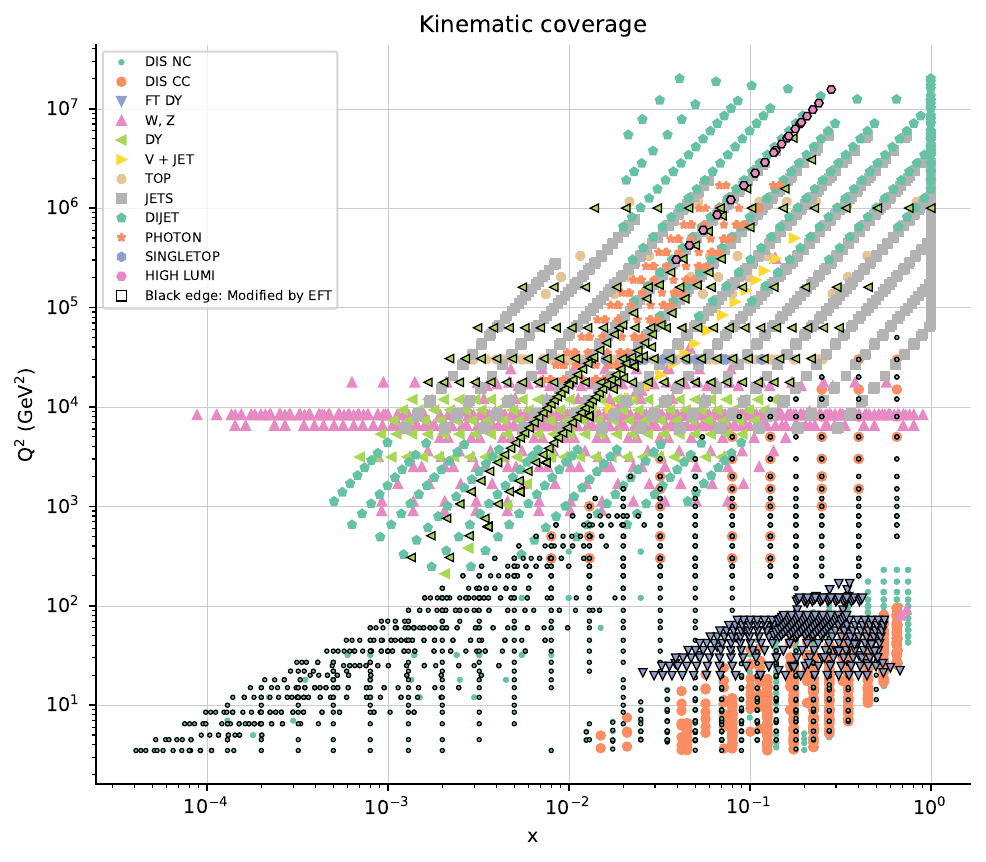}
  \caption{Kinematic coverage of data points included in the fit. The EFT corrections
  for this study have been computed for the points which are highlighted with a black
  edge. The values of $x$ have been computed using a linear order approximation. \label{fig:xq2}}
\end{figure}

The kinematic coverage of the data points used in
this study are shown in Fig.~\ref{fig:xq2}. The points are shown
in \((x,Q^2)\) space with the data points that are modified by the EFT operators
highlighted with a black border; such points thus constrain the Wilson
coefficients as well as the PDFs. We note that, while DIS theory predictions are
modified by the operators we consider in the two benchmark scenarios,
the change in the HERA DIS cross sections upon the variation of the Wilson
coefficients under consideration is minimal, as is explicitly assessed in Ref.~\cite{DYpaper}.

In what follows, we will assess the impact of the injection of NP in the data on the 
fitted PDFs by looking at the integrated luminosity for the parton pair $i,j$, which is defined as:
\begin{equation}
\label{eq:integrated_luminosity}
  {\cal L}_{ij}(m_X,\sqrt{s}) = \frac{1}{s}\int\limits_{-y}^{y}\,d\tilde{y}\,
  \left[
  f_{i}\left(\frac{m_X}{\sqrt{s}}e^{\tilde{y}},m_X\right)\, 
  f_j\left(\frac{m_X}{\sqrt{s}}e^{-\tilde{y}},m_X\right)
  +
  (i \leftrightarrow j)
  \right]
  ,
  \end{equation}
 where $f_i \equiv f_i(x,Q)$ is the PDF corresponding to the parton flavour $i$ 
 , and the integration limits are defined by:
 \begin{equation}
 \label{eq:rapidity}
  y=\ln\left(\frac{\sqrt{s}}{m_X}\right).
  \end{equation}
In particular we will focus on the luminosities that are most constrained by the Neutral Current (NC) and Charged Current (CC) Drell-Yan data respectively, namely   
  \begin{align}
  \label{eq:wzlumi}
    {\cal L}^{\rm NC}(m_X,\sqrt{s}) &= {\cal L}_{u\bar{u}}(m_X,\sqrt{s}) + {\cal L}_{d\bar{d}}(m_X,\sqrt{s}),\\[1.5ex]
      {\cal L}^{\rm CC}(m_X,\sqrt{s}) &= {\cal L}_{u\bar{d}}(m_X,\sqrt{s}) + {\cal L}_{d\bar{u}}(m_X,\sqrt{s}).
    \end{align}

\subsection{Effects of new heavy bosons in PDF fits}
\label{sec:contaminated_pdfs}

\begin{figure}[b!]
\centering
  \includegraphics[width=0.6\linewidth]{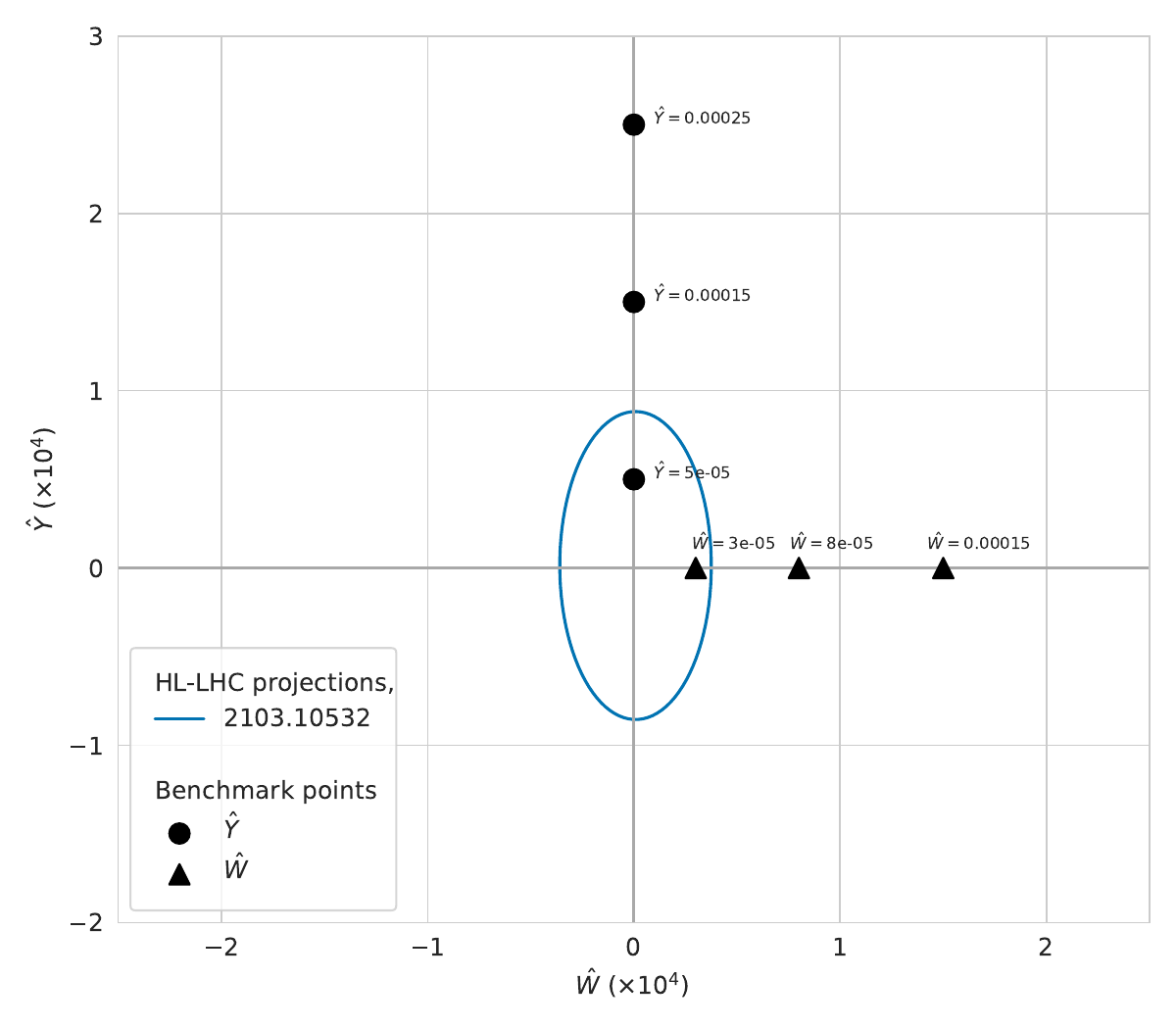}
	\caption{\label{fig:bps} Benchmark $\hat{Y}$ and $\hat{W}$ points explored in this analysis compared to 
	the constraints at 95\% CL as given by the analysis of fully-differential Drell-Yan
 projections given in Ref.~\cite{Panico:2021vav}.}
      \end{figure}

In Fig.~\ref{fig:bps} we display the benchmark points that we
consider, corresponding to the two scenarios described in
Sect.~\ref{sec:scenarios}. Namely, the points along the vertical axis
correspond to the flavour-universal $Z'$ model (Scenario I), while
those along the horizontal axis correspond to the flavour-universal
$W'$ model (Scenario II). The benchmark points are compared to projected constraints
from the HL-LHC.  In particular, we consider the most up-to-date constraints 
from the analysis of a fully-differential Drell-Yan projection in the HL-LHC regime, as 
given by Ref.~\cite{Panico:2021vav}.
%
%

In order to estimate the effect of a heavy $Z'$ ($W'$) in Nature and the ability of PDFs
to fit it away, we inject new physics in the artificial Monte Carlo data by setting $\hat{Y}\neq
0$ ($\hat{W}\neq 0$) to the values that we consider in our benchmark (see
Fig.~\ref{fig:bps}) and we measure the effect on the fit
quality and on the PDFs. To assess the fit quality, we generate
{\it L1 pseudodata}, as in Eq.~\eqref{eq:observed_data}, according to
1000 variations of the random seed $k$ and compare the distributions of the corresponding $\chi^2$-statistic per data point, 
$\chi^{2(k)}/n_{\rm dat}$, and the number of standard deviations from the expected $\chi^2$, $n^{(k)}_{\sigma}$, across the 1000
random seed variations for the
baseline and the 3 benchmark values in each of the two
scenarios.
If the distributions shift above the critical levels defined in
Sect.~\ref{sec:methodology}, then the PDFs have {\it not} been able to absorb
the effects of new physics and the datasets that display a bad
data-theory agreement would be excluded from a PDF fit. If instead the
distributions remain statistically compatible with those of the baseline
PDF fit, then the PDFs have been able to absorb new physics.

Note that in this exercise the distribution across random seed values is calculated by keeping the PDF fixed
to the value obtained with a given random seed, while if we were
refitting them for each random seed, we would obtain slightly
different PDFs. A comparison at the level of PDFs and parton luminosities is then performed to
assess whether the absorption of new physics shifts them significantly with
respect to the baseline PDFs. We have verified that the effect is negligible and
does not modify the results. A more detailed account of the contaminated PDF's random
seed dependence is given in App.~\ref{app:random}. 
The goal of this exercise is to estimate the maximum strength of new physics
effects beyond which PDFs are no longer able to absorb the effect, and
subsequently assess whether the effect is significant or not.\\

\subsubsection*{(i) Scenario I}

\begin{figure}[t!]
  \includegraphics[width=0.48\linewidth, page=1]{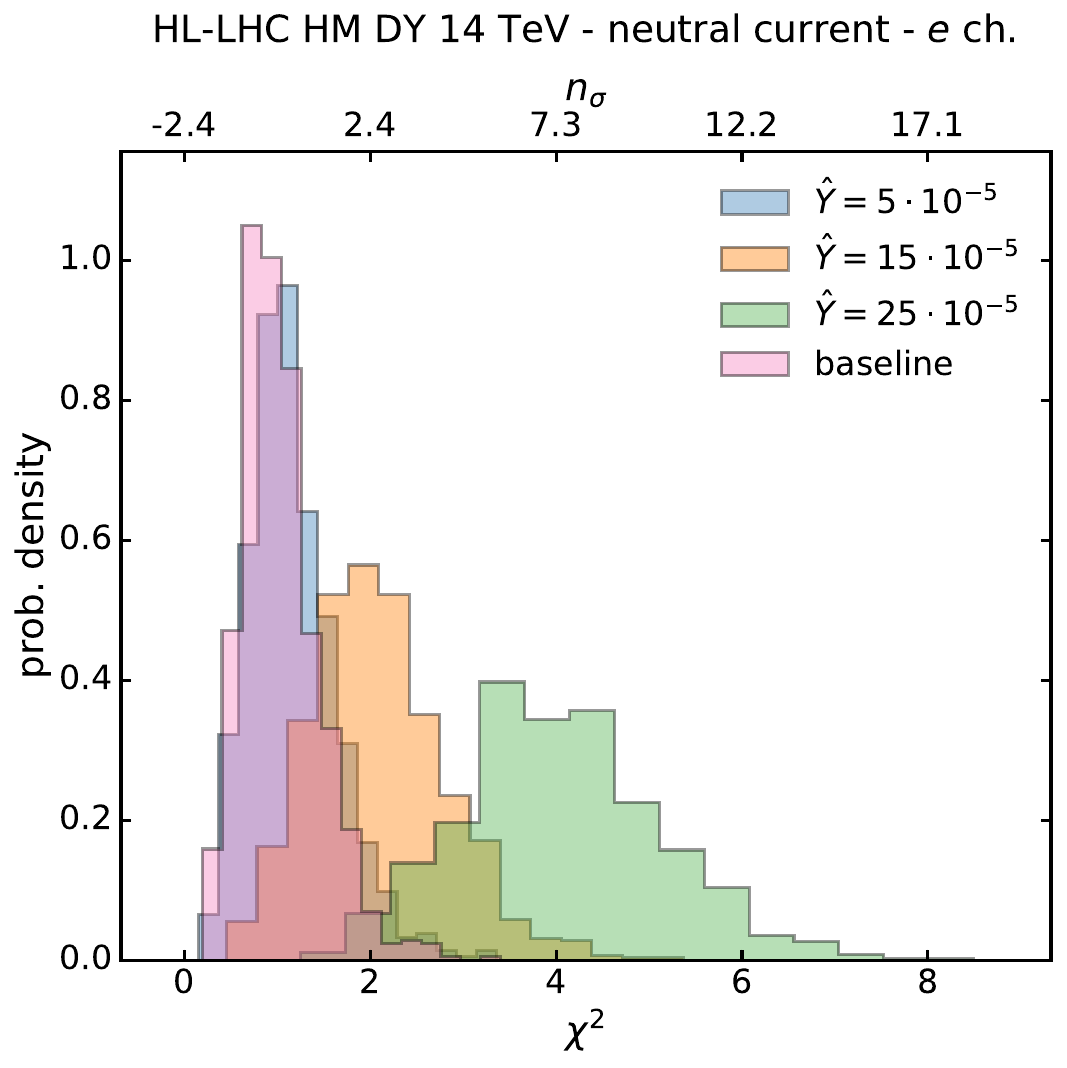}
  \includegraphics[width=0.48\linewidth, page=3]{Figures/chi2_nsigma_Y.pdf}\\
  \includegraphics[width=0.48\linewidth, page=6]{Figures/chi2_nsigma_Y.pdf}
  \includegraphics[width=0.48\linewidth, page=10]{Figures/chi2_nsigma_Y.pdf}\\
  \caption{Distribution of $\chi^2$ and $n_\sigma$ for selected datasets in the $\hat{Y}$ contamination scenarios.}
\label{fig:Y_chi2_nsigma_dist}
\end{figure}

\noindent In the case of the flavour-universal $Z'$ model, we inject three non-zero
values of $\hat{Y}=5\cdot 10^{-5},\, 15\cdot 10^{-5},\, 25\cdot
10^{-5}$. In Fig.~\ref{fig:Y_chi2_nsigma_dist} we display 
the $\chi^{2(k)}$ and $n_{\sigma}^{(k)}$ distributions
across the 1000 $k$ random seeds for a selection of the datasets
included in each of the fits. In particular, we display the datasets in
which a shift occurs either because of the direct effect of the
non-zero Wilson coefficients in the partonic cross sections (such as
the high-mass Drell-Yan in the HL-LHC projections) or
because of the indirect effect of the change of PDFs, which can alter the behaviour of other datasets
that probe the large-$x$ light quark and antiquark distributions. Full
details about the trend in the fit quality for all datasets are
given in App.~\ref{app:fit}.

As far as the quality of the fit is concerned, we observe that, for
$\hat{Y}=5\cdot 10^{-5}$, the global fit is
    equivalent to the SM baseline, while as $\hat{Y}$ is increased to
    $15\cdot 10^{-5}$ the quality of the fit deteriorates. This is due
    mostly to a worse description of the HL-LHC neutral current data (top left panel in Fig.~\ref{fig:Y_chi2_nsigma_dist}) data, while the other datasets remain
    roughly equivalent. This is an indication that there is a bulk of data points 
    in the global dataset that constrains the ${\cal L}^{\rm NC}$ luminosity behaviour at high-$x$ and 
    does not allow the PDF to shift and accommodate the HL-LHC Drell-Yan NC data.
    According to the selection criteria outlined in
    Sect.~\ref{subsec:postfit}, the deterioration of both the $\chi^2$ and the $n_{\sigma}$ indicators
    would single out the high-mass Drell-Yan data and indicate that they
    are incompatible with the rest of the data included in the PDF fit.
    As a consequence, they would be excluded from the fit and no contamination would occur. 
    Hence, in this scenario, $\hat{Y}=15\cdot10^{-5}$ falls in the interval of NP values in which the disagreement in
    the data metrics would flag the incompatibility of the high-mass Drell-Yan
    tails with the rest of the datasets.  
%

\begin{figure}[t!]
  \includegraphics[width=0.49\linewidth]{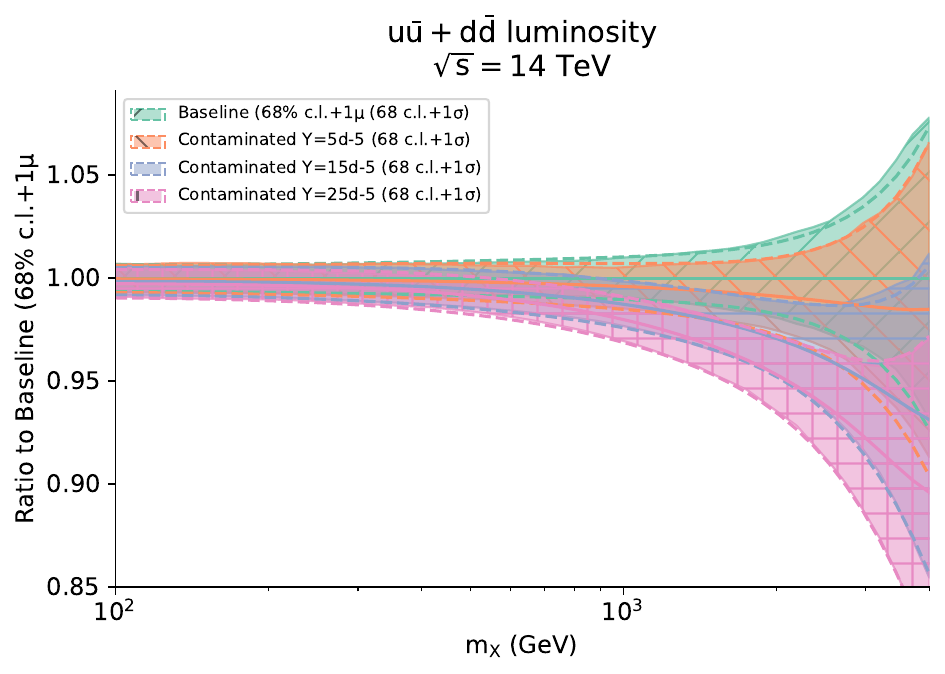}
  \includegraphics[width=0.49\linewidth]{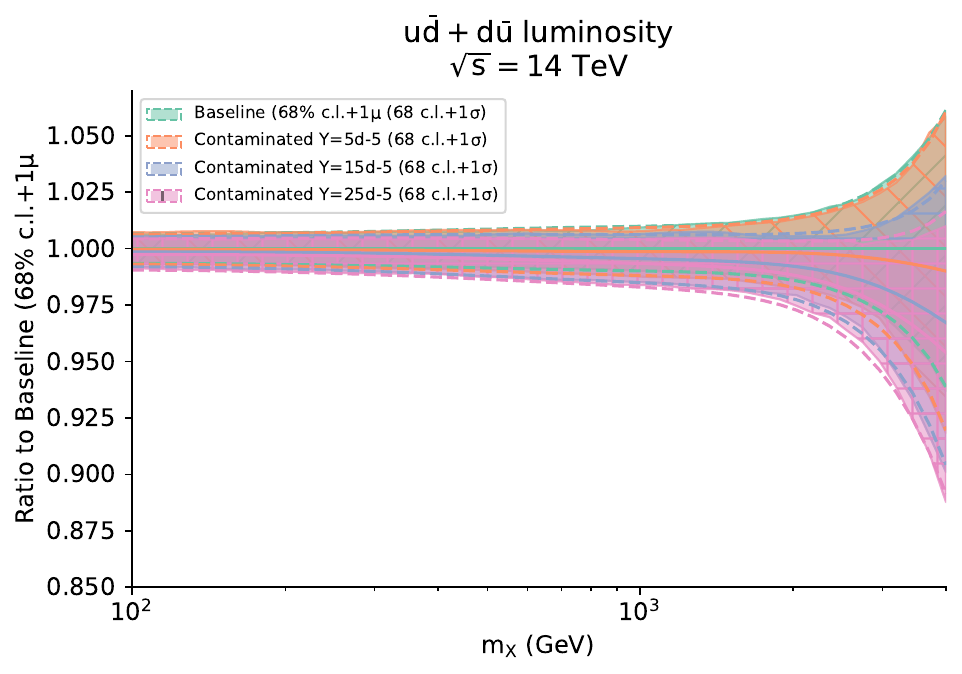}
	\caption{Contaminated versus baseline ${\cal L^{\rm NC}}$ and ${\cal L^{\rm CC}}$ (defined in Eq.~\eqref{eq:wzlumi},
        at $\sqrt{s}=$ 14 TeV in the central rapidity region. The results are normalised to the baseline SM luminosities and the 68\% C.L. band 
        is displayed. Contaminated PDFs have been obtained by fitting the MC data in which $\hat{Y}=5\cdot 10^{-5}$ (orange line),
        $\hat{Y}=15\cdot 10^{-5}$ (blue line) and $\hat{Y}=25\cdot 10^{-5}$ (pink line) has been injected.}
	\label{fig:Ylumi}
      \end{figure}

We now want to check whether, for such values, there is any significant
shift in the relevant NC and CC parton luminosities at the HL-LHC centre-of-mass energy of $\sqrt{s}=14$ TeV. They are displayed in 
Fig.~\ref{fig:Ylumi}. 
We observe that in general the PDFs do not manage to shift much to accommodate the $Z'$ induced contamination. The plots of the individual PDFs are displayed in App.~\ref{app:pdfs}. 
In general the CC luminosity remains compatible with the baseline SM one up to large values of $\hat{Y}$, while, 
as soon as the NC luminosity manages to shift beyond the 1$\sigma$ level, the fit quality of the NC high-mass data 
deteriorates. For the maximum value of new physics contamination that the PDFs can absorb in this scenario, $Y=5\cdot 10^{-5}$ (corresponding to a $Z'$ mass above 30 TeV), 
the parton luminosity shift is contained within the baseline 1$\sigma$ error bar. 
Overall, we see that there is a certain sturdiness in the fit, such that even in the presence of big $\hat{Y}$ values, the parton luminosity does not deviate much from the underlying law.

\subsubsection*{(ii) Scenario II}
\begin{figure}[t!]
  \includegraphics[width=0.48\linewidth, page=1]{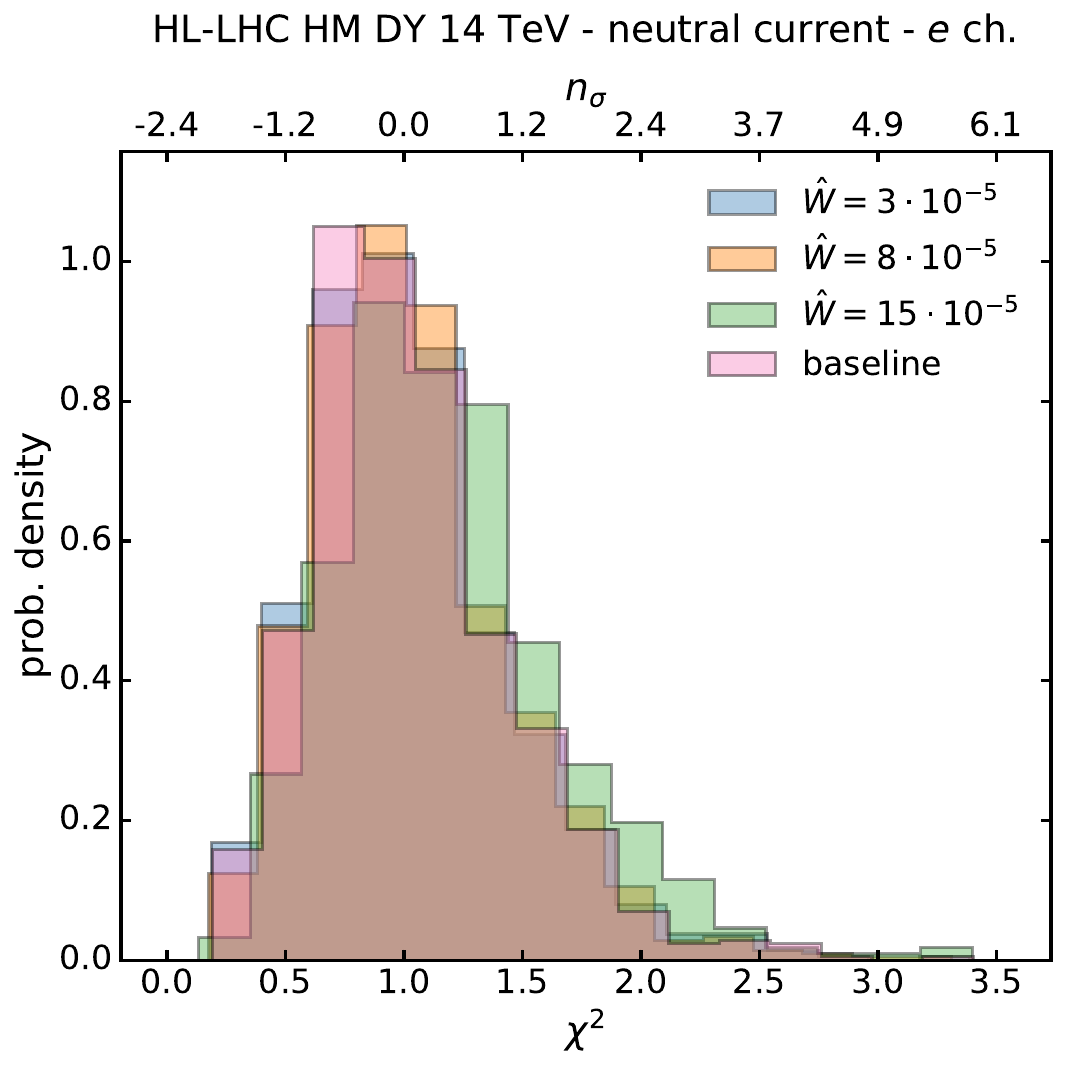}
  \includegraphics[width=0.48\linewidth, page=3]{Figures/chi2_nsigma_W.pdf}\\
  \includegraphics[width=0.48\linewidth, page=6]{Figures/chi2_nsigma_W.pdf}
  \includegraphics[width=0.48\linewidth, page=10]{Figures/chi2_nsigma_W.pdf}\\
  \caption{Distribution of $\chi^2$ and $n_\sigma$ for selected datasets in the $\hat{W}$ contamination scenarios.}
\label{fig:W_chi2_nsigma_dist}
\end{figure}

\noindent  In the flavour-universal $W'$ model we inject three non-zero
values of $\hat{W}=3\cdot 10^{-5},\, 8\cdot 10^{-5},\, 15\cdot
10^{-5}$. In Fig.~\ref{fig:W_chi2_nsigma_dist} we display 
the $\chi^{2(k)}$ and $n_{\sigma}^{(k)}$ distributions
across the 1000 random seeds $k$ for a selection of the datasets
included in each of the fits. In particular we display the datasets in
which a shift occurs either because of the direct effect of the
non-zero Wilson coefficients in the partonic cross sections (such as
the high-mass Drell-Yan in the HL-LHC projections) or
because of the indirect effect of the change of PDFs on other datasets
that probe the large-$x$ light quark and antiquark distributions. Full
details about the trend in the fit quality for all datasets is
given in App.~\ref{app:fit}.

In this case, concerning the quality of the fit, we observe that
up to $\hat{W}=8\cdot 10^{-5}$, the global fit shows
    equivalent behaviours to the SM baseline, while as $\hat{W}$ is increased to
    $15\cdot 10^{-5}$, the quality of the fit markedly deteriorates. This is due
    mostly to a worse description of the HL-LHC charged current
    $e\nu_{e}$ (top right panel in Fig.~\ref{fig:W_chi2_nsigma_dist}) as
    well as the $\mu\nu_\mu$ data. It is interesting to observe that
    also the low-mass fixed-target Drell-Yan data from the E886
    experiment experiences a deterioration in the fit quality due to
    the shift that occurs in the large-$x$ quark and antiquark PDFs. 
    For this largest value, $\hat{W} = 15 \cdot 10^{-5}$, according to the selection criteria outlined in
    Sect.~\ref{subsec:postfit}, the deterioration of both the $\chi^2$ and the $n_{\sigma}$ indicators
    would highlight the high-mass Drell-Yan data as being incompatible with the bulk of the data included in the PDF fit,
    thus excluding them from the fit; thus, no contamination would occur. 
    Hence, in this scenario, $\hat{W}=15\cdot10^{-5}$ falls in the NP parameter region 
    in which the disagreement between the data and theory predictions would unveil the presence of incompatibility of the high-mass Drell-Yan
    tails with the rest of the data; on the other hand, the contamination would go undetected for $\hat{W} = 8 \cdot 10^{-5}$.

\begin{figure}[t!]
  \includegraphics[width=0.49\linewidth]{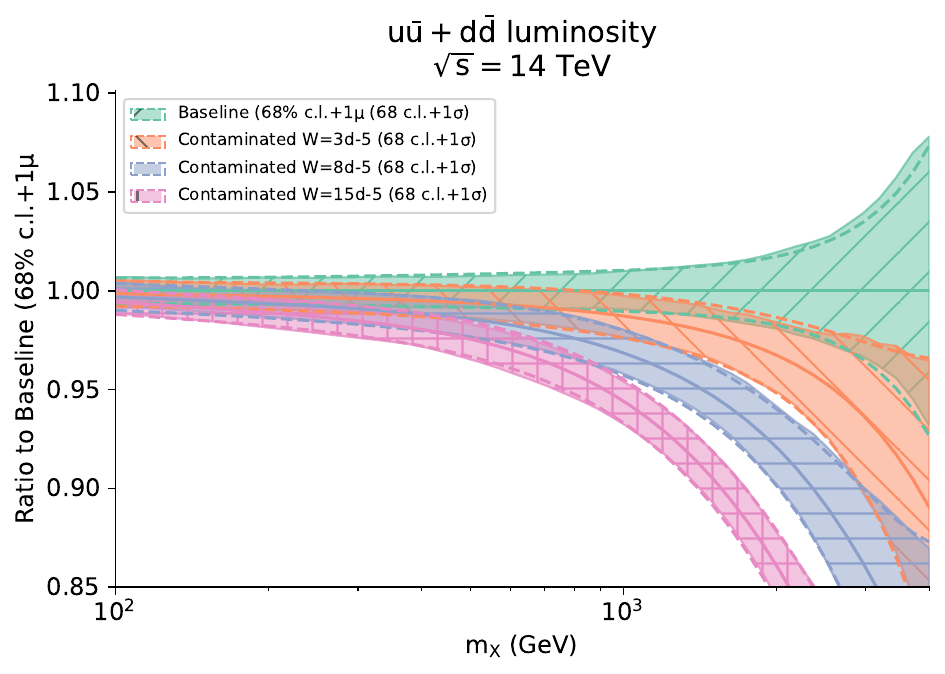}
  \includegraphics[width=0.49\linewidth]{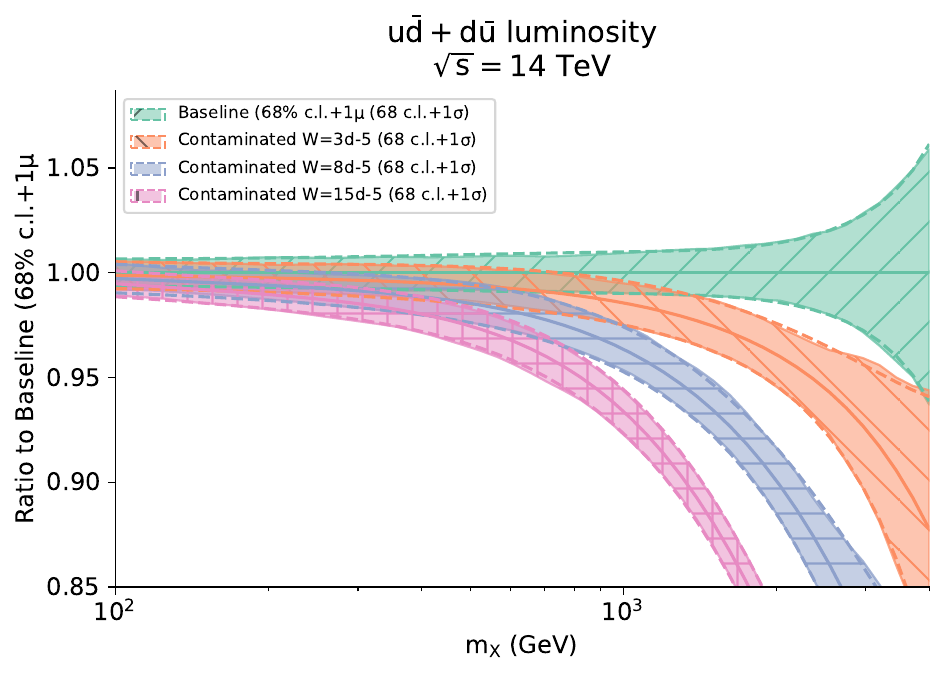}
  \caption{Same as Fig.~\ref{fig:Ylumi} for $\hat{W}=3\cdot 10^{-5}$ (orange line),
    $\hat{W}=8\cdot 10^{-5}$ (blue line) and $\hat{W}=15\cdot
    10^{-5}$ (pink line). }
\label{fig:Wlumi}
\end{figure}
We now check whether, for such $\hat{W}$ values, there is any significant
shift in the PDFs and in the parton luminosities. Individual PDFs are displayed in App.~\ref{app:pdfs}.
In Fig.~\ref{fig:Wlumi} we observe that in this scenario the
NC and CC luminosities defined in Eq.~\eqref{eq:wzlumi} can both shift significantly in the high-mass region, even for low values of $\hat{W}$ ($m_{W'}$ above 20 TeV).
Contrary to the case outlined in the $\hat{Y}$ scenario, the fit \textit{does} have enough flexibility to absorb significant deviations in the high-mass Drell-Yan without impacting 
the rest of the dataset. In particular, until the deviations become too large, the NC and CC sectors, which are both affected by the $W'$ boson, manage to compensate each other. \\

\subsubsection*{(iii) Summary}
Overall, we find that in Scenario I the presence of a new
heavy $Z'$ of about 18 TeV would affect the high-energy tails of the Drell-Yan
  distributions in such a way that they are no longer compatible with the
  bulk of the data included in a PDF analysis.
    On the other hand, in Scenario II, a model of new physics involving a $W'$ of about
  14 TeV  would affect the high-energy tails of the Drell-Yan
  distributions in a way that can be compensated by the PDFs. As a result, if there
  is such a $W'$ in Nature, then this would yield a good $\chi^2$ for
  the high-mass Drell-Yan tails that one includes in a PDF fit as well
  as for the bulk of the data included in a PDF fit, but it 
  would significantly modify PDFs.  Thus, in this case new physics
  contamination does occur.

These results are in agreement with the results of Ref.~\cite{DYpaper}, which generalises the analysis of
Ref.~\cite{Torre:2020aiz} by allowing the PDFs to vary along with the
$\hat{Y}$ and $\hat{W}$ coefficients, finding less stringent
constraints from the same HL-LHC projections.  In particular, it was found that $\hat{W}=8\cdot 10^{-5}$ would have been excluded by the HL-LHC under the assumption of SM PDFs, but that this value of $\hat{W}$ was allowed by the constraints at 95\% CL obtained by varying the PDFs along with the SMEFT.
Ref.~\cite{DYpaper} also indicated that the impact of varying the PDFs along with the
$\hat{W}$ coefficient was more significant than the impact in the $\hat{Y}$ direction, indicating 
a greater possibility to absorb the effects of new physics into the PDFs in the $\hat{W}$ direction.


%

Comparing the two scenarios considered in this section, one might wonder why 
the $Z'$ scenario does not yield any contamination, while the $W'$ does. 
%
%
Looking at the effect of the $Z'$ and $W'$ bosons on the observables included in a PDF fit (see Eqs.~\eqref{eq:zprimeL} and \eqref{eq:wprimeL} respectively), 
we see that the main difference lies in the fact that the $Z'$ scenario only affects the NC DY high-mass data, while the $W'$ scenario affects both the
NC and the CC DY high-mass data. Hence, in the former scenario, the shift required in ${\cal L}^{\rm NC}\equiv (u\bar{u}+d\bar{d})$ to accommodate the 
effect of a $Z'$ in the tail of the $m_{ll}$ distribution would cause a shift in ${\cal L}^{\rm CC}\equiv (u\bar{d}+d\bar{u})$, thus spoiling its agreement with the data, 
in particular the tails of the $m_T$ distribution -- which is unaffected by the presence of a $Z'$. 

On the other hand, in the $W'$ scenario, the shift in the $(u\bar{u}+d\bar{d})$ parton channel that accommodates the effect of a $W'$ in the tail of the 
NC DY $m_{ll}$ distribution is compensated by the shift in the $(u\bar{d}+d\bar{u})$ parton channel that accommodates the presence of a $W'$ in the tail of the 
CC DY $m_T$ distribution (as, in this scenario, they are both affected by new physics). It is as if there is a flat direction in the luminosity versus the 
matrix element space. This continues until, for sufficiently large $\hat{W}$, a critical 
point is reached in which the two effects do not manage to compensate each other as they start affecting significantly the luminosities at lower $\tau=M/\sqrt{s}$, 
hence spoiling the agreement with the other less precise datasets included in a PDF fit which are sensitive to large-$x$ antiquarks. 

\begin{figure}[tb]
  \begin{center}
    \includegraphics[width=0.49\linewidth]{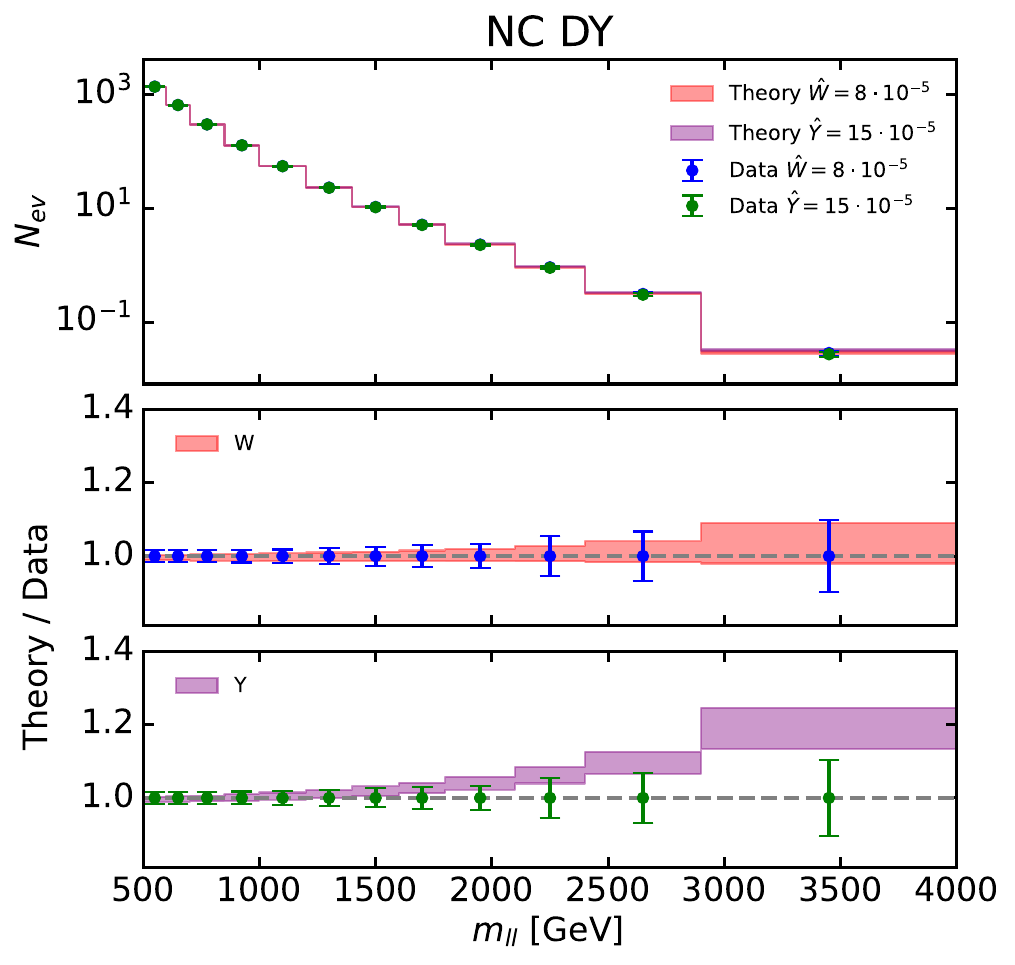}
    \includegraphics[width=0.49\linewidth]{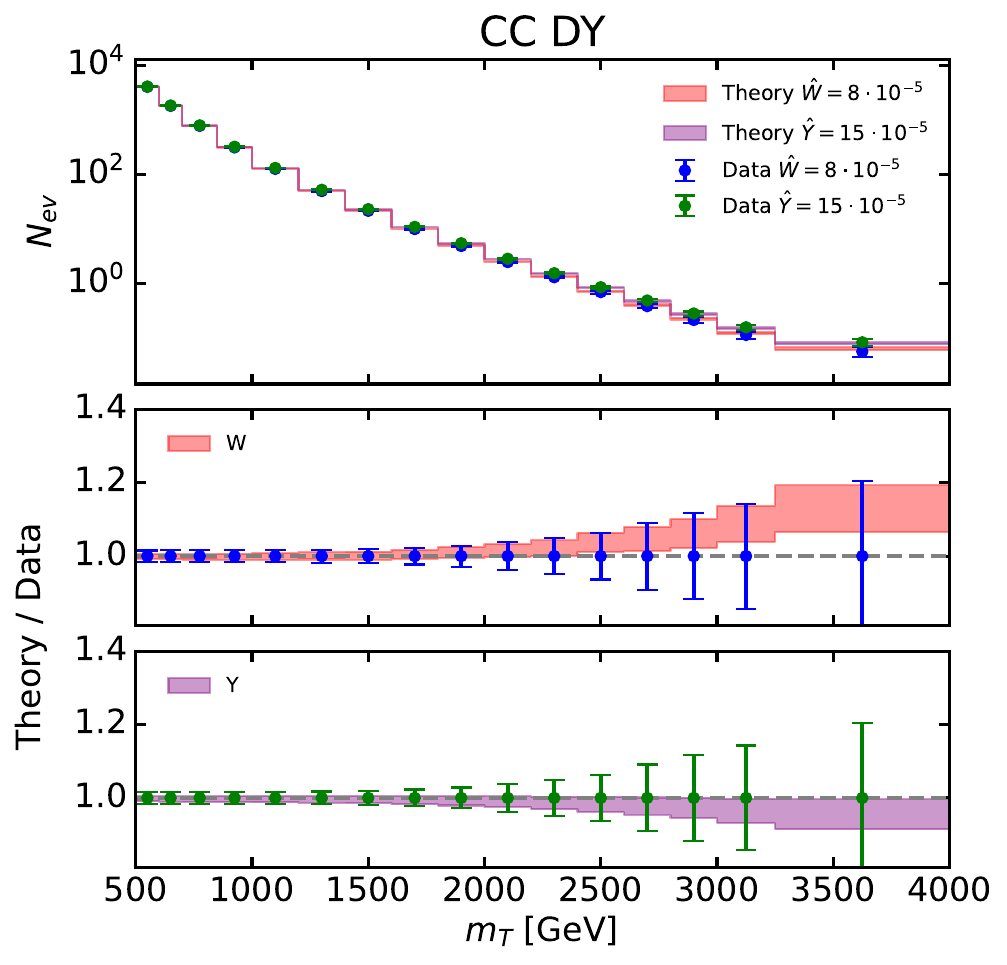}
  \end{center}
	  \caption{For two of the 
      representative scenarios that we consider, $\hat{W}=8\cdot 10^{-5}$ and $\hat{Y}=15\cdot 10^{-5}$, we show the comparison between the data expected in the presence of new physics (``Data'' points) and the SM theory predictions obtained with the potentially contaminated PDFs (``Theory'' bands). Left panel: NC Drell-Yan $m_{ ll}$ distribution. 
      Right panel: CC Drell-Yan $m_T$ distribution. 
    }
  \label{fig:wy_dy_prod}
\end{figure}

To see this more clearly, we plot in Fig.~\ref{fig:wy_dy_prod} the data-theory comparison for the HL-LHC NC and CC Drell-Yan Monte Carlo data that we include in the fit. 
%
The points labelled as ``Data'' correspond to the `truth' in the presence of the new physics, namely they are obtained by convolving the DY prediction with non-zero $\hat{Y},\hat{W}$ parameters with a non-contaminated PDF set.
The bands labelled as ``Theory'' represent the theoretical predictions for pure SM DY production, but obtained with the 
PDFs fitted with the inclusion of the DY data modified by the effect of non-zero $\hat{Y},\hat{W}$ parameters. 
We observe that the SM predictions obtained with the contaminated PDFs do fit the data well in the case of $\hat{W}=8\cdot 10^{-5}$, 
because the significant depletion of the $(u\bar{u}+d\bar{d})$ and $(u\bar{d}+d\bar{u})$ parton luminosities observed in Fig.~\ref{fig:Wlumi} compensates 
the enhancement in the partonic cross section observed in Fig.~\ref{fig:Wp_DY_Comp}. This is not the case for $\hat{Y}=15\cdot 10^{-5}$, where instead the 
much milder modification of the parton luminosities observed in Fig.~\ref{fig:Ylumi} does not manage to compensate the enhancement of the partonic cross 
section observed in Fig.~\ref{fig:Zp_DY_Comp}. We can also notice that $\hat{W}=8\cdot 10^{-5}$ is within a region in the $\hat{W}$ parameter space 
beyond which the parton luminosities do not manage to move enough to compensate the shift in the matrix elements of the $m_T$ distribution. 
To find the exact critical value of $\hat{W}$ one would need a finer scan.
Analogously, $\hat{Y}=15\cdot 10^{-5}$ is in the region of $\hat{Y}$ such that  contamination in the PDFs does not occur.
However, these values have been determined assuming a given statistical uncertainty in the distributions;
the regions in which these values fall clearly depend on the actual statistical 
uncertainty that the $m_T$ and $m_{ll}$ distributions will reach in the HL-LHC phase.

\subsection{Consequence of new physics contamination in PDF fits}
\label{sec:projections}
In the previous section, we showed that in the presence of heavy new physics effects in 
DY observables, the flexible PDF parametrisation is able to accommodate the deviations
and absorb the effects coming from the new interactions. In particular, we observe that when
data are contaminated with the presence of a $W^\prime$, we generally find good fits and are able to accommodate
even large deviations from the SM. However, it is worth reminding the reader that
the leading source of contaminated data are the HL-LHC projections, as present data would not be as susceptible to the $W^\prime$ effects.
Hence, from now on we will focus on the scenario in which data include the presence of a heavy $W^\prime$ that induces 
a modified interaction parametrised by the $\hat{W}$ parameter with value $\hat{W}=8\cdot 10^{-5}$.

\begin{figure}[htb!]
\centering
\includegraphics[width=0.9\textwidth]{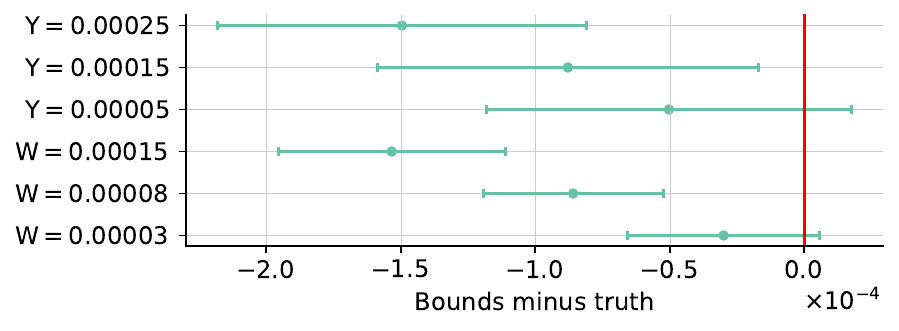}
\caption{A comparison of the 95\% C.~L.~bounds obtained using different contaminated PDFs to fit the $\hat{W}$, $\hat{Y}$ parameters 
to HL-LHC high-mass Drell-Yan projected data, relative to the true values of $\hat{W}, \hat{Y}$. In some cases, the true value is not 
contained in the 95\% C.~L.~bounds.}
\label{fig:bounds_comparison}
\end{figure}

In this section we examine the consequences of
using unknowingly contaminated PDFs, and the implications of this for possible new physics searches.
%
The first interesting consequence is that, if we use the contaminated PDF as an input set in a SMEFT study 
of HL-LHC projected data to gather knowledge on the $\hat{W}$ parameter, we find that the analysis excludes 
the ``true'' value of the SMEFT coefficients that the data should reveal.
Indeed, in Fig.~\ref{fig:bounds_comparison} we observe that, in both scenarios under consideration, and in particular for the one
corresponding to $\hat{W}=8\cdot10^{-5}$, the 95\% C.L. bounds on the Wilson coefficients that one would extract from the precise
HL-LHC data discussed in Sect.~\ref{sec:contaminated_pdfs} would agree with the SM and would not contain the true 
``values'' of the underlying law of Nature that the data should reveal.
In fact, the measured value would exclude the true value with a significance that ranges from  $\sim 1.5 \, \sigma$ to $\sim 4.5 \, \sigma$. A comparison of whether the bounds generated by the different contaminated PDFs considered in this study contain the true value is shown in Fig.~\ref{fig:bounds_comparison}.
This is all very expected, as the quark-antiquark luminosity for this specific scenario does exhibit signs of
new physics absorption in a significant amount, as can be seen in Fig.~\ref{fig:Wlumi}. 
As a matter of fact, we expect all data that entered in the PDF fit to be well described by the combination of the PDF set and the SM theory.
This simple fact is once again reminding us that it might be dangerous to perform SMEFT studies on overlapping datasets and that simultaneous studies should be preferred or, at least, a conservative approach with disjoint datasets should be undertaken. 
This was discussed in Ref.~\cite{Iranipour:2022iak}, for instance,
where it was shown that by means of a simultaneous study, one is able to recover 
both the underlying true PDFs and the presence of a new interaction. It is worth 
mentioning that the use of conservative PDF sets, while appealing given the simplicity, 
might also come with its own shortcomings, see Ref.~\cite{DYpaper} and Ref.~\cite{Kassabov:2023hbm} for detailed studies on the matter in the 
Drell-Yan sector and the top quark sector respectively. In particular, 
the extrapolated PDFs might both underestimate the error band and
have a significant bias.\\

\noindent We now turn to study the effects of the contaminated PDFs in observables and processes that did not enter the PDF fit.
We focus in particular on the EW sector, given its relevance for NP searches and the fact that the contaminated PDFs show deviations from 
the true PDFs mostly in the quark-antiquark luminosities, which are particularly relevant for theoretical predictions involving EW interactions.
The study is performed by producing projected data according to the true laws of Nature, i.e. the true PDFs of choice and 
the SM + $\hat{W}=8\cdot 10^{-5}$ in the matrix elements. 

In particular, we 
produce MC data for several diboson processes, including $H$ production in association with EW bosons. 
Given that the $\hat{W}$ operator 
induces only four-fermion interactions, $\hat{W}$ does not have an effect on these observables, and the hard scattering 
amplitudes are given by the SM ones. For each observable we build HL-LHC projections and devise bins with the objective 
of probing the high-energy tails of the distributions, scouting for new physics effects, although we know these do not exist in the ``true'' law of Nature for these observables.
We then produce predictions by convolving the contaminated PDF set obtained with a value of $\hat{W}=8\cdot 10^{-5}$ and the SM matrix elements. 
Given our knowledge of the ``true'' law of Nature, the possible deviations between theory and data are therefore
only a consequence of the shift in the PDFs coming from the contaminated Drell-Yan data.
Whenever in the presence of $W$ bosons, we decided to split the contributions of $W^+ X$ and $W^- X$ as they probe different luminosities and in particular, 
from the contaminated fits, we know that the luminosity $u \bar{d}$ deviates more severely than $d \bar{u}$ from the true luminosity.

Both SM theory and data have been produced at NLO in QCD making use of the Monte Carlo generator \madgraph{}.
In the case of $ZH$ production, the gluon fusion channel has also been taken into account.
Data are obtained by fluctuating around the central value, assuming a Gaussian distribution with total 
covariance matrix given by the sum of the statistical, luminosity and systematic covariance matrices. 
Regarding the theory predictions, we also provide an estimate 
of the PDF uncertainty.
We assume a luminosity of $3$ ab$^{-1}$ and we estimate the systematic uncertainties on each observable 
by referring to the experimental papers~\cite{ATLAS:2020osn, ATLAS:2019bsc, CMS:2019efc} .
These systematic uncertainties can be experimental or come from other additional sources such as background and signal theoretical calculations.
We also include estimates of the luminosity uncertainty by taking as a reference the CMS measurement at 13 TeV~\cite{CMS:2018mdl}. 
%
Statistical uncertainties are given by $\sqrt{N}$, where $N$ is the number of expected events in each bin. Performing a fully realistic simulation, with acceptance cuts and detector effects, is beyond the scope of the current study, and we simply simulate events at parton level and apply the branching ratios into relevant decay channels. 
Specifically, in the case of $W$ bosons we apply a branching ratio of $Br(W \to l \nu) = 0.213$ with $l=e, \, \mu$, for the $H$ boson we use $Br(H \to b \bar{b})=0.582$ and for the $Z$ boson we use $Br(Z \to l^+ l^-) = 0.066$ with $l=e, \, \mu$~\cite{Workman:2022ynf}.
Multiple sources of uncertainty are simply added in quadrature.
\begin{table}[t!]
        \small
        \centering
        \begin{tabular}{l|c|c|c|c|c|}
                \toprule
       & \multicolumn{2}{c}{HL-LHC} & \multicolumn{2}{c}{Stat. improved}\\
          \midrule
       Dataset  & $\chi^2/n_{\rm dat}$ &  $n_\sigma$ & $\chi^2/n_{\rm dat}$ & $n_\sigma$ \\
          \midrule
 $W^+ H$         & 1.17 & 0.41 & 1.77 & 1.97 \\
 $W^- H$         & 1.08 & 0.19 & 1.08 & 0.19 \\
 $W^+ Z$         & 1.08 & 0.19 & 1.49 & 1.20\\
 $W^- Z$         & 0.99 & -0.03 & 1.02 & 0.05\\
 $Z H$           & 1.19 & 0.44 & 1.67 & 1.58\\
 $W^+ W^-$       & 2.19 & 3.04 & 2.69 & 4.31\\
 VBF $\to$ H        & 0.70 & -0.74 & 0.62 & -0.90\\
                \bottomrule
        \end{tabular}
        \caption{Values of the \(\chi^2\) and $n_\sigma$ for the projected observables at HL-LHC in the EW sector. In the left column we report the values from a realistic estimate of the statistical uncertainties, while in the right columns we show what would be obtained if statistics were to improve by a factor $10$. \label{tab:chi2_hllhc}}
\end{table}

In Table~\ref{tab:chi2_hllhc}, for each process considered, we collect the computed $\chi^2$ and the corresponding value of $n_\sigma$. These numbers are obtained by performing several fluctuations of the data and then taking the average $\chi^2$ from all the replicas. As a consequence, the quoted $\chi^2$ are considered the expected $\chi^2$ and are not associated to a specific random fluctuation. The numbers are provided both in a realistic scenario, with a reasonable estimate of the statistical uncertainties, and in a scenario in which the statistics are improved by a factor $10$. 
The latter could be both the result of an increased luminosity and/or additional decay channels of the EW bosons, e.g. decays into jets. 
As it can be seen by inspection of the table, the processes that would lead to the most notable deviations between data and theory are $W^+ H$ and $W^+ W^-$, with the latter being in significant tension already in the scenario of a realistic uncertainty estimation. With improved statistics, slight tensions start to appear in $Z H$ and $W^+ Z$, both exhibiting a deviation just above 1$\sigma$.
Interestingly, the clear smoking gun process here seems to be $W^+ W^-$, which just by 
itself would point towards a significant tension with the SM, which could potentially and erroneously be interpreted in terms of new interactions.
\begin{figure}[t!]
  \includegraphics[scale=0.45]{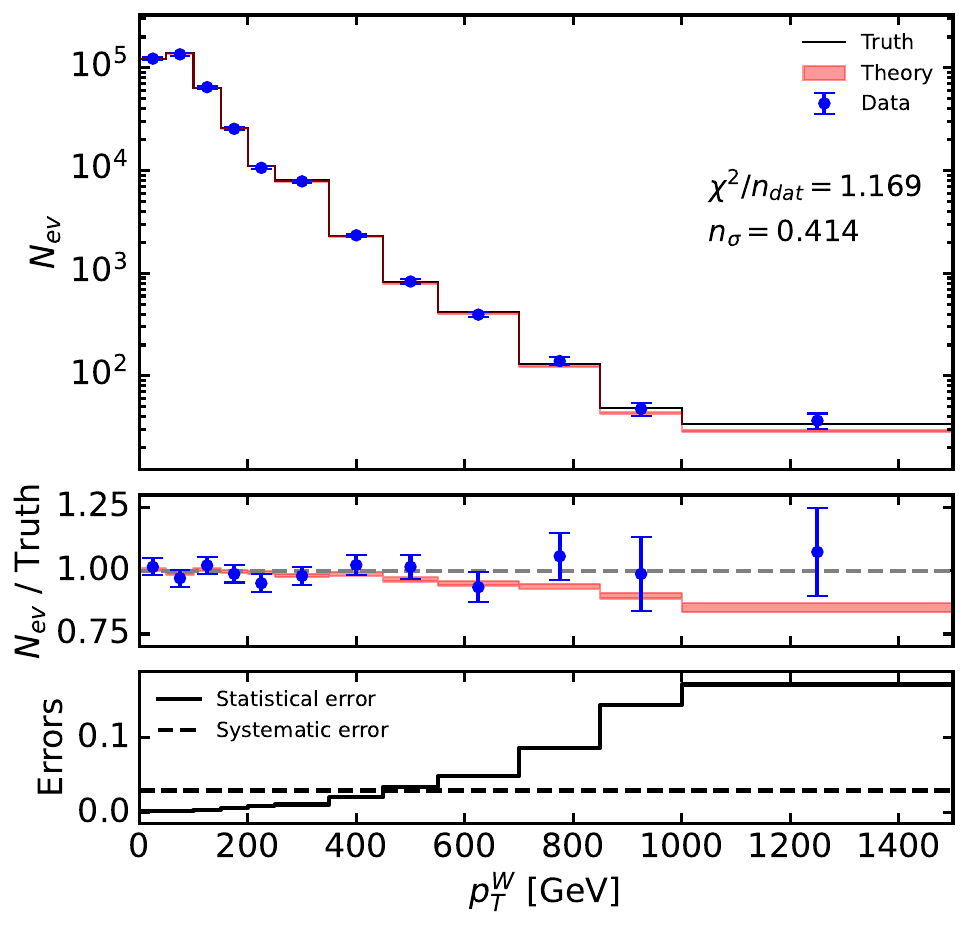}
  \includegraphics[scale=0.45]{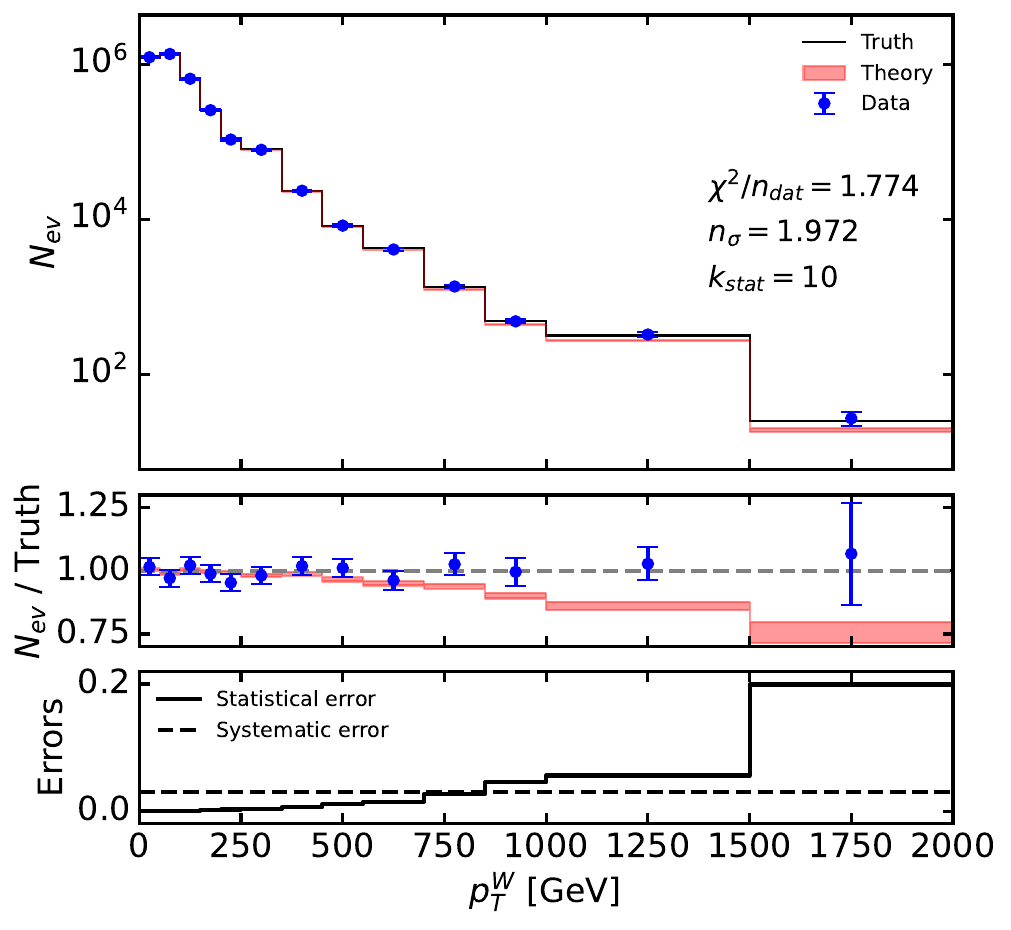}
  \caption{Predictions (with the contaminated PDF) for $W^+ H$ at the HL-LHC compared with the projected data. Left: HL-LHC projection. Right: statistics improved by a factor 10 (futuristic scenario). In the latter, an additional bin is added at high energy to take advantage of the additional expected events.}
  \label{fig:wph_hllhc}
      \end{figure}

\begin{figure}[t!]
  \includegraphics[scale=0.45]{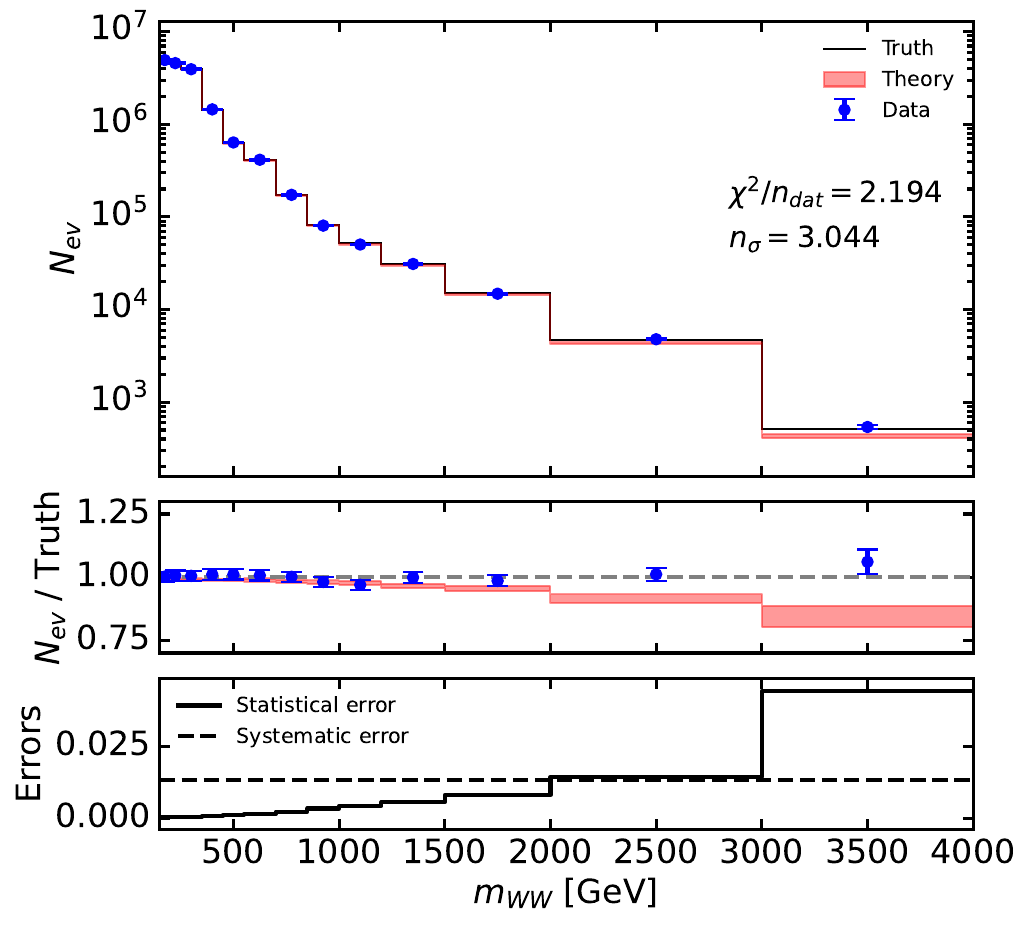}
  \includegraphics[scale=0.45]{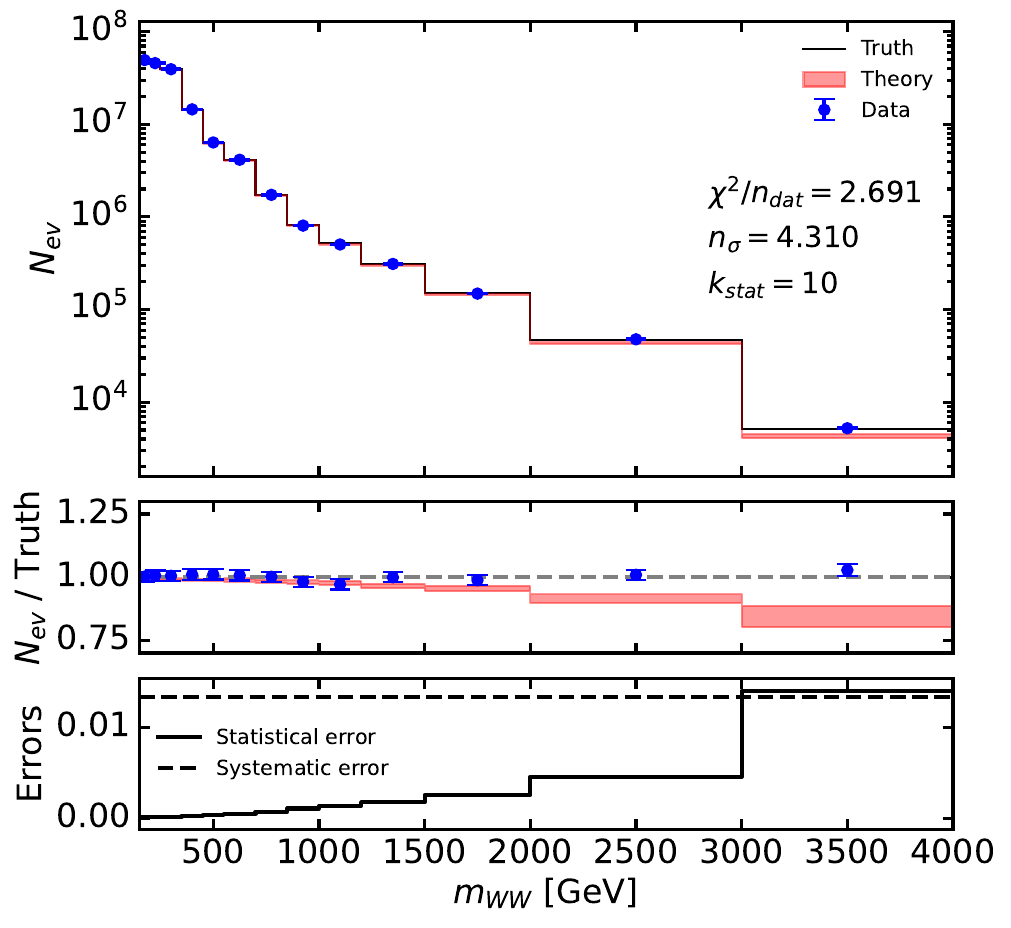}
  \caption{Predictions (with contaminated PDF) for $W^+ W^-$ at the HL-LHC compared with the projected data. Left: HL-LHC projection. Right: statistics improved by a factor 10 (futuristic scenario).}
  \label{fig:ww_hllhc}
      \end{figure}



In Figs.~\ref{fig:wph_hllhc} and ~\ref{fig:ww_hllhc}, we show plots of the two most affected observables considered in 
this section, namely $W^+ H$ and $W^+ W^-$ respectively.
While in all other processes the deviations between the true central values and the theoretical predictions obtained with the contaminated PDFs are limited, in the case of $W^+ H$ and $W^+ W^-$, they are substantial. 
It is clear that the true limiting factor is that as soon as we are in the high energy tails of the distributions and potentially sensitive to the PDF contamination, the pseudodata become statistically dominated and therefore we lose resolution. This is particularly true in the case of $W^+ H$, while $W^+ W^-$ is predicted to have a higher number of events and could potentially probe higher energies.

We also assess the ratios $W^+ Z / W^- Z$ and $W^+ H / W^- H$, and observe that in 
this case the deviations resulting from contaminated PDFs are 
no longer visible. In general the ratios cancel the effect of 
any possible contamination in the parton luminosities if they are correlated. The fact 
that the effect disappears is a proof that the $u\bar{d}$ and $d\bar{u}$ luminosities are highly correlated and 
the contamination effects are compatible.
%
%
%
%
%
%
%
%
%

In summary, the PDF contamination has the potential to generate substantial deviations in observables and processes generally considered to be good portals to new physics, which could nonetheless be unaffected by the presence of heavy states at the current probed energies, as in the scenarios considered in this work.

\section{How to disentangle new physics effects}
\label{sec:solution}

In this section we discuss several strategies that might be proposed in order to disentangle 
new physics effects in a global fit of PDFs. In Sect.~\ref{subsec:forward} we start by 
assessing the potential of precise on-shell forward vector boson production 
data in the HL-LHC phase and check whether their inclusion in a PDF fit helps to disentangle 
new physics effects in the high-mass Drell-Yan tails. 
In Sect.~\ref{subsec:cut} we scrutinise whether the data-theory agreement displays a deterioration 
that scales with the maximum energy probed by the data included in the fit. We will see that 
neither of these strategies helps to disentangle the contamination that arises in the scenario that we have 
highlighted in our study and we outline the reason for this.  We then turn to analyse the behaviour of suitable observable ratios in Sect.~\ref{subsec:ratio}; we will see that such ratios \textit{do} correctly indicate the 
presence of new physics in the observables that are affected by it, although they would not be able to distinguish between the two observables that 
enter the ratio. Finally in Sect.~\ref{subsec:lowE} we will determine the  observables in current PDF fits that are correlated to the 
large-$x$ antiquarks and we will highlight the signs of tension with the ``contaminated'' high-mass Drell-Yan data via suitably devised weighted fits. 
The result of these tests points to the need for the inclusion of independent low-energy/large-$x$ constraints in future PDF analyses, 
if one wishes to safely exploit the constraining power of high-energy data without inadvertently absorbing signs of new physics 
in the high-energy tails.

\subsection{On-shell forward boson production}
\label{subsec:forward}
The most obvious way to disentangle any possible contamination effects in the PDF is the inclusion 
of observables that probe the large-$x$ region in the PDFs at low energies, where NP-induced energy growing effects are not present.  
In this section we assess whether the inclusion of precise forward LHCb distributions measured at the $W$ and $Z$ on-shell energy 
at the HL-LHC might help spotting NP-induced inconsistencies in the high-mass distributions measured by ATLAS and CMS. 

\begin{figure}[htb]
  \centering
  \includegraphics[width=0.49\linewidth]{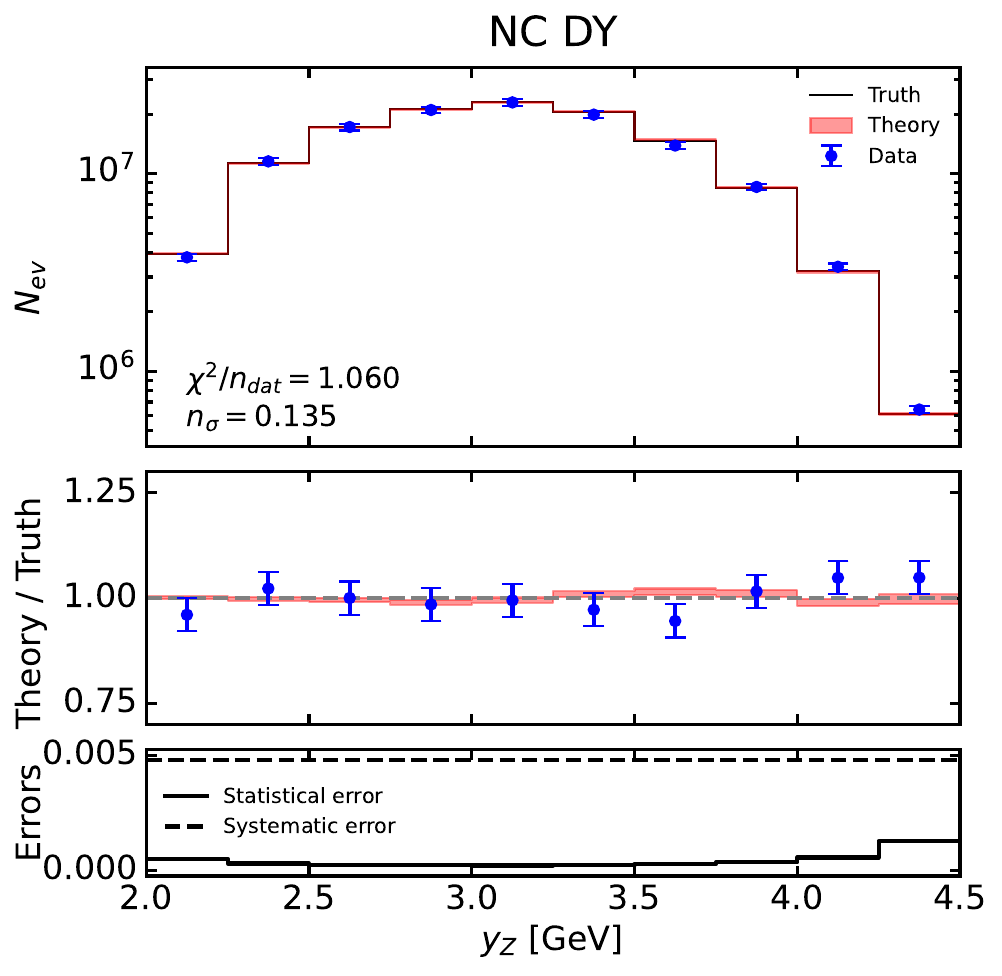}
  \includegraphics[width=0.49\linewidth]{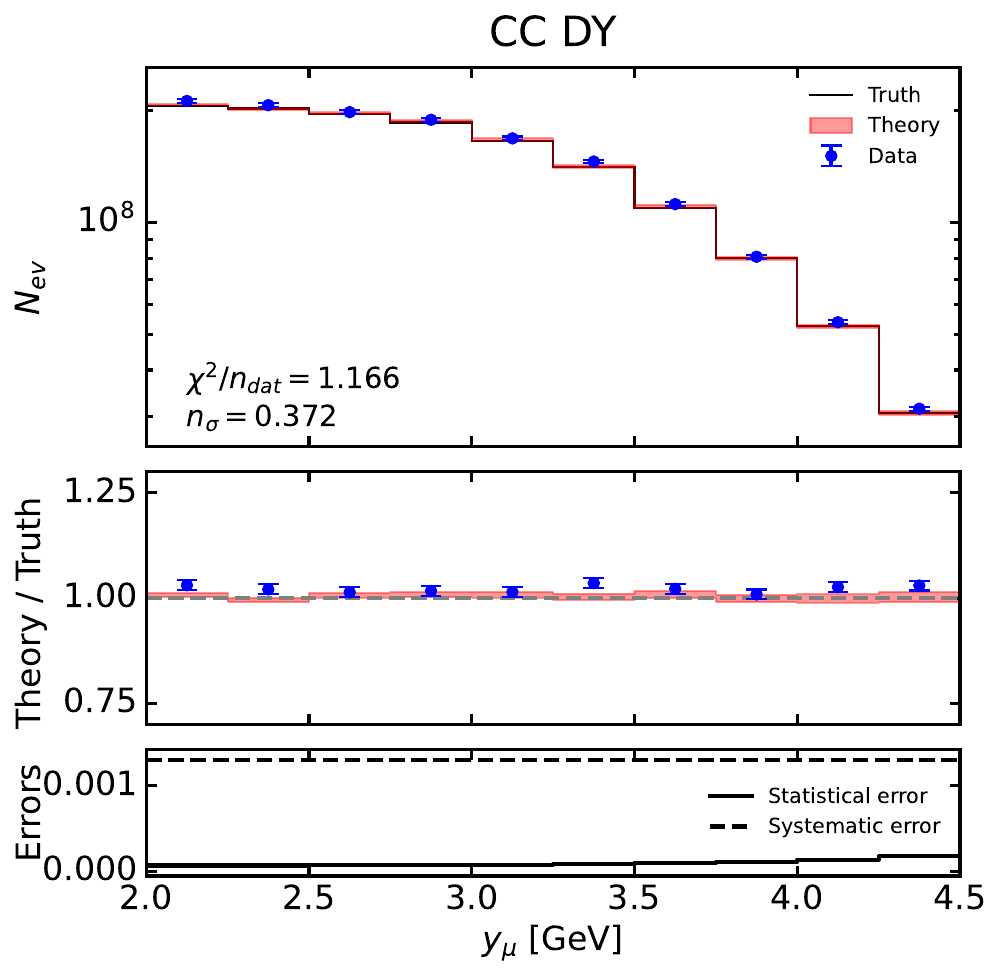}
  \caption{Predictions (with $\hat{W}=8\cdot 10^{-5}$ contaminated PDF) for forward vector bosons production 
  in the HL-LHC phase at LHCb compared with the projected data. Left panel: on-shell $Z$ production cross section as a 
  function of the $Z$ boson rapidity $y_Z$. Right panel: $W$ production cross section as a function of the final-state muon
  pseudo-rapidity $y_\mu$.}
  \label{fig:forwardZ_hllhc}
\end{figure}

In order to test this, we compute HL-LHC projections for LHCb, taking 0.3 ab$^{-1}$ as benchmark luminosity~\cite{Azzi:2019yne} 
and focusing on the forward production of $W/Z$. The $Z$ boson is produced on-shell 
($60 \text{ GeV} < m_{ll} < 120 \text{ GeV}$), while no explicit cuts are applied on  the transverse mass $m_T$ in 
the case of a produced $W$ boson decaying into a muon and a muonic neutrino, which is dominated by the mass-shell region. 
We impose the LHCb forward cuts on the lepton transverse momentum ($p^l_T > 20 \text{ GeV}$) 
and on both the $Z$ rapidity and pseudo-rapidity of the $\mu$ originated by $W$ ($2.0 < |y_{Z,\mu}| < 4.5$).
Fig.~\ref{fig:forwardZ_hllhc} shows a comparison between the pseudodata generated with the ``true'' PDFs and NP-corrected matrix elements,\footnote{Note that at the energy probed by the 
forward $W/Z$ production the NP contribution associated to the presence of a $W'$ boson is negligible} and the theory predictions 
obtained with the $\hat{W}=8\cdot 10^{-5}$ contaminated fit and the SM matrix element, for each of the two processes.
We observe that there are no significant deviations between the theory predictions obtained from a 
contaminated PDF set and the true underlying law. 
Intuitively this can be understood, as the produced leptons are in the forward region measured at LHCb, and one of the initial 
partons must have more longitudinal momentum than the other. 

To visualise more precisely the regions in $x$ that are constrained by a measurement of a given final state at 
the energy $E\sim m_X$ and at a given rapidity $y$, we display the scatter plot for $x_{1,2} = m_X/\sqrt{s}\,\exp(\pm y)$ in the
large-$m_X$ and central region, namely $|y|<2.0$ and $1\,{\rm TeV}<m_X<4\,{\rm TeV}$, and compare it to the low-to-intermediate-$m_X$ 
and forward rapidity region, namely $2.0<|y|<4.5$ and $10\,{\rm GeV} < m_X < 1\,{\rm TeV}$.
\begin{figure}[htb]
  \begin{center}
    \includegraphics[width=0.8\linewidth]{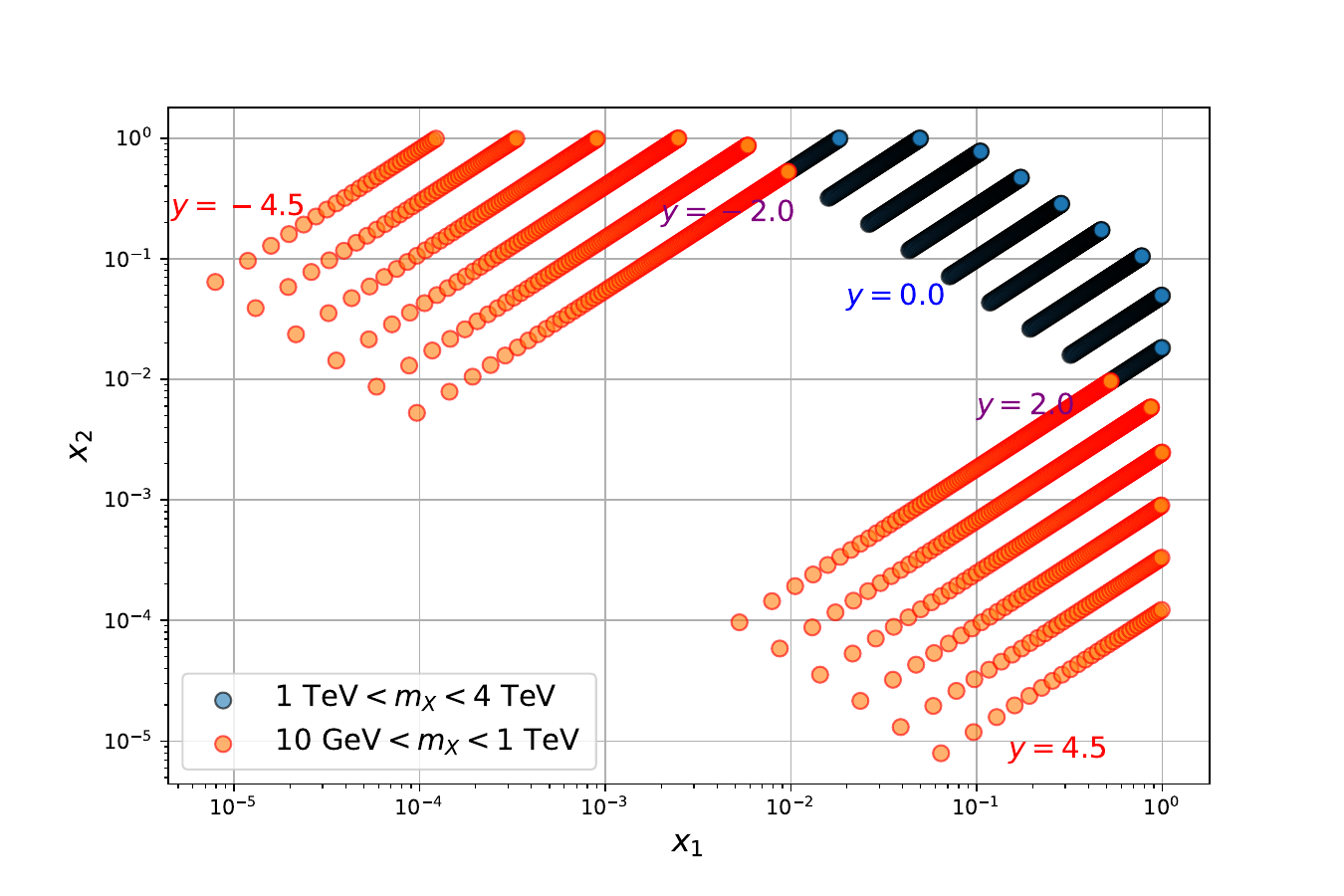}
    \end{center}
	\caption{Leading order kinematic plot of $x_{1,2} = m_ X/\sqrt{s}\,\exp(\pm y)$ in the
large-$M$ and central region, $|y|<2.0,\quad 1\,{\rm TeV}<m_X<4\,{\rm TeV}$, (black dots) and in the low-intermediate-$m_X$ and forward region, $2.0<|y|<4.5\quad 10\,{\rm GeV} < m_X < 1\,{\rm TeV}$ (red dots). Here $\sqrt{s}=14$~TeV.}
        \label{fig:kinplot}
\end{figure}
We can see that, while the measurements of large-invariant mass objects in the central rapidity region constrain solely the large-$x$ region, and
where both partons carry a fraction $x$ of the proton's momentum in the region $0.01\lesssim x \lesssim 0.8$, the low-to-intermediate invariant
mass region in the forward rapidity region constrains both the small and the large $x$ region, given that at $|y|\approx 4.0$ the $x$-region
probed is around $0.1\lesssim x \lesssim 0.8$ for one parton and around $10^{-5}\lesssim x \lesssim 10^{-4}$ for the other parton. 
Given that the valence quarks are much more abundant 
at large $x$ than the sea quarks, in most collisions the up or down quarks will be the partons carrying a large fraction $x$ of the proton's momentum, 
while the antiquarks will carry a small fraction $x$. Hence, this observable will not be sensitive to the shift in the large-$x$ anti-up and 
anti-down that the global PDF fit yields in order to compensate the effect of NP in the tails. 

\subsection{Sliding ${m_{ll}}$ cut}
\label{subsec:cut}
Another way to potentially disentangle EFT effects in PDF fits was explored in Ref.~\cite{DYpaper} where, in the context of the $\hat{W}$ and $\hat{Y}$ parameters, the authors exploited the energy-growing effects of the EFT operators at high invariant dilepton mass. 
They introduced a ratio evaluation metric $R_{\chi^2}$ (in the notation of the original reference) which described how much the PDF fit quality deteriorated when data-points at high dilepton invariant mass were included in the computation of the $\chi^2$, with respect to a fit that only used low invariant mass bins.
In this way, $R_{\chi^2} \sim 1 $ indicated a fit quality similar to a purely SM scenario, as the sensitivity to energy-growing EFT effects is suppressed. 

They found that, when including higher dilepton invariant mass measurements (where the effect of EFT operators is enhanced) in the computation of the $\chi^2$, the fit quality deteriorated and $R_{\chi^2}$ growth almost monotonically away from $1$. With this metric other sources of PDF deterioration are minimised and the tension arises fundamentally because of the EFT effects.

\begin{figure}[htb]
  \centering
  \includegraphics[width=0.49\linewidth]{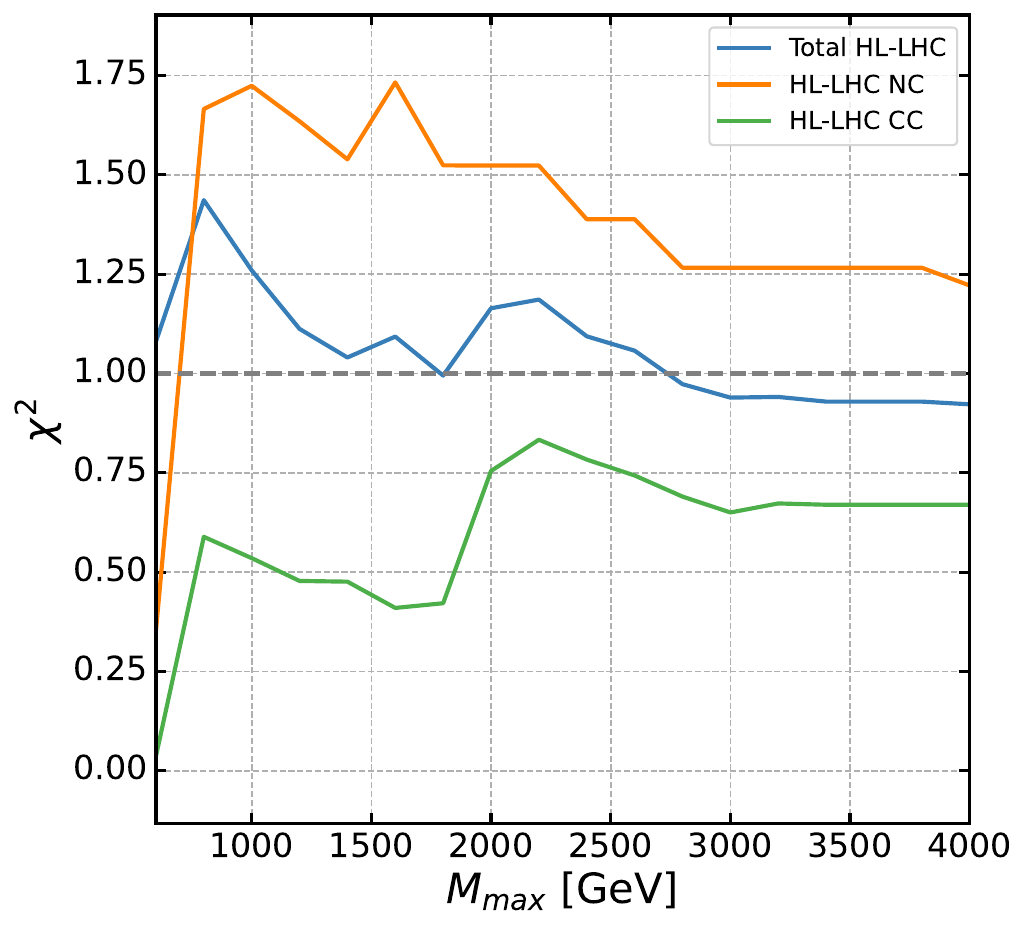}
  \includegraphics[width=0.49\linewidth]{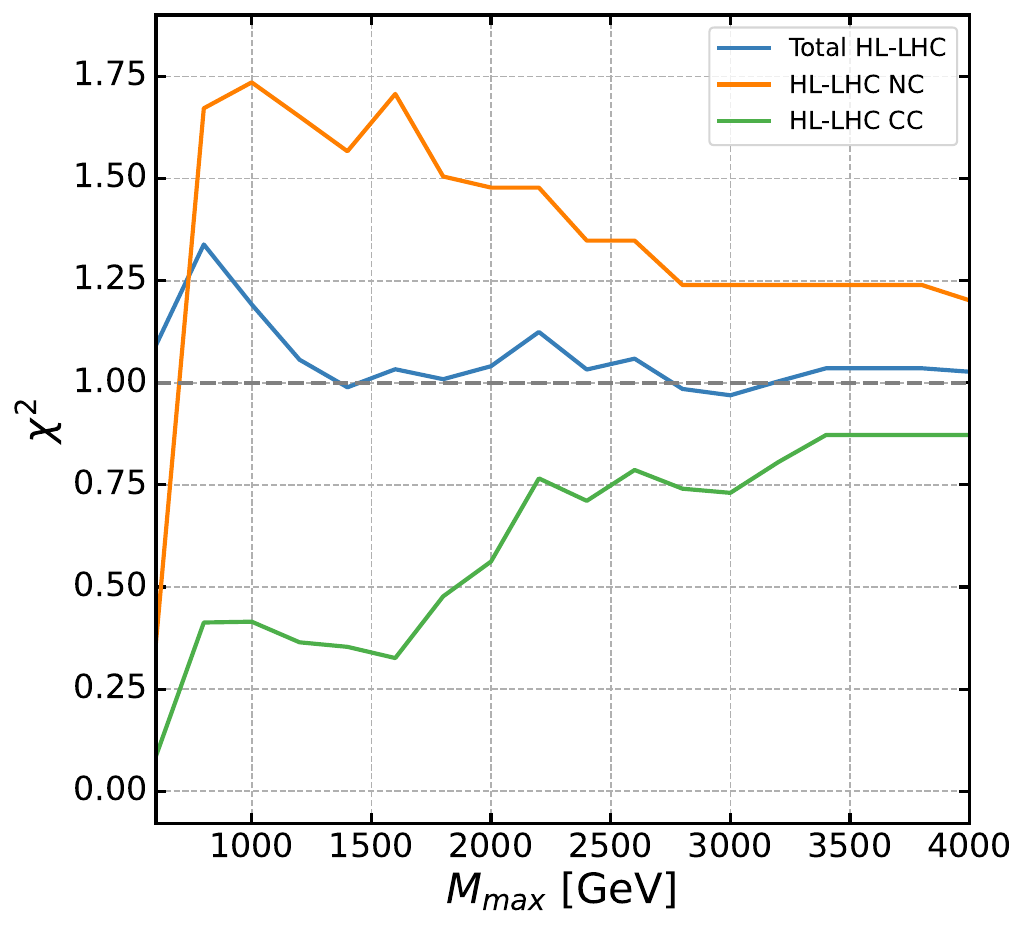}
  \caption{Value of the $\chi^2$ per HL-LHC dataset as a function of the maximum value of the invariant mass allowed in the kinematics $M_{\rm max}$. This parameter is analogous to $m_{ll}^{({\rm max})}$ of Sect. 4 of~\cite{DYpaper} . Left: baseline SM fit. Right: contaminated $\hat{W}=8\cdot 10^{-5}$.}
  \label{fig:sliding_chi2}
\end{figure}

For our study, we show in Fig. ~\ref{fig:sliding_chi2} the $\chi^2$ values for the NC and CC HL-LHC datasets, combining the two lepton channels, in the SM and $\hat{W} = 8 \cdot 10^{-5}$ contaminated cases. $M_{\rm max}$ is the maximum value of the dilepton invariant mass that we include in the computation of the $\chi^2$.
We see that after $M_{\rm max} \sim 2$ TeV, the $\chi^2$ values in both the SM and contaminated scenario stagnate and no persistent deterioration is observed. This means that it is not possible to isolate the EFT effects, which are enhanced at higher masses, in the worsening of the fit quality. In this way, it is not possible to disentangle the EFT effects on the PDF fit and other options have to be explored.

\subsection{Observable ratios}
\label{subsec:ratio}
In order to disentangle PDF contamination, another quantity worth studying is the ratio between observables whose processes have similar parton channels. Indeed, in this case the impact of the PDF is much reduced and any discrepancy between data and theory predictions can be more confidently attributed to new physics in the partonic cross-section. Practically, a deviation would mean that one of the two datasets involved in the ratio is ``contaminated'' by new physics and should therefore be excluded from the PDF fit. 

We have studied the ratio between the number of events in $WW$ production and Neutral Current Drell-Yan (NC DY), as well as between $WH$ production and Charged Current Drell-Yan (CC DY). In each pair both processes are initiated from the same parton channels. 

\begin{figure}[!h]
  \begin{center}
    \includegraphics[width=0.49\linewidth]{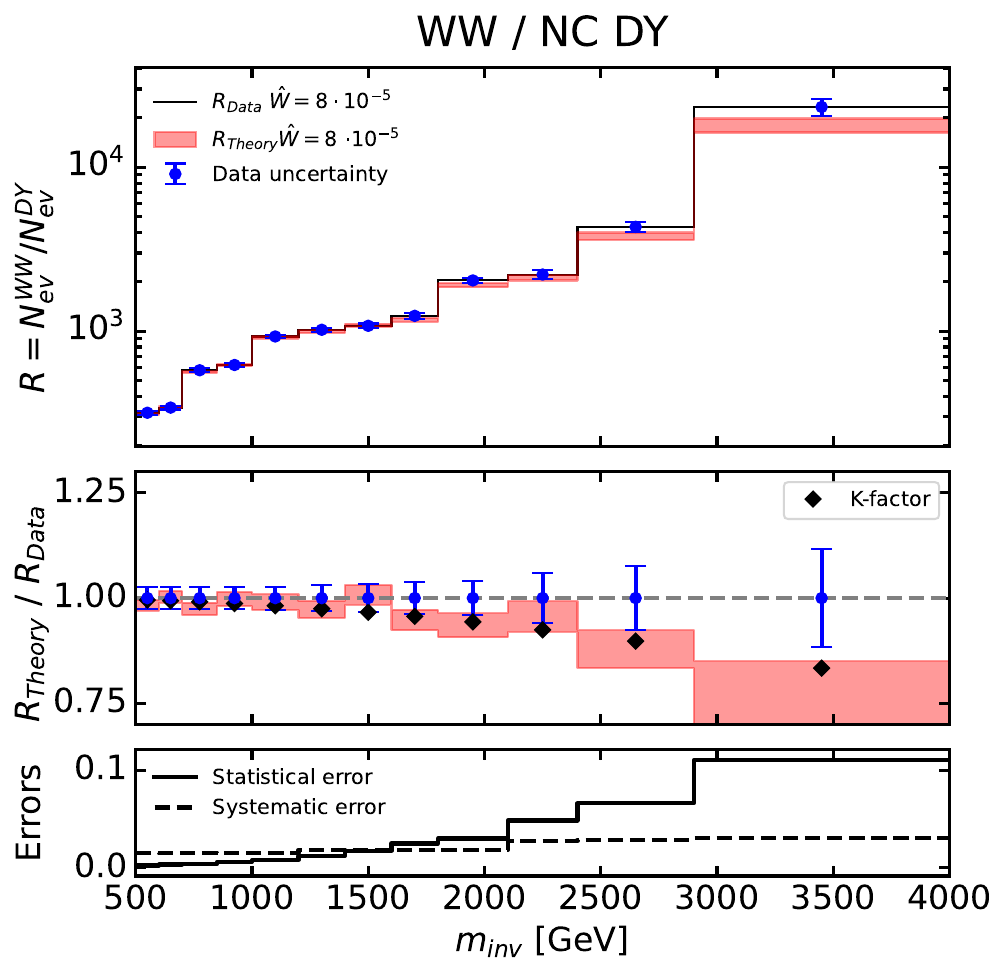}
    \includegraphics[width=0.49\linewidth]{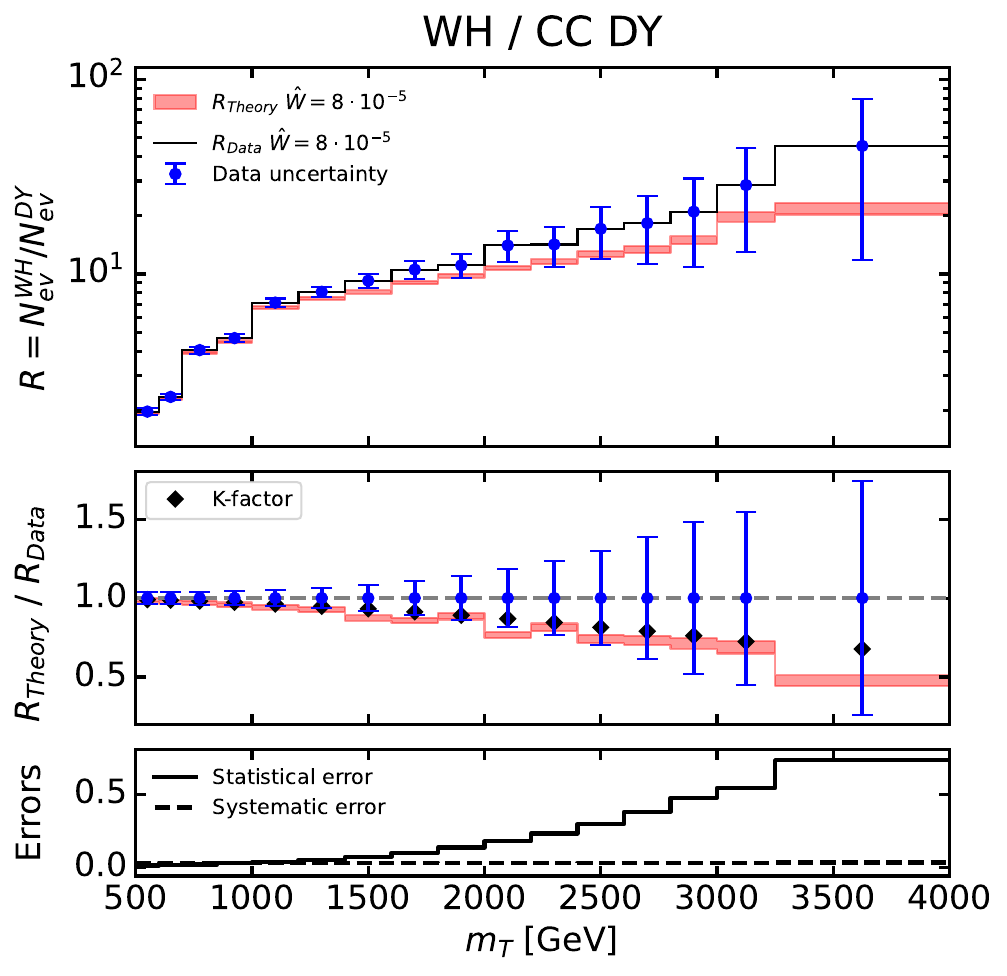}
  \end{center}
	\caption{
    Ratio between diboson production and Drell-Yan processes for the HL-LHC predictions. On the left, we show the ratio of $W^+ W^-$ to NC DY binned in invariant mass and on the right, we show the ratio of $WH$ to CC DY binned in transverse mass.
    In the top panel we plot the ratios of number of events for data and theory predictions.
    In the middle panel, we plot the ratio of those ratios (theory over data) alongside the $K$-factors. The lower panel displays the uncertainties.
  }
  \label{fig:diboson_dy_ratio}
\end{figure}

The Drell-Yan events we use are displayed in Fig. ~\ref{fig:wy_dy_prod}. The diboson events can be seen in Fig.~\ref{fig:ww_hllhc} for $WW$ and in Fig.~\ref{fig:wph_hllhc} for $W^{+}H$. However, note that we also include the $W^{-}H$ channel to measure the ratio of $WH$ and CC DY here. We plot the ratio of those quantities in Fig.~\ref{fig:diboson_dy_ratio}. We compare data and theory predictions where, as in Fig.~\ref{fig:wy_dy_prod}, data corresponds to a baseline PDF and a BSM partonic cross-section ($f_{\rm Baseline} \otimes \hat{\sigma} _{BSM}$) and theory is computed from a contaminated PDF and a SM partonic cross-section ($f_{\rm Cont} \otimes \hat{\sigma}_{SM}$). We also compare those results to $K$-factors which are obtained by taking the ratio of Drell-Yan BSM predictions over the SM ones. Practically the $K$-factors are a ratio of their respective partonic cross-section ($K = \hat{\sigma}^{DY} _{BSM} / \hat{\sigma}^{DY} _{SM}$).

We see in both cases a deviation between data and theory predictions growing with the energy. The uncertainties are smaller in the ratio $\text{WW} / \text{NC DY}$, which allows the discrepancy to be over $1 \sigma$ in the last bin. Furthermore, we also witness that the deviation follows the $K$-factors, which reinforces our initial assumption that using ratios greatly diminishes the impact of the PDFs.

As we mentioned earlier, the lesson we can get from this plot is that there is some new physics in either the DY or the diboson datasets. Unfortunately, without further information it is not possible to identify in which of those datasets the new physics is. Therefore, with just this plot in hand, the only reasonable decision would be to exclude the two datasets involved in the ratio where the deviation is observed from the fit. The downside of this disentangling method is that it might worsen the overall quality of the fit and increase the PDF uncertainties in certain regions of the parameter space. However, it proves to be an efficient solution against the sort of contamination we studied. Indeed, by excluding the DY datasets in this case, one would exclude the contamination we manually introduced there from the PDF fit.

\subsection{Alternative constraints on large-$x$ antiquarks}
\label{subsec:lowE}

In Sect.~\ref{subsec:forward}, it was shown that the inclusion of precise on-shell forward $W$ and $Z$ 
production measurements does not disentangle the contamination that new physics in the high-energy tails 
might yield. In this section, we ask ourselves whether there are any other future low-energy observables that 
might constrain large-$x$ antiquarks and show tension with the high-energy data in case the latter are 
affected by NP-induced incompatibilities.

\begin{figure}[t!]
  \begin{center}
    \includegraphics[width=0.49\linewidth]{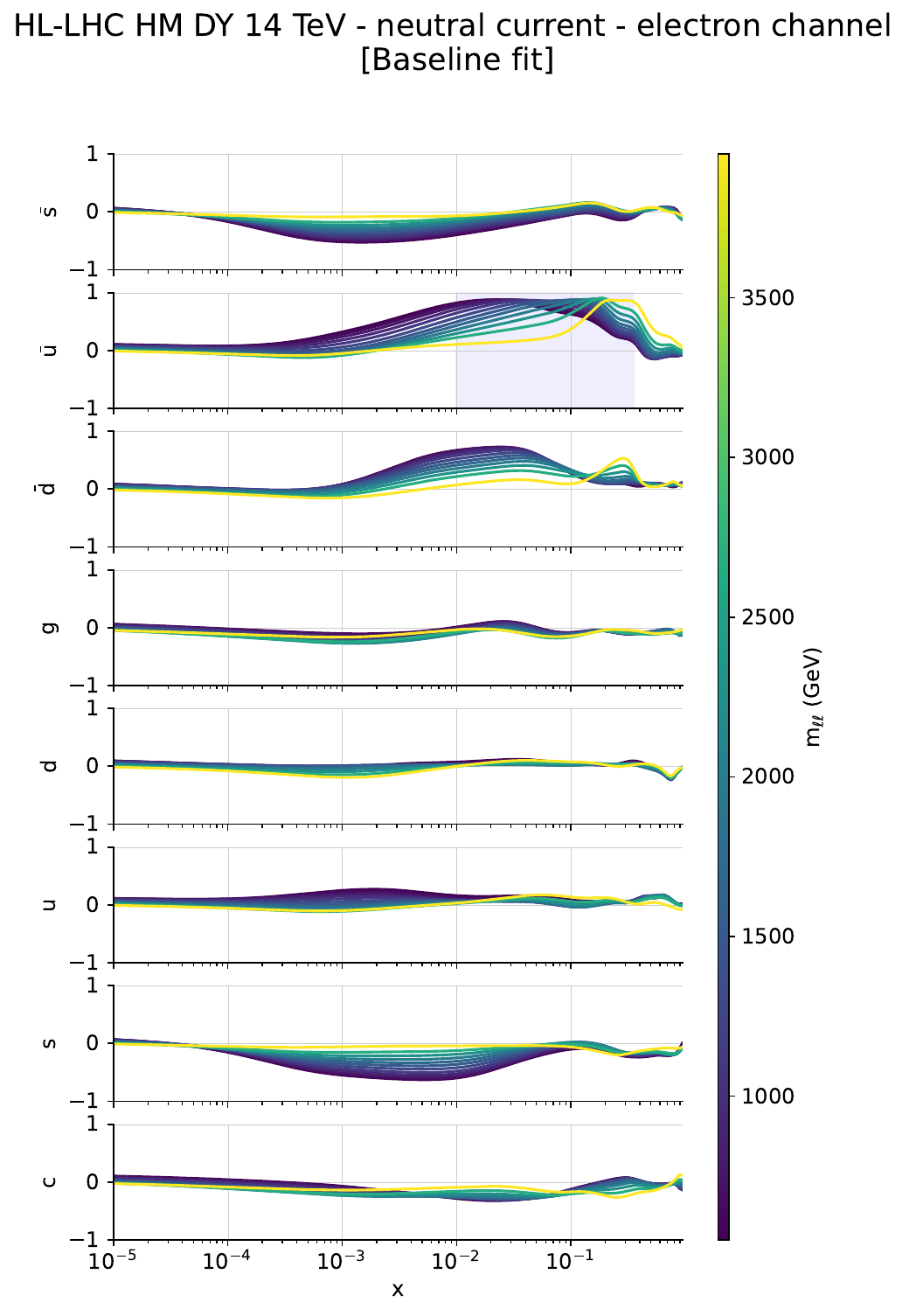}
    \includegraphics[width=0.49\linewidth]{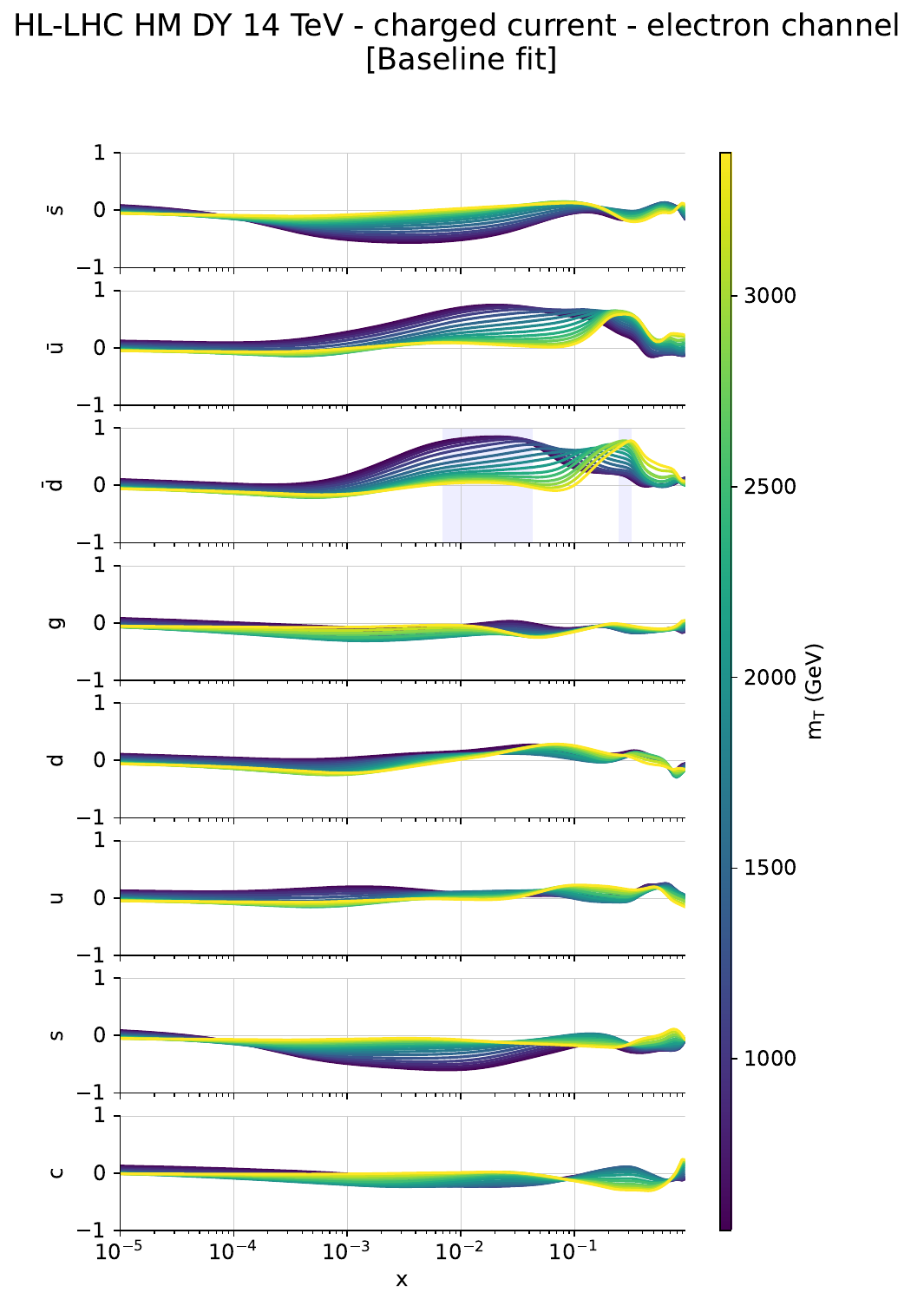}
    \end{center}
	\caption{\label{fig:smpdfplots_hl}Correlation coefficient $\rho$ defined in Eq.~\eqref{eq:correlations} between the flavour PDFs of the baseline  
        set and the HL-LHC neutral current Drell-Yan data (left panel); the HL-LHC charged current Drell-Yan data (right panel). The highlighted 
        region corresponds to $\rho>0.9\,\rho_{\rm max}$.}
\end{figure}

\begin{figure}[htb]
  \begin{center}
    \includegraphics[width=0.3\linewidth]{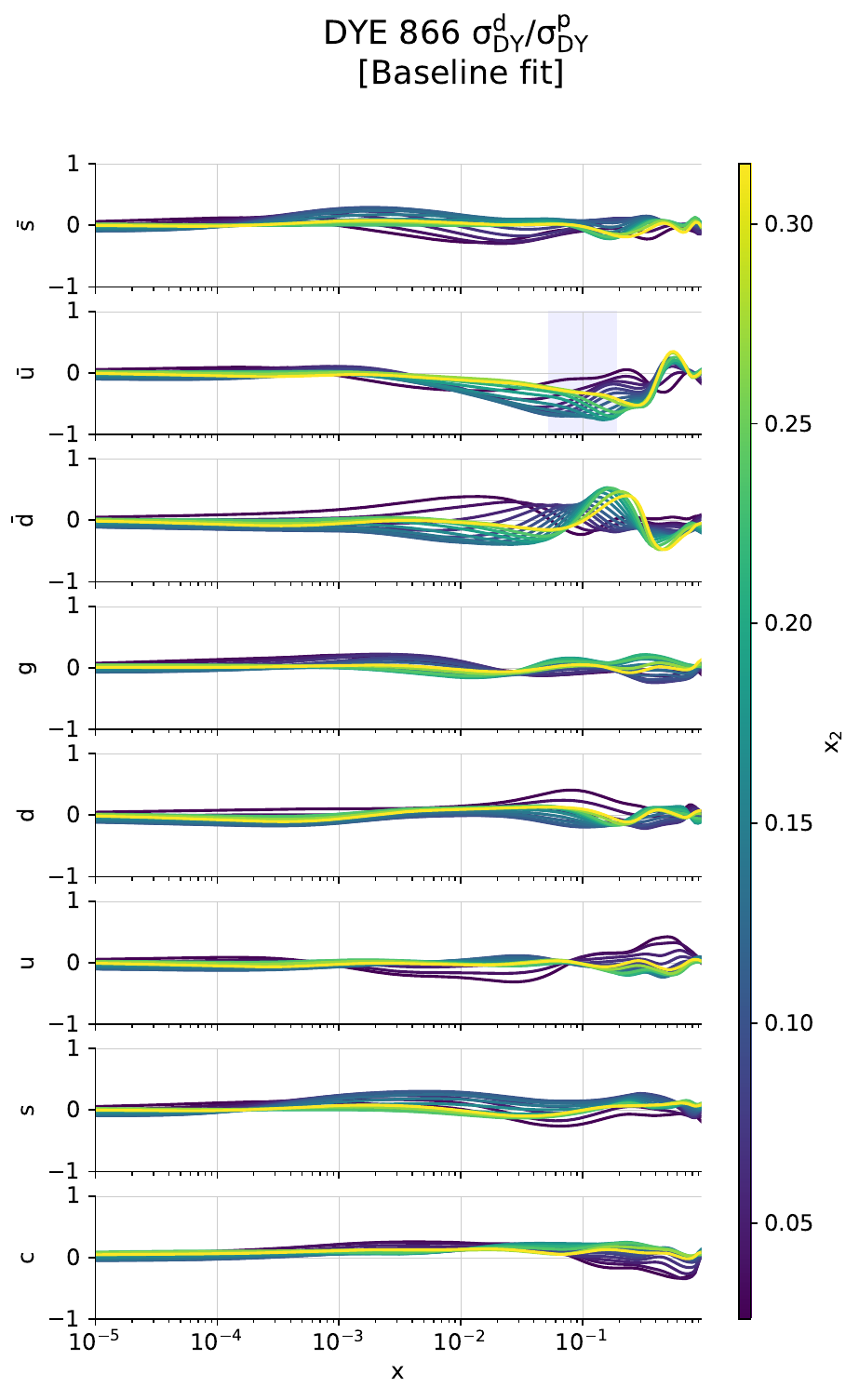}
    \includegraphics[width=0.3\linewidth]{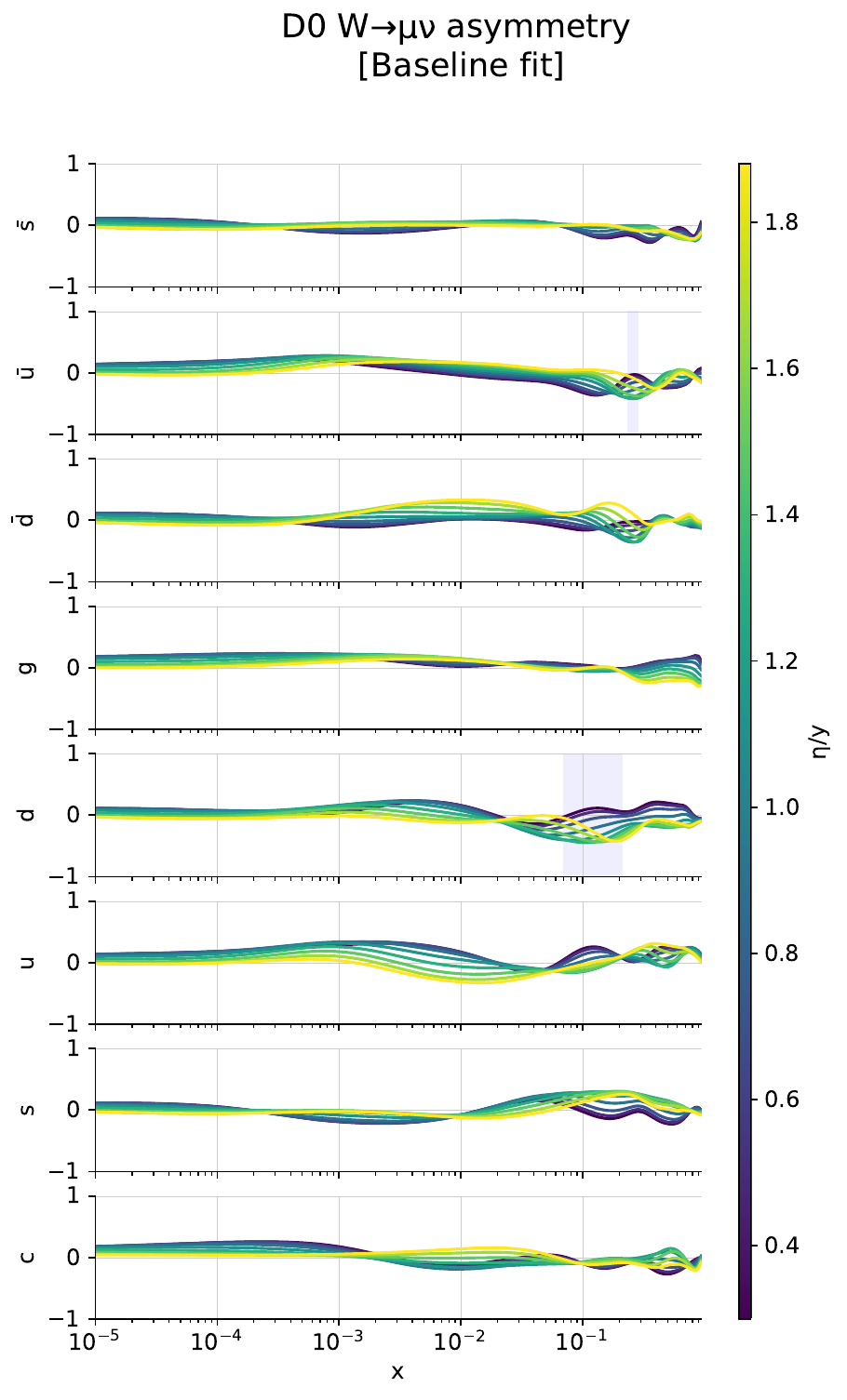}
    \includegraphics[width=0.3\linewidth]{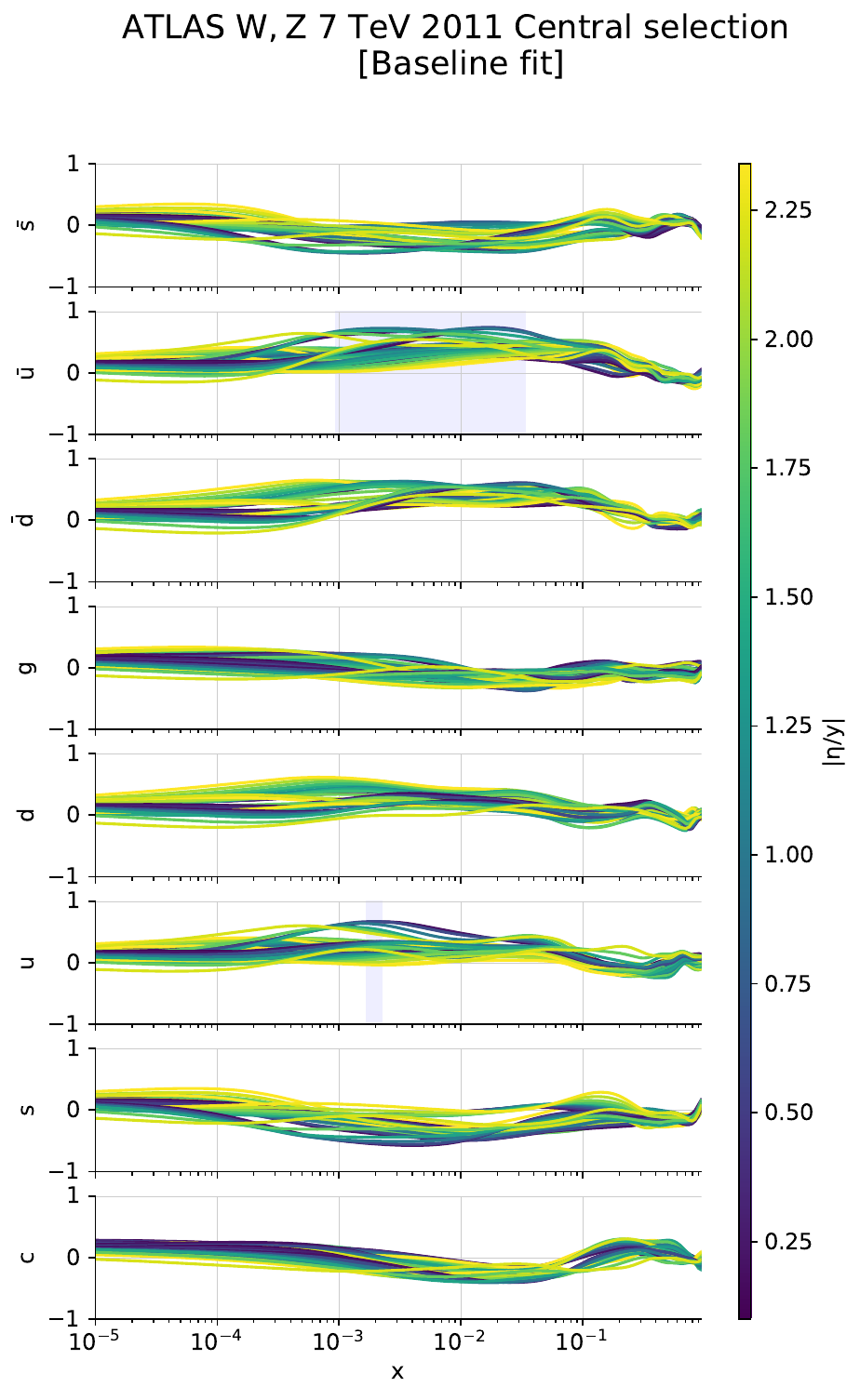}
    \end{center}
	\caption{\label{fig:smpdfplots} Same as Fig.~\ref{fig:smpdfplots_hl} for the FNAL E866 data measuring the ratio between low energy Drell-Yan muon pair production on 
 proton and deuteron targets~\cite{NuSea:2001idv} (left panel), the Tevatron D0 muon charge asymmetry~\cite{D0:2013xqc} (central panel) and the ATLAS 
      measurements of the $W$ and $Z$ differential cross-section at $\sqrt{s}=7$ TeV in the central rapidity region\cite{ATLAS:2016nqi} (right panel).}
\end{figure}

We start by looking at the correlation between the data that are currently included in our baseline PDF fit 
and the various PDF flavours. To assess the level of correlation, we plot the correlation defined in Ref.~\cite{Carrazza:2016htc}. 
The correlation function is defined as:
\begin{equation}
\label{eq:correlations}
    \rho(j,x,{\cal O})\equiv\frac{N_{\rm rep}}{N_{\rm rep}-1}
    \left(\frac{\langle f_j(x,Q)\,{\cal O}\rangle_{\rm reps}-\langle f_j(x,Q)\rangle_{\rm reps}\langle {\cal O}\rangle_{\rm reps}}{\Delta_{\rm PDF}f(x,Q)\,\Delta_{\rm PDF}{\cal O}}\right),
\end{equation}
where the PDFs are evaluated at a given scale $Q$ and the observable ${\cal O}$ is computed with the set of PDFs $f$, $j$ is the PDF 
flavour, $N_{\rm rep}$ is the number of replicas in the baseline PDF set and $\Delta_{\rm PDF}$ are the PDF uncertainties. 
In Figs.~\ref{fig:smpdfplots_hl} and~\ref{fig:smpdfplots} we show the correlation between the PDFs in the flavour basis 
and the observables which are strongly correlated with the antiquark distributions. 
The region highlighted in blue is the region in $x$ such that the correlation 
coefficient defined in Eq.~\eqref{eq:correlations} is larger than $0.9 \, \rho_{\rm max}$, where $\rho_{\rm max}$ is 
the maximum value that the correlation coefficient takes over the grid of points in $x$ and over the flavours $j$.
From Fig.~\ref{fig:smpdfplots_hl} we observe that while the largest invariant mass bins of 
the HL-LHC NC are most strongly correlated with the up antiquark distribution in the $10^{-2}\lesssim x \lesssim 3\cdot 10^{-1}$ region, 
the HL-LHC CC, particularly the lowest invariant mass bins, are most strongly correlated with the down antiquark distribution in 
the $ 7\cdot 10^{-3}\lesssim x \lesssim 5\cdot 10^{-2}$ region. This observation is quite interesting as it gives us a further insight into the 
difference between the $Z'$ and the $W'$ scenarios discussed at the end of Sect.~4.2. Indeed the $W'$ scenario affecting both the 
NC and CC distributions manages to compensate the $\bar{u}$ shift with the $\bar{d}$ shift in a slightly smaller region, hence the 
successful contamination. 

We now ask ourselves whether there are other observables that display a similar correlation pattern with the light antiquark distributions. 
In Fig.~\ref{fig:smpdfplots} we show the three most interesting showcases. In the left panel, we see that that the FNAL E866/NuSea measurements 
of the Drell-Yan muon pair production cross section from an 800 \rm{GeV} proton beam incident on proton and deuterium targets~\cite{NuSea:2001idv} 
yields constraints on the 
the ratio of anti-down to anti-up quark distributions in the proton in the large Bjorken-$x$ region, and the correlation is 
particularly strong with the anti-up in the $5\cdot 10^{-2}\lesssim x \lesssim 3\cdot 10^{-1}$ region.
The central panel shows that the Tevatron D0 muon charge asymmetry~\cite{D0:2013xqc} exhibits a strong correlation with the 
up antiquark around $x\approx 0.3$ and the down quark around $x\approx 0.1$. This is understood, as by charge conjugation the 
anti-up distribution of the proton corresponds to the up distribution of the anti-proton. Finally, on the right 
panel we see that the precise ATLAS measurements of the $W$ and $Z$ differential cross-section at $\sqrt{s}=7$ TeV~\cite{ATLAS:2016nqi} 
have a strong constraining power on the up antiquark in a slightly lower $x$ region around $3\cdot 10^{-3}\lesssim x \lesssim 2\cdot 10^{-2}$.

The results presented in Sect.~\ref{sec:dy} show that the tension with the low-energy datasets that 
constrain the same region in $x$ as the high-mass Drell-Yan HL-LHC data is not strong enough to flag the 
HL-LHC datasets. Hence, the conditions highlighted in Sect.~\ref{subsec:postfit} are necessary in order to determine a bulk of maximally
consistent datasets, but they are not sufficient, as they still allow
new physics contamination to go undetected. 
A way to emphasise the tension is to produce {\it weighted fits} which give a larger weight to the high-energy data that are affected by 
new physics effects. The rationale behind this is that, if some energy-growing
effect associated to the presence of new physics in the data shows up in the tails
of the distributions, PDFs might accommodate this effect without deteriorating the agreement with the other datasets 
only up to a point. If the datasets that are affected by new
physics are given a large weight in the fit, the tension with the datasets constraining the large-$x$ region 
that are perfectly described by the SM could in principle get worse. Hence by giving a
larger weight to a specific NP-affected dataset, the disagreement with the datasets highlighted 
above should become more apparent. 
Depending on the kind of new physics model that Nature might display in the data, 
the effect might affect either of the three
classes of the high energy processes entering PDF fits, namely: (i)
jets and dijets, (ii) top and (iii) Drell-Yan. In our example, in order to emphasise the tension with the 
low-energy Drell-Yan data and the Tevatron data, we would have to give more weight to the HL-LHC high-mass 
Drell Yan data. However, performing this exercise we observe that, although the $\chi^2$ of the HL-LHC Drell-Yan further improves and the 
ones of the highlighted data deteriorates, the level of deterioration is never strong enough to flag the tension. 
The result of this test points to the fact that one should include independent and more precise 
low-energy/large-$x$ constraints in future PDF analyses\footnote{Note that some of this data have the disadvantage of being affected by 
additional uncertainties associated with nuclear corrections, target mass corrections, higher twists and other low-energy effects.}
if one wants to safely exploit the constraining power of high-energy 
data without inadvertently absorbing signs of new physics 
in the high-energy tails. In this sense the EIC programme~\cite{Khalek:2021ulf,Abir:2023fpo}, as well as 
other low-energy data which are not exploited in the standard PDF global fits, such 
as JLAB~\cite{Accardi:2023chb,Accardi:2021ysh} or STAR and SeaQuest measurements~\cite{Accardi:2023gyr,Guzzi:2021fre,Alekhin:2023uqx}, will be a precious 
input in future PDF analyses, alongside the constraints from lattice data~\cite{Hou:2022onq}. 



\section{Summary}
\label{sec:summary}

In this work we have analysed two concrete new physics scenarios
that would distort the high energy Drell-Yan invariant mass distributions that
enter PDF fits. We considered scenarios of flavour universal NP, manifested in the form of a heavy 
$Z^\prime$ and $W^\prime$ coupled to both quarks and leptons. For simplicity, we parametrised the modified interactions
by means of an EFT, where the corresponding Wilson coefficients are denoted $\hat{Y}$ and $\hat{W}$ respectively.
By generating pseudodata in different benchmark scenarios, we assessed the ability of the PDF fitting framework to
absorb the modified interactions in the PDF parametrisation, effectively hiding NP inside the modelling of the proton.
For this exploratory study we chose values of the Wilson coefficients that are not strongly disfavoured from global EFT fits.
In terms of data, we considered the state-of-the-art {\tt NNPDF4.0} dataset and its extension with
future HL-LHC projections. 

The first important conclusion from our study is that current Drell-Yan data is
not precise enough to lead to a contaminated PDF set. The uncertainties on the Drell-Yan tails of the distribution
are big enough to render the NP effects sub-leading corrections with respect to the PDF uncertainty.
On the other had, when HL-LHC projections are included, we see that more interesting effects occur.
In particular, while the heavy $Z'$ scenario does not 
lead to any contamination, a flexible enough PDF parametrization would be able to fit away signs of a heavy $W'$ boson when performing a global PDF determination and thus
introduce spurious contamination in the large-$x$ structure of the proton. 

In particular, this has been assessed by 
looking at possible contamination in both the neutral and charged currents,
finding that it is when both charged current and neutral current observables are effected by NP
that we have the highest freedom
of parametrisation. This is ultimately traced back to the lack of data constraining the large-$x$
antiquark PDFs, in particular the $\bar{d}$ PDF. For this reason, we observe that, when
data are contaminated with the presence of a $W^\prime$, we generally find good fits and are able to accommodate
even large deviations from the SM. It is however worth noting that
the leading source of contaminated data are the high-statistics data expected from the HL-LHC, as present data would not be as susceptible to the $W^\prime$ effects.

One could argue that, for a safe use of  PDFs in the context of searches for new physics, 
datasets possibly susceptible to contamination from new physics should be systematically left out of PDF fits. 
But this would be naive: if a deviation from a SM projection based on available PDF fits were found in the large-$x$ tails of some distribution, 
the first systematics to double check would be the robustness of the PDF parametrisation in the region of the anomaly. 
This should be assessed by means such as those presented in this work, namely checking whether the deviation can be washed 
away by modifying the PDFs in a way that does not significantly impact the overall quality of the global fit. 
If it can, the deviation should be attributed to the PDF systematics, rather than to new physics. 

We discussed possible consequences of having a contaminated PDF set, finding that such NP contamination would have consequences
in the interpretation of the LHC data. On the one hand, by computing predictions with the aforementioned set of PDFs, one would likely hinder searches for new physics
affected by the presence of the $W^\prime$, leading to biased exclusion bounds. On the other hand, one would also potentially see deviations
from the SM predictions in processes not affected by the presence of the heavy state, purely as a consequence of the spurious
behaviour of the PDFs at large-$x$.

We discussed possible strategies to disentangle the NP effects from the PDF determination. In particular, we verified that a post-fit sliding invariant mass cut on the data entering the $\chi^2$ calculation would not highlight any trend, indicating that the PDF set describes the high invariant mass data-points across the dataset well. We also checked whether the contaminated PDF set could be flagged by HL-LHC projections of on-shell forward $Z/W$ production, an observable that would 
probe high-$x$ but not high-$Q$. However, as previously mentioned, the contamination seems to be related more to the flexibility in the antiquark PDF flavours, which in the case of forward EW boson production are mostly probed in the low-$x$ regime.

However, a more effective strategy of disentangling PDFs and NP is given by the study of the ratio of differential cross sections. By exploiting the fact that different processes might be dependent on the same parton luminosity channels, one could devise observables that remove the dependence on the PDF set by taking ratios of cross sections in similar kinematical regions. In the case of $W^\prime$ contamination, a promising test would be to take for instance the ratio of neutral/charged current Drell-Yan against neutral/charged diboson production.  We found that this test could in principle ascertain the presence of NP independently of the PDF set, with the ultimate caveat that the source of the deviation could either be affecting the numerator, the denominator or both.

Finally, we discussed the potential for low-energy observables to provide complementary constraints on the large-$x$ sea quarks.  Such low-energy observables would not be as susceptible to the effects of new physics, and could therefore show a tension with the NP-affected high-energy data.  We found that, although the PDF fit 
includes low-energy observables that exhibit a large correlation with the large-$x$ $\bar{d}$ and $\bar{u}$ PDFs, the precision offered by these measurements is not sufficient to create a tension with the high-energy data.  Future precision measurements of the large-$x$ sea quarks, for example the EIC programme, will provide important inputs in PDF fits to avoid the contamination studied here.

To conclude, in this work we tackled the problem of inconsistent data in PDF determination.
Although it is well known that inconsistencies between experimental data entering
a global PDF determination can distort the statistical interpretation of
PDF uncertainties and although there are mechanisms to select
a sufficiently consistent bulk of data to use as an input of a global PDF analysis,
the inconsistency of any individual dataset with the bulk of the global fit may suggest that its understanding, 
either from the theoretical or experimental point of view, is not complete. How can we establish whether the inconsistency 
comes from missing higher order uncertainties, data inconsistencies or unaccounted new physics effects?
In this work we tackled the latter, trying to take advantage of the fact that NP contamination has a different 
energy-scaling behaviour compared to effects that might arise from missing higher orders or from experimental inconsistencies. 
We tackled it by setting up a set of tools and analyses built on statistical closure tests in a specific and simple 
test-case scenario. 

In summary, we have provided a concrete example of the issues that can emerge from possible new physics effects 
in the data used in a fit of PDFs, and how to possibly address those issues, when trying to expose new physics from 
departures of SM predictions in the tails of kinematical distributions at large energy. 
The tools that were developed to analyse the problem of PDF contamination can be extended to deal with more general scenarios than the ones we studied here. 
However dedicated studies must be considered to scrutinise different scenarios of new physics contamination of PDFs that arise from other distributions and 
objects, including e.g. high-$E_T$ jets, tops, vector or Higgs bosons.  We hope that this work will trigger analogous studies. To facilitate such studies, 
we make public the  analyses and tools used in this work along with a detailed set of instructions at
\begin{center}
    \url{https://www.pbsp.org.uk/contamination/}
\end{center}
so that any users can utilise and modify the available scripts in conjunction with the public {\tt NNPDF} code~\cite{NNPDF:2021uiq} 
to test the robustness of the signal associated to a specific BSM model against any possible absorption by the PDFs. 
If the NP scenario affects datasets that are already included in the {\tt NNPDF4.0} analysis, the injection of new physics in the 
MC data is straightforward. If users wish to test the effect of a given NP scenario by including more MC data in the analysis, 
instructions can be found in the public {\tt NNPDF} documentation\footnote{https://docs.nnpdf.science/tutorials/buildmaster.html}.

\section*{Acknowledgements}
We thank Juan Rojo, Giulia Zanderighi, Marius Wiesemann and Admir Greljo for initial discussions about the direction of the project. 
We thank Emanuele Nocera for his precious insight into the generation of the Monte Carlo artificial data.
We thank the members of the NNPDF collaboration and of the Pheno working group at the University of Cambridge for useful
discussion during the project.
M.~U., J.~M.~M.~, E.~H.~, L.~M., M.~M.~A. and Z.~K. are supported by the European Research Council under the
European Union’s Horizon 2020 research and innovation Programme (grant agreement n.950246). M.~U., J.~M.~M.~, E.~H.~,L.~M., M.~M.~A. and Z.~K. are also 
partially supported by the STFC grant ST/T000694/1. J.~M.~M. is also supported by the Sims Fund Studentship.
The research of M.~M. is supported by the Deutsche Forschungsgemeinschaft (DFG, German Research Foundation) under grant 396021762 – TRR 257 Particle Physics Phenomenology
after the Higgs Discovery.

\appendix
\section{Fit quality}
\label{app:fit}

\begin{figure}[htb]

  \includegraphics[width=0.98\linewidth]{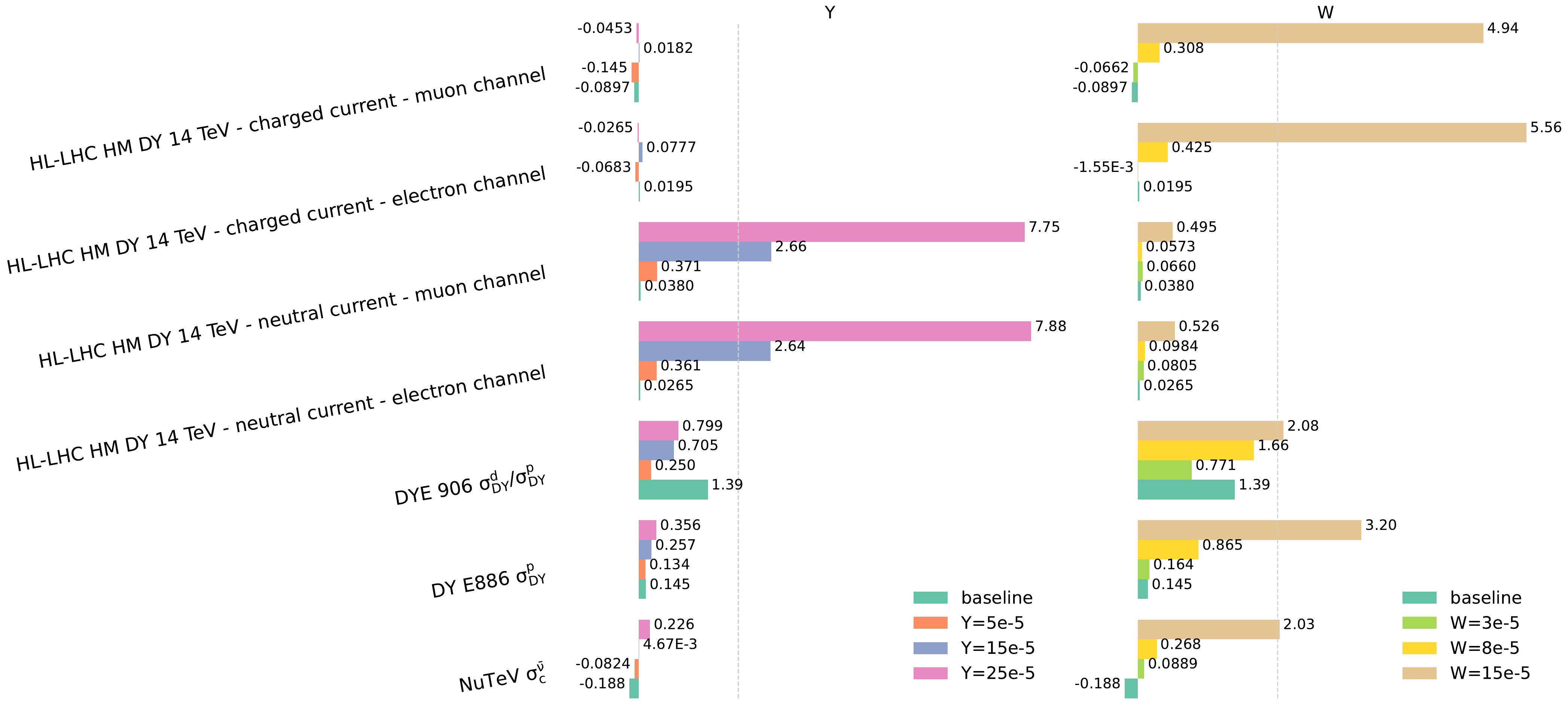}
  \caption{
    \label{fig:nsigma_multi}
  Value of $n_{\sigma}$, defined in Eq.~\eqref{eq:nsigma}
    for all datasets that pass the threshold criterion of
    $n_{\sigma}>2$ discussed in Sect.~\ref{subsec:postfit} in each of the
    three fits performed by injecting various degrees of new physics.
    The figure on the left, new physics signals in the data are added
    according to Scenario I (flavour-universal $Z^{'}$ model), namely the baseline $\hat{Y}=0$ (green bars),
    $\hat{Y}=5\cdot 10^{-5}$ (orange bars), $\hat{Y}=15\cdot 10^{-5}$
    (blue bars) and $\hat{Y}=25\cdot 10^{-5}$ (pink bars).
    In the figure on the right, signals are added according to Scenario II
    (flavour-universal $W^{'}$ model), namely the baseline $\hat{W}=0$ (again, green
    bars), $\hat{W}=3\cdot 10^{-5}$ (light green bars), $\hat{W}=8\cdot 10^{-5}$
    (yellow bars) and $\hat{W}=15\cdot 10^{-5}$ (brown bars)}
\label{fig:Ynsigma}
\end{figure}

\begin{figure}[htb]
  
  \includegraphics[width=0.90\linewidth]{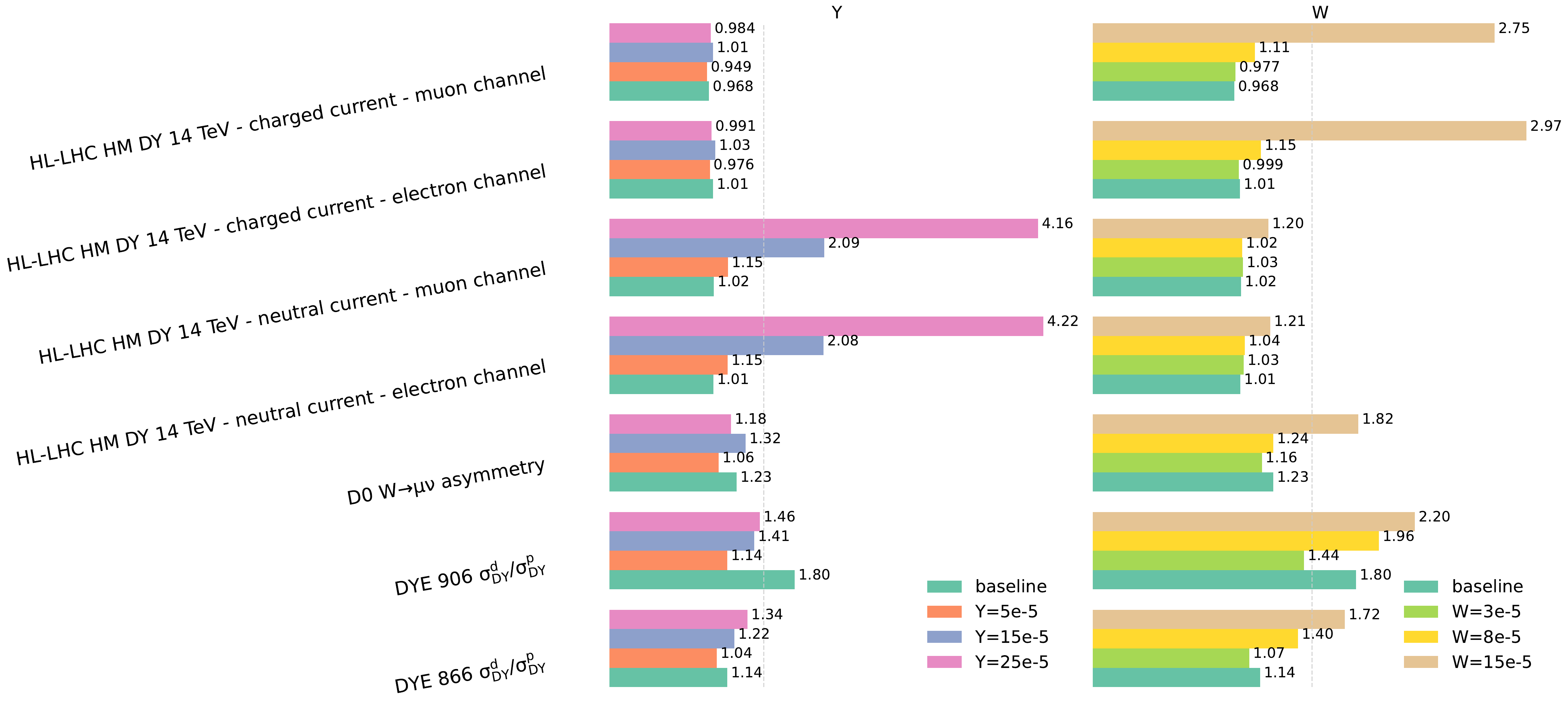}
  \caption{
  \label{fig:chi2_multi}
  $\chi^2/n_{\rm dat}$ distribution  for all datasets that pass the threshold criterion of
    $\chi^2/n_{\rm dat}>1.5$ discussed in Sect.~\ref{subsec:postfit} in each of the
    three fits performed by injecting various degrees of new physics
    signals in the data according to Scenario I (left panel) and
    Scenario II (right panel)  }
\end{figure}

In this appendix we give details about the fit-quality of the closure tests presented in this work.
In Table~\ref{tab:chi2_y} and~\ref{tab:chi2_w}, we list the value of the reduced $\chi^2/n_{\rm dat}$ 
as well as of the $n_\sigma$ estimator (see Sect.~\ref{subsec:postfit} for details) 
for each dataset included in the fit, under all the contamination scenarios we have tested. 
We have highlighted the datasets whose fit quality deteriorates the most in Figs.~\ref{fig:nsigma_multi} and~\ref{fig:chi2_multi}. 
In particular, the two figures showcase the tension between the fixed-target datasets and the HL-LHC projected data as the value of the $\hat{W}$ increases.

\begin{table}[t]
        \tiny
        \centering
\label{tab:chi2_y}
\begin{tabular}{lrrrrrrrr}
\toprule
 & \multicolumn{2}{r}{baseline} & \multicolumn{2}{r}{Y=5e-5} & \multicolumn{2}{r}{Y=15e-5} & \multicolumn{2}{r}{Y=25e-5} \\
 & $\chi^2$ & $n_\sigma$ & $\chi^2$ & $n_\sigma$ & $\chi^2$ & $n_\sigma$ & $\chi^2$ & $n_\sigma$ \\
\midrule
NMC $d/p$ & 1.02 & 0.14 & 1.02 & 0.12 & 1.03 & 0.24 & 1.06 & 0.45 \\
NMC $p$ & 1.03 & 0.26 & 1.02 & 0.23 & 1.02 & 0.18 & 1.02 & 0.18 \\
SLAC $p$ & 1.02 & 0.06 & 1.01 & 0.05 & 1.01 & 0.03 & 1.02 & 0.07 \\
SLAC $d$ & 1.00 & -0.01 & 0.98 & -0.07 & 0.99 & -0.02 & 0.99 & -0.04 \\
BCDMS $p$ & 1.02 & 0.20 & 1.00 & 0.06 & 1.02 & 0.21 & 1.01 & 0.11 \\
BCDMS $d$ & 1.01 & 0.07 & 1.00 & 0.01 & 1.01 & 0.08 & 1.00 & 0.03 \\
CHORUS $\sigma_{CC}^{\nu}$ & 1.00 & 0.02 & 1.00 & -0.06 & 1.00 & 0.06 & 1.00 & 0.01 \\
CHORUS $\sigma_{CC}^{\bar{\nu}}$ & 0.99 & -0.13 & 0.99 & -0.13 & 1.00 & -0.03 & 0.99 & -0.08 \\
NuTeV $\sigma_{c}^{\nu}$ & 0.99 & -0.06 & 0.99 & -0.05 & 0.99 & -0.02 & 1.00 & 0.02 \\
NuTeV $\sigma_{c}^{\bar{\nu}}$ & 0.96 & -0.19 & 0.98 & -0.08 & 1.00 & 0.00 & 1.05 & 0.23 \\
HERA I+II inclusive NC $e^-p$ & 1.00 & -0.02 & 1.01 & 0.12 & 1.00 & 0.02 & 1.02 & 0.17 \\
HERA I+II inclusive NC $e^+p$ 460 GeV & 1.01 & 0.08 & 1.01 & 0.12 & 1.01 & 0.13 & 1.02 & 0.18 \\
HERA I+II inclusive NC $e^+p$ 575 GeV & 0.98 & -0.21 & 1.00 & 0.01 & 0.99 & -0.17 & 1.01 & 0.07 \\
HERA I+II inclusive NC $e^+p$ 820 GeV & 1.00 & -0.00 & 1.01 & 0.07 & 1.00 & -0.01 & 1.01 & 0.09 \\
HERA I+II inclusive NC $e^+p$ 920 GeV & 1.02 & 0.29 & 1.05 & 0.63 & 1.03 & 0.35 & 1.05 & 0.67 \\
HERA I+II inclusive CC $e^-p$ & 0.99 & -0.05 & 1.03 & 0.13 & 0.99 & -0.03 & 1.03 & 0.15 \\
HERA I+II inclusive CC $e^+p$ & 1.02 & 0.08 & 1.02 & 0.07 & 1.03 & 0.11 & 1.04 & 0.18 \\
HERA comb. $\sigma_{c\bar c}^{\rm red}$ & 1.00 & 0.02 & 1.02 & 0.09 & 1.01 & 0.05 & 1.03 & 0.11 \\
HERA comb. $\sigma_{b\bar b}^{\rm red}$ & 1.12 & 0.43 & 1.13 & 0.45 & 1.14 & 0.49 & 1.13 & 0.48 \\
DYE 866 $\sigma^d_{\rm DY}/\sigma^p_{\rm DY}$ & 1.14 & 0.40 & 1.04 & 0.12 & 1.22 & 0.59 & 1.34 & 0.94 \\
DY E886 $\sigma^p_{\rm DY}$ & 1.02 & 0.14 & 1.02 & 0.13 & 1.04 & 0.26 & 1.05 & 0.36 \\
DY E605 $\sigma^p_{\rm DY}$ & 1.08 & 0.53 & 1.07 & 0.43 & 1.06 & 0.42 & 1.06 & 0.39 \\
DYE 906 $\sigma^d_{\rm DY}/\sigma^p_{\rm DY}$ & 1.80 & 1.39 & 1.14 & 0.25 & 1.41 & 0.70 & 1.46 & 0.80 \\
CDF $Z$ rapidity (new) & 1.06 & 0.21 & 1.03 & 0.12 & 1.06 & 0.21 & 1.01 & 0.06 \\
D0 $Z$ rapidity & 1.03 & 0.10 & 1.02 & 0.08 & 1.04 & 0.17 & 1.03 & 0.11 \\
D0 $W\to \mu\nu$ asymmetry & 1.23 & 0.50 & 1.06 & 0.13 & 1.32 & 0.69 & 1.18 & 0.38 \\
ATLAS $W,Z$ 7 TeV 2010 & 1.05 & 0.20 & 1.04 & 0.17 & 1.05 & 0.20 & 1.05 & 0.20 \\
ATLAS HM DY 7 TeV & 1.02 & 0.04 & 1.05 & 0.12 & 1.01 & 0.02 & 1.03 & 0.09 \\
ATLAS low-mass DY 2011 & 0.90 & -0.17 & 1.04 & 0.07 & 0.87 & -0.22 & 1.01 & 0.01 \\
ATLAS $W,Z$ 7 TeV 2011 Central selection & 1.06 & 0.28 & 1.07 & 0.36 & 1.07 & 0.31 & 1.07 & 0.35 \\
ATLAS $W,Z$ 7 TeV 2011 Forward selection & 0.91 & -0.25 & 1.33 & 0.90 & 0.90 & -0.29 & 1.32 & 0.87 \\
ATLAS DY 2D 8 TeV high mass & 1.02 & 0.11 & 1.03 & 0.13 & 1.02 & 0.08 & 1.03 & 0.12 \\
ATLAS DY 2D 8 TeV low mass & 1.03 & 0.16 & 1.00 & 0.00 & 1.02 & 0.13 & 0.99 & -0.04 \\
ATLAS $W,Z$ inclusive 13 TeV & 1.07 & 0.09 & 1.08 & 0.10 & 1.09 & 0.11 & 1.13 & 0.16 \\
ATLAS $W^+$+jet 8 TeV & 1.17 & 0.46 & 0.96 & -0.11 & 1.17 & 0.48 & 0.95 & -0.13 \\
ATLAS $W^-$+jet 8 TeV & 1.19 & 0.51 & 0.97 & -0.09 & 1.20 & 0.56 & 0.97 & -0.08 \\
ATLAS $Z$ $p_T$ 8 TeV $(p_T^{ll},M_{ll})$ & 1.01 & 0.03 & 0.98 & -0.07 & 1.01 & 0.04 & 0.99 & -0.05 \\
ATLAS $Z$ $p_T$ 8 TeV $(p_T^{ll},y_{ll})$ & 0.98 & -0.10 & 0.94 & -0.31 & 0.98 & -0.10 & 0.93 & -0.36 \\
ATLAS $\sigma_{tt}^{\rm tot}$ & 1.03 & 0.02 & 1.14 & 0.10 & 1.05 & 0.03 & 1.19 & 0.14 \\
ATLAS $\sigma_{tt}^{\rm tot}$ 8 TeV & 1.31 & 0.22 & 1.12 & 0.08 & 1.27 & 0.19 & 1.08 & 0.06 \\
ATLAS $\sigma_{tt}^{\rm tot}$ 13 TeV Run II full lumi & 0.92 & -0.06 & 0.93 & -0.05 & 0.97 & -0.02 & 0.98 & -0.01 \\
ATLAS $t\bar{t}$ $y_t$ & 1.03 & 0.05 & 1.06 & 0.08 & 1.04 & 0.06 & 1.04 & 0.05 \\
ATLAS $t\bar{t}$ $y_{t\bar{t}}$ & 1.04 & 0.05 & 1.04 & 0.05 & 1.06 & 0.08 & 1.05 & 0.07 \\
ATLAS $t\bar{t}$ normalised $|y_t|$ & 1.13 & 0.21 & 1.13 & 0.20 & 1.15 & 0.24 & 1.17 & 0.26 \\
ATLAS jets 8 TeV, R=0.6 & 0.83 & -1.53 & 0.94 & -0.58 & 0.83 & -1.55 & 0.94 & -0.60 \\
ATLAS dijets 7 TeV, R=0.6 & 1.03 & 0.19 & 1.00 & -0.00 & 1.04 & 0.24 & 1.01 & 0.09 \\
ATLAS direct photon production 13 TeV & 0.97 & -0.16 & 1.03 & 0.14 & 0.98 & -0.11 & 1.04 & 0.21 \\
ATLAS single top $R_{t}$ 7 TeV & 1.14 & 0.10 & 1.26 & 0.18 & 1.05 & 0.03 & 1.15 & 0.11 \\
ATLAS single top $R_{t}$ 13 TeV & 0.91 & -0.07 & 1.01 & 0.01 & 0.93 & -0.05 & 1.04 & 0.03 \\
ATLAS single top $y_t$ (normalised) & 0.94 & -0.07 & 1.07 & 0.09 & 0.93 & -0.09 & 1.04 & 0.04 \\
ATLAS single antitop $y$ (normalised) & 0.92 & -0.10 & 0.91 & -0.11 & 0.97 & -0.04 & 0.98 & -0.03 \\
CMS $W$ asymmetry 840 pb & 0.99 & -0.03 & 0.98 & -0.04 & 0.98 & -0.04 & 1.00 & -0.01 \\
CMS $W$ asymmetry 4.7 fb & 0.97 & -0.07 & 0.97 & -0.06 & 0.97 & -0.08 & 1.00 & 0.00 \\
CMS Drell-Yan 2D 7 TeV 2011 & 1.01 & 0.05 & 1.01 & 0.07 & 1.00 & 0.04 & 1.01 & 0.10 \\
CMS $W$ rapidity 8 TeV & 1.06 & 0.21 & 1.12 & 0.39 & 1.07 & 0.22 & 1.13 & 0.42 \\
CMS $Z$ $p_T$ 8 TeV $(p_T^{ll},y_{ll})$ & 1.03 & 0.12 & 1.03 & 0.11 & 1.03 & 0.12 & 1.04 & 0.14 \\
CMS dijets 7 TeV & 0.97 & -0.15 & 1.05 & 0.24 & 0.97 & -0.14 & 1.05 & 0.25 \\
CMS jets 8 TeV & 0.99 & -0.11 & 1.00 & -0.03 & 0.99 & -0.09 & 1.00 & -0.01 \\
CMS $\sigma_{tt}^{\rm tot}$ 7 TeV & 0.86 & -0.10 & 0.95 & -0.03 & 0.86 & -0.10 & 1.00 & 0.00 \\
CMS $\sigma_{tt}^{\rm tot}$ 8 TeV & 1.18 & 0.13 & 1.09 & 0.06 & 1.21 & 0.15 & 1.07 & 0.05 \\
CMS $\sigma_{tt}^{\rm tot}$ 13 TeV & 0.98 & -0.01 & 1.11 & 0.08 & 0.99 & -0.00 & 1.12 & 0.09 \\
CMS $t\bar{t}$ rapidity $y_{t\bar{t}}$ & 1.06 & 0.12 & 1.04 & 0.08 & 1.04 & 0.09 & 1.01 & 0.02 \\
CMS $\sigma_{tt}^{\rm tot}$ 5 TeV & 0.86 & -0.10 & 0.77 & -0.17 & 0.82 & -0.13 & 0.75 & -0.18 \\
CMS $t\bar{t}$ double differential $(m_{t\bar{t}},y_{t\bar{t}})$ & 0.99 & -0.04 & 1.00 & 0.01 & 1.02 & 0.04 & 1.03 & 0.07 \\
CMS $t\bar{t}$ absolute $y_t$ & 1.01 & 0.03 & 1.02 & 0.04 & 1.02 & 0.05 & 1.03 & 0.06 \\
CMS $t\bar{t}$ absolute $|y_t|$ & 0.98 & -0.05 & 0.99 & -0.03 & 0.97 & -0.06 & 0.96 & -0.09 \\
CMS single top $\sigma_{t}+\sigma_{\bar{t}}$ 7 TeV & 0.93 & -0.05 & 0.91 & -0.06 & 0.89 & -0.08 & 0.86 & -0.10 \\
CMS single top $R_{t}$ 8 TeV & 0.64 & -0.26 & 1.21 & 0.15 & 0.63 & -0.26 & 1.14 & 0.10 \\
CMS single top $R_{t}$ 13 TeV & 1.50 & 0.35 & 1.44 & 0.31 & 1.46 & 0.33 & 1.42 & 0.30 \\
LHCb $Z$ 940 pb & 1.08 & 0.17 & 0.96 & -0.09 & 1.11 & 0.24 & 0.97 & -0.06 \\
LHCb $Z\to ee$ 2 fb & 1.03 & 0.08 & 1.04 & 0.13 & 1.01 & 0.02 & 1.04 & 0.11 \\
LHCb $W,Z \to \mu$ 7 TeV & 0.98 & -0.07 & 0.97 & -0.10 & 1.02 & 0.06 & 1.01 & 0.05 \\
LHCb $W,Z \to \mu$ 8 TeV & 1.08 & 0.32 & 1.13 & 0.51 & 1.09 & 0.35 & 1.18 & 0.68 \\
LHCb $Z\to \mu\mu$ & 1.09 & 0.26 & 1.05 & 0.15 & 1.10 & 0.27 & 1.05 & 0.13 \\
LHCb $Z\to ee$ & 1.07 & 0.19 & 1.03 & 0.08 & 1.07 & 0.19 & 1.04 & 0.10 \\
CMS HM DY 8 TeV & 0.98 & -0.09 & 0.98 & -0.08 & 0.97 & -0.13 & 0.98 & -0.10 \\
CMS HM DY 13 TeV - combined channel & 0.97 & -0.14 & 0.97 & -0.16 & 0.97 & -0.14 & 0.97 & -0.15 \\
HL-LHC HM DY 14 TeV - neutral current - electron channel & 1.01 & 0.03 & 1.15 & 0.36 & 2.08 & 2.64 & 4.22 & 7.88 \\
HL-LHC HM DY 14 TeV - neutral current - muon channel & 1.02 & 0.04 & 1.15 & 0.37 & 2.09 & 2.66 & 4.16 & 7.75 \\
HL-LHC HM DY 14 TeV - charged current - electron channel & 1.01 & 0.02 & 0.98 & -0.07 & 1.03 & 0.08 & 0.99 & -0.03 \\
HL-LHC HM DY 14 TeV - charged current - muon channel & 0.97 & -0.09 & 0.95 & -0.14 & 1.01 & 0.02 & 0.98 & -0.05 \\
\bottomrule
\end{tabular}
\caption{\label{tab:chi2_y} Fit quality in fits contaminated with the $\hat{Y}$ parameter.}
\end{table}

\begin{table}[t]
        \tiny
        \centering
\begin{tabular}{lrrrrrrrr}
\toprule
 & \multicolumn{2}{r}{baseline} & \multicolumn{2}{r}{W=3e-5} & \multicolumn{2}{r}{W=8e-5} & \multicolumn{2}{r}{W=15e-5} \\
 & $\chi^2$ & $n_\sigma$ & $\chi^2$ & $n_\sigma$ & $\chi^2$ & $n_\sigma$ & $\chi^2$ & $n_\sigma$ \\
\midrule
NMC $d/p$ & 1.02 & 0.14 & 1.01 & 0.04 & 1.04 & 0.31 & 1.05 & 0.42 \\
NMC $p$ & 1.03 & 0.26 & 1.03 & 0.27 & 1.02 & 0.22 & 1.03 & 0.28 \\
SLAC $p$ & 1.02 & 0.06 & 1.02 & 0.07 & 1.01 & 0.03 & 1.02 & 0.06 \\
SLAC $d$ & 1.00 & -0.01 & 0.98 & -0.07 & 0.99 & -0.05 & 1.00 & 0.02 \\
BCDMS $p$ & 1.02 & 0.20 & 1.01 & 0.07 & 1.02 & 0.24 & 1.01 & 0.11 \\
BCDMS $d$ & 1.01 & 0.07 & 1.00 & 0.01 & 1.01 & 0.10 & 1.00 & 0.02 \\
CHORUS $\sigma_{CC}^{\nu}$ & 1.00 & 0.02 & 1.00 & -0.07 & 1.00 & 0.06 & 1.00 & 0.04 \\
CHORUS $\sigma_{CC}^{\bar{\nu}}$ & 0.99 & -0.13 & 0.99 & -0.13 & 1.00 & -0.00 & 1.00 & -0.02 \\
NuTeV $\sigma_{c}^{\nu}$ & 0.99 & -0.06 & 0.99 & -0.05 & 1.01 & 0.05 & 1.00 & 0.01 \\
NuTeV $\sigma_{c}^{\bar{\nu}}$ & 0.96 & -0.19 & 1.02 & 0.09 & 1.06 & 0.27 & 1.47 & 2.03 \\
HERA I+II inclusive NC $e^-p$ & 1.00 & -0.02 & 1.01 & 0.13 & 1.00 & 0.03 & 1.02 & 0.19 \\
HERA I+II inclusive NC $e^+p$ 460 GeV & 1.01 & 0.08 & 1.01 & 0.12 & 1.01 & 0.12 & 1.02 & 0.20 \\
HERA I+II inclusive NC $e^+p$ 575 GeV & 0.98 & -0.21 & 1.00 & 0.01 & 0.98 & -0.18 & 1.01 & 0.10 \\
HERA I+II inclusive NC $e^+p$ 820 GeV & 1.00 & -0.00 & 1.01 & 0.07 & 1.00 & -0.02 & 1.02 & 0.10 \\
HERA I+II inclusive NC $e^+p$ 920 GeV & 1.02 & 0.29 & 1.06 & 0.76 & 1.04 & 0.54 & 1.09 & 1.23 \\
HERA I+II inclusive CC $e^-p$ & 0.99 & -0.05 & 1.03 & 0.13 & 1.00 & -0.00 & 1.03 & 0.15 \\
HERA I+II inclusive CC $e^+p$ & 1.02 & 0.08 & 1.02 & 0.08 & 1.04 & 0.19 & 1.10 & 0.45 \\
HERA comb. $\sigma_{c\bar c}^{\rm red}$ & 1.00 & 0.02 & 1.02 & 0.08 & 1.01 & 0.02 & 1.01 & 0.04 \\
HERA comb. $\sigma_{b\bar b}^{\rm red}$ & 1.12 & 0.43 & 1.13 & 0.45 & 1.13 & 0.48 & 1.13 & 0.47 \\
DYE 866 $\sigma^d_{\rm DY}/\sigma^p_{\rm DY}$ & 1.14 & 0.40 & 1.07 & 0.20 & 1.40 & 1.11 & 1.72 & 1.98 \\
DY E886 $\sigma^p_{\rm DY}$ & 1.02 & 0.14 & 1.02 & 0.16 & 1.13 & 0.87 & 1.48 & 3.20 \\
DY E605 $\sigma^p_{\rm DY}$ & 1.08 & 0.53 & 1.07 & 0.44 & 1.07 & 0.47 & 1.08 & 0.50 \\
DYE 906 $\sigma^d_{\rm DY}/\sigma^p_{\rm DY}$ & 1.80 & 1.39 & 1.44 & 0.77 & 1.96 & 1.66 & 2.20 & 2.08 \\
CDF $Z$ rapidity (new) & 1.06 & 0.21 & 1.03 & 0.12 & 1.06 & 0.22 & 1.02 & 0.07 \\
D0 $Z$ rapidity & 1.03 & 0.10 & 1.02 & 0.07 & 1.04 & 0.16 & 1.02 & 0.07 \\
D0 $W\to \mu\nu$ asymmetry & 1.23 & 0.50 & 1.16 & 0.33 & 1.24 & 0.50 & 1.82 & 1.73 \\
ATLAS $W,Z$ 7 TeV 2010 & 1.05 & 0.20 & 1.04 & 0.17 & 1.06 & 0.22 & 1.05 & 0.18 \\
ATLAS HM DY 7 TeV & 1.02 & 0.04 & 1.05 & 0.12 & 1.01 & 0.04 & 1.03 & 0.06 \\
ATLAS low-mass DY 2011 & 0.90 & -0.17 & 1.04 & 0.07 & 0.87 & -0.23 & 0.99 & -0.02 \\
ATLAS $W,Z$ 7 TeV 2011 Central selection & 1.06 & 0.28 & 1.07 & 0.35 & 1.06 & 0.28 & 1.08 & 0.37 \\
ATLAS $W,Z$ 7 TeV 2011 Forward selection & 0.91 & -0.25 & 1.33 & 0.90 & 0.90 & -0.29 & 1.31 & 0.84 \\
ATLAS DY 2D 8 TeV high mass & 1.02 & 0.11 & 1.03 & 0.14 & 1.02 & 0.10 & 1.04 & 0.20 \\
ATLAS DY 2D 8 TeV low mass & 1.03 & 0.16 & 1.00 & 0.00 & 1.03 & 0.16 & 0.99 & -0.04 \\
ATLAS $W,Z$ inclusive 13 TeV & 1.07 & 0.09 & 1.07 & 0.09 & 1.09 & 0.11 & 1.08 & 0.10 \\
ATLAS $W^+$+jet 8 TeV & 1.17 & 0.46 & 0.96 & -0.10 & 1.17 & 0.48 & 0.96 & -0.12 \\
ATLAS $W^-$+jet 8 TeV & 1.19 & 0.51 & 0.97 & -0.10 & 1.21 & 0.58 & 0.98 & -0.06 \\
ATLAS $Z$ $p_T$ 8 TeV $(p_T^{ll},M_{ll})$ & 1.01 & 0.03 & 0.98 & -0.07 & 1.01 & 0.03 & 0.99 & -0.05 \\
ATLAS $Z$ $p_T$ 8 TeV $(p_T^{ll},y_{ll})$ & 0.98 & -0.10 & 0.94 & -0.29 & 0.99 & -0.06 & 0.96 & -0.21 \\
ATLAS $\sigma_{tt}^{\rm tot}$ & 1.03 & 0.02 & 1.14 & 0.10 & 1.04 & 0.03 & 1.17 & 0.12 \\
ATLAS $\sigma_{tt}^{\rm tot}$ 8 TeV & 1.31 & 0.22 & 1.12 & 0.09 & 1.30 & 0.21 & 1.13 & 0.09 \\
ATLAS $\sigma_{tt}^{\rm tot}$ 13 TeV Run II full lumi & 0.92 & -0.06 & 0.93 & -0.05 & 0.93 & -0.05 & 0.97 & -0.02 \\
ATLAS $t\bar{t}$ $y_t$ & 1.03 & 0.05 & 1.06 & 0.09 & 1.03 & 0.04 & 1.06 & 0.08 \\
ATLAS $t\bar{t}$ $y_{t\bar{t}}$ & 1.04 & 0.05 & 1.04 & 0.06 & 1.05 & 0.08 & 1.09 & 0.12 \\
ATLAS $t\bar{t}$ normalised $|y_t|$ & 1.13 & 0.21 & 1.13 & 0.21 & 1.14 & 0.22 & 1.18 & 0.28 \\
ATLAS jets 8 TeV, R=0.6 & 0.83 & -1.53 & 0.94 & -0.57 & 0.83 & -1.53 & 0.94 & -0.54 \\
ATLAS dijets 7 TeV, R=0.6 & 1.03 & 0.19 & 1.00 & 0.00 & 1.03 & 0.18 & 1.01 & 0.10 \\
ATLAS direct photon production 13 TeV & 0.97 & -0.16 & 1.03 & 0.14 & 0.98 & -0.13 & 1.03 & 0.16 \\
ATLAS single top $R_{t}$ 7 TeV & 1.14 & 0.10 & 1.25 & 0.18 & 1.06 & 0.04 & 1.16 & 0.11 \\
ATLAS single top $R_{t}$ 13 TeV & 0.91 & -0.07 & 1.01 & 0.01 & 0.94 & -0.04 & 1.05 & 0.03 \\
ATLAS single top $y_t$ (normalised) & 0.94 & -0.07 & 1.07 & 0.09 & 0.94 & -0.08 & 1.04 & 0.04 \\
ATLAS single antitop $y$ (normalised) & 0.92 & -0.10 & 0.91 & -0.11 & 0.94 & -0.07 & 0.98 & -0.03 \\
CMS $W$ asymmetry 840 pb & 0.99 & -0.03 & 0.99 & -0.02 & 0.97 & -0.08 & 1.05 & 0.12 \\
CMS $W$ asymmetry 4.7 fb & 0.97 & -0.07 & 0.97 & -0.06 & 0.97 & -0.06 & 0.97 & -0.06 \\
CMS Drell-Yan 2D 7 TeV 2011 & 1.01 & 0.05 & 1.01 & 0.07 & 1.01 & 0.04 & 1.01 & 0.08 \\
CMS $W$ rapidity 8 TeV & 1.06 & 0.21 & 1.11 & 0.38 & 1.07 & 0.25 & 1.11 & 0.38 \\
CMS $Z$ $p_T$ 8 TeV $(p_T^{ll},y_{ll})$ & 1.03 & 0.12 & 1.03 & 0.12 & 1.04 & 0.13 & 1.06 & 0.21 \\
CMS dijets 7 TeV & 0.97 & -0.15 & 1.05 & 0.24 & 0.97 & -0.13 & 1.05 & 0.28 \\
CMS jets 8 TeV & 0.99 & -0.11 & 1.00 & -0.02 & 0.99 & -0.05 & 1.01 & 0.06 \\
CMS $\sigma_{tt}^{\rm tot}$ 7 TeV & 0.86 & -0.10 & 0.95 & -0.03 & 0.86 & -0.10 & 0.99 & -0.00 \\
CMS $\sigma_{tt}^{\rm tot}$ 8 TeV & 1.18 & 0.13 & 1.08 & 0.06 & 1.22 & 0.16 & 1.07 & 0.05 \\
CMS $\sigma_{tt}^{\rm tot}$ 13 TeV & 0.98 & -0.01 & 1.11 & 0.08 & 0.99 & -0.01 & 1.13 & 0.09 \\
CMS $t\bar{t}$ rapidity $y_{t\bar{t}}$ & 1.06 & 0.12 & 1.04 & 0.08 & 1.03 & 0.07 & 1.02 & 0.05 \\
CMS $\sigma_{tt}^{\rm tot}$ 5 TeV & 0.86 & -0.10 & 0.77 & -0.16 & 0.81 & -0.13 & 0.73 & -0.19 \\
CMS $t\bar{t}$ double differential $(m_{t\bar{t}},y_{t\bar{t}})$ & 0.99 & -0.04 & 1.01 & 0.02 & 1.01 & 0.04 & 1.03 & 0.08 \\
CMS $t\bar{t}$ absolute $y_t$ & 1.01 & 0.03 & 1.02 & 0.04 & 1.02 & 0.04 & 1.05 & 0.11 \\
CMS $t\bar{t}$ absolute $|y_t|$ & 0.98 & -0.05 & 0.99 & -0.03 & 0.98 & -0.05 & 0.96 & -0.10 \\
CMS single top $\sigma_{t}+\sigma_{\bar{t}}$ 7 TeV & 0.93 & -0.05 & 0.91 & -0.06 & 0.88 & -0.09 & 0.86 & -0.10 \\
CMS single top $R_{t}$ 8 TeV & 0.64 & -0.26 & 1.21 & 0.15 & 0.65 & -0.25 & 1.15 & 0.10 \\
CMS single top $R_{t}$ 13 TeV & 1.50 & 0.35 & 1.44 & 0.31 & 1.46 & 0.32 & 1.40 & 0.28 \\
LHCb $Z$ 940 pb & 1.08 & 0.17 & 0.95 & -0.10 & 1.12 & 0.25 & 0.96 & -0.08 \\
LHCb $Z\to ee$ 2 fb & 1.03 & 0.08 & 1.04 & 0.12 & 1.01 & 0.02 & 1.01 & 0.03 \\
LHCb $W,Z \to \mu$ 7 TeV & 0.98 & -0.07 & 0.96 & -0.17 & 1.07 & 0.26 & 1.13 & 0.48 \\
LHCb $W,Z \to \mu$ 8 TeV & 1.08 & 0.32 & 1.12 & 0.45 & 1.17 & 0.65 & 1.32 & 1.22 \\
LHCb $Z\to \mu\mu$ & 1.09 & 0.26 & 1.05 & 0.14 & 1.10 & 0.28 & 1.05 & 0.15 \\
LHCb $Z\to ee$ & 1.07 & 0.19 & 1.03 & 0.08 & 1.08 & 0.22 & 1.04 & 0.11 \\
CMS HM DY 8 TeV & 0.98 & -0.09 & 0.98 & -0.08 & 0.99 & -0.05 & 1.00 & -0.02 \\
CMS HM DY 13 TeV - combined channel & 0.97 & -0.14 & 0.97 & -0.16 & 0.97 & -0.14 & 0.97 & -0.12 \\
HL-LHC HM DY 14 TeV - neutral current - electron channel & 1.01 & 0.03 & 1.03 & 0.08 & 1.04 & 0.10 & 1.21 & 0.53 \\
HL-LHC HM DY 14 TeV - neutral current - muon channel & 1.02 & 0.04 & 1.03 & 0.07 & 1.02 & 0.06 & 1.20 & 0.49 \\
HL-LHC HM DY 14 TeV - charged current - electron channel & 1.01 & 0.02 & 1.00 & -0.00 & 1.15 & 0.42 & 2.97 & 5.56 \\
HL-LHC HM DY 14 TeV - charged current - muon channel & 0.97 & -0.09 & 0.98 & -0.07 & 1.11 & 0.31 & 2.75 & 4.94 \\
\bottomrule
\end{tabular}
\caption{\label{tab:chi2_w} Fit quality in fits contaminated with the $\hat{W}$ parameter.}
\end{table}

\section{PDF comparison}
\label{app:pdfs}
\begin{figure}[t!]
  \includegraphics[width=0.49\linewidth]{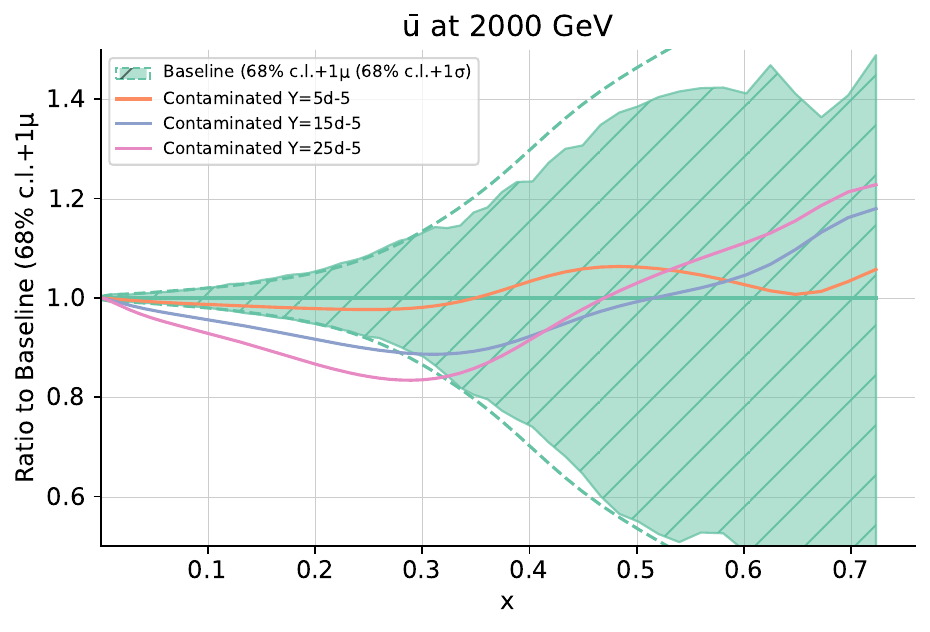}
  \includegraphics[width=0.49\linewidth]{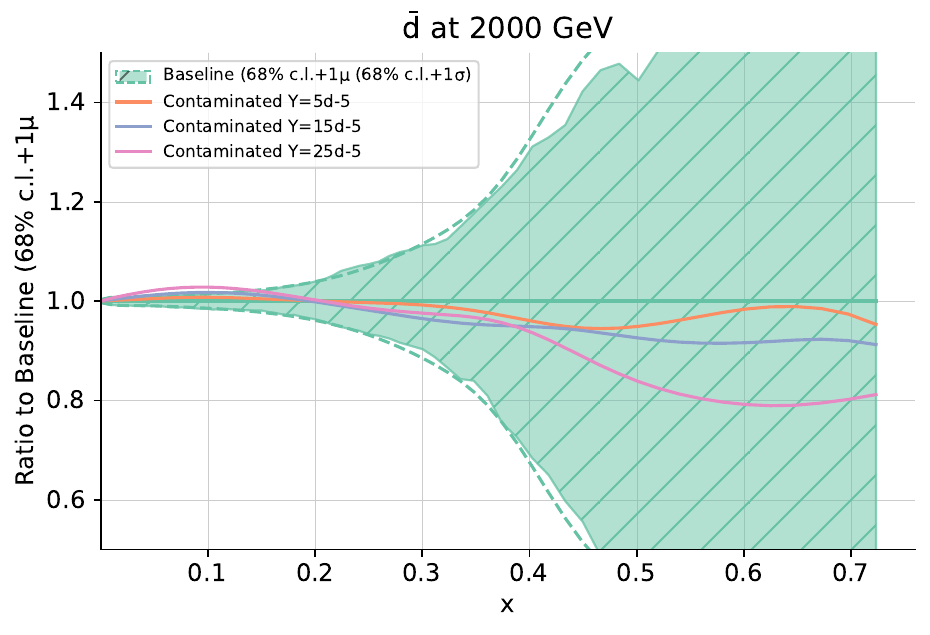}
  \includegraphics[width=0.49\linewidth]{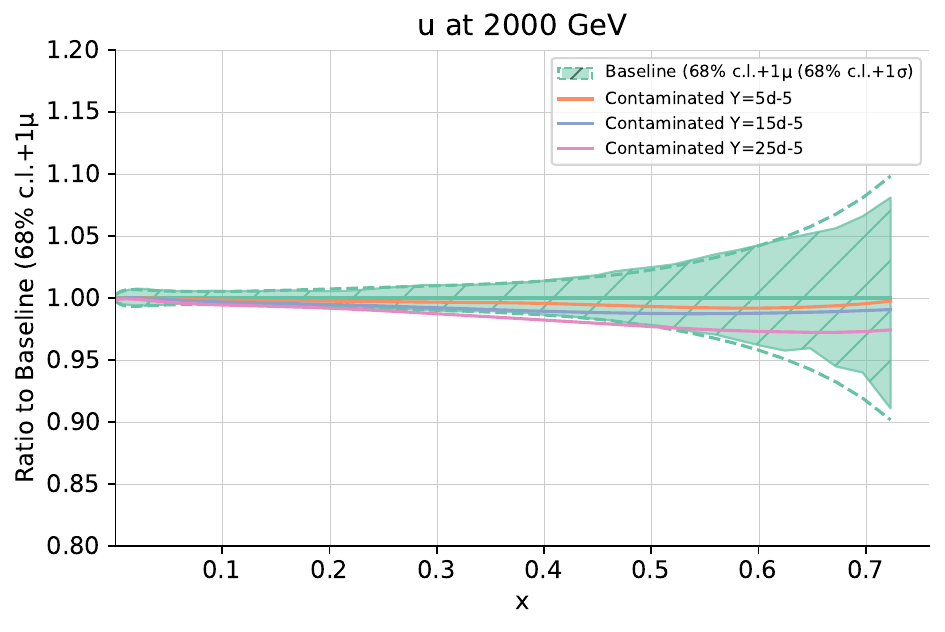}
  \includegraphics[width=0.49\linewidth]{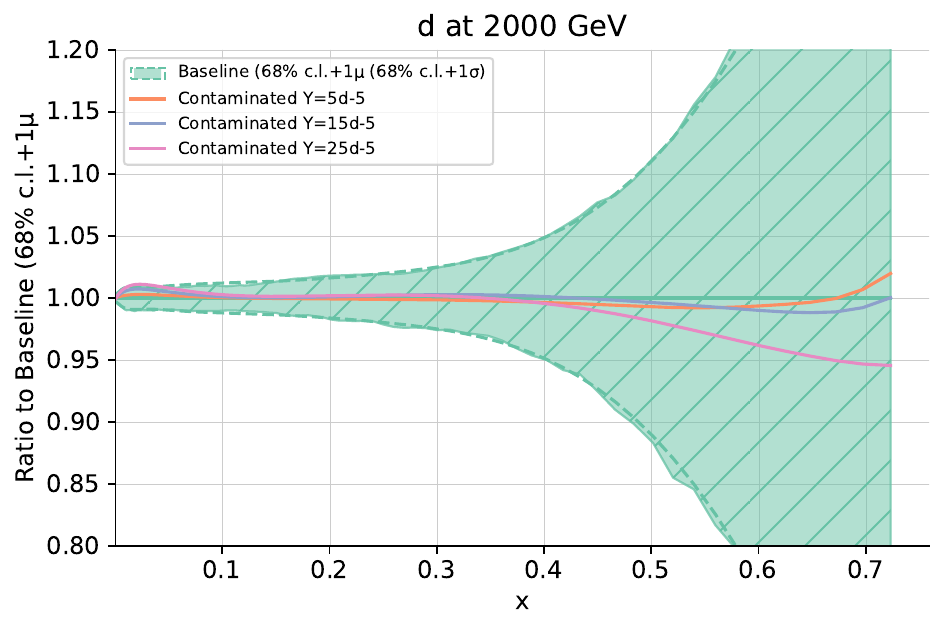}
  \caption{Contaminated versus baseline anti-up (top-left panel), anti-down (top-right panel), up (bottom-left panel) and down (bottom-right panel) PDFs at
    $Q=2$ TeV. The results are normalised to the baseline SM PDFs and
    the 68\% C.L. band is displayed. Contaminated PDFs have been
    obtained by fitting the Monte Carlo pseudodata
    produced with $\hat{Y}=5\cdot 10^{-5}$ (orange line),
    $\hat{Y}=15\cdot 10^{-5}$ (blue line) and $\hat{Y}=25\cdot
    10^{-5}$ (pink line) assuming the SM in the theory predictions. }
	\label{fig:Ypdfs}
      \end{figure}

In Fig.~\ref{fig:Ypdfs}, we display the PDFs that are mostly affected
by the new physics contamination in Scenario I, namely the anti-up and anti-down
distributions at $Q=$ 2 TeV in the large-$x$ region. We see that for
$\hat{Y} = 5\cdot 10^{-5}$, PDFs are statistically equivalent to the baseline ones.

      \begin{figure}[t!]
    \includegraphics[width=0.49\linewidth]{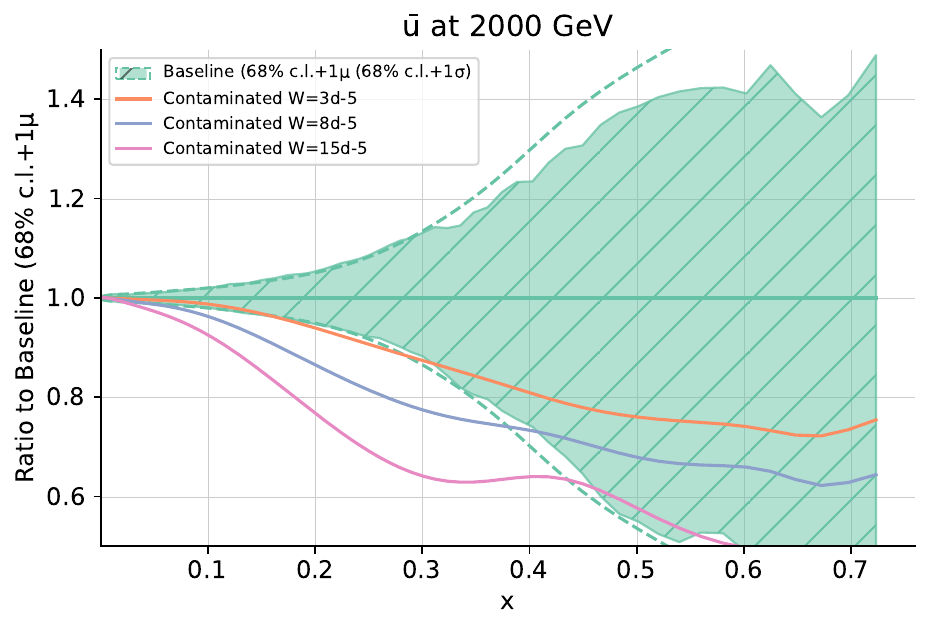}
  \includegraphics[width=0.49\linewidth]{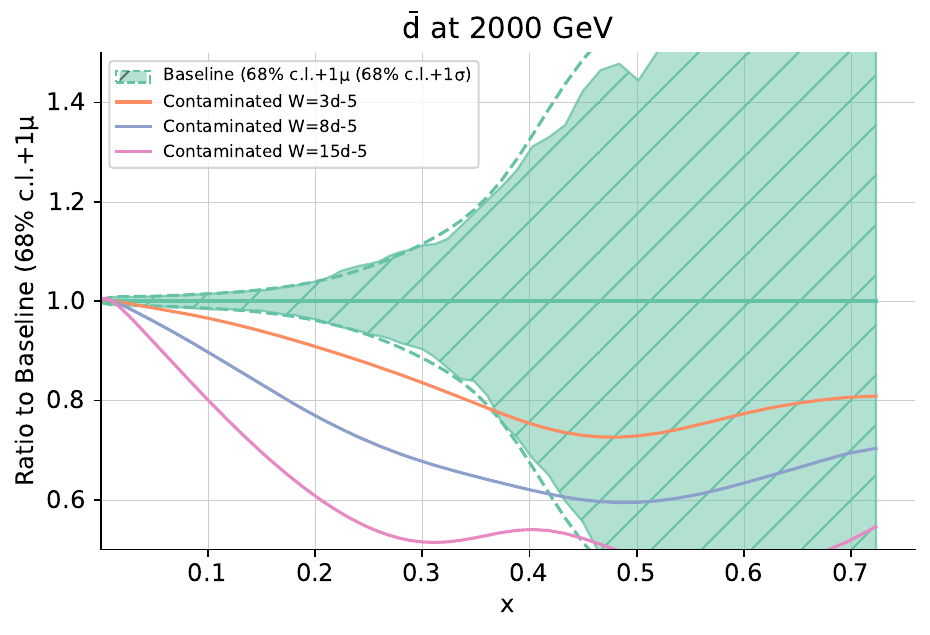}\\
  \includegraphics[width=0.49\linewidth]{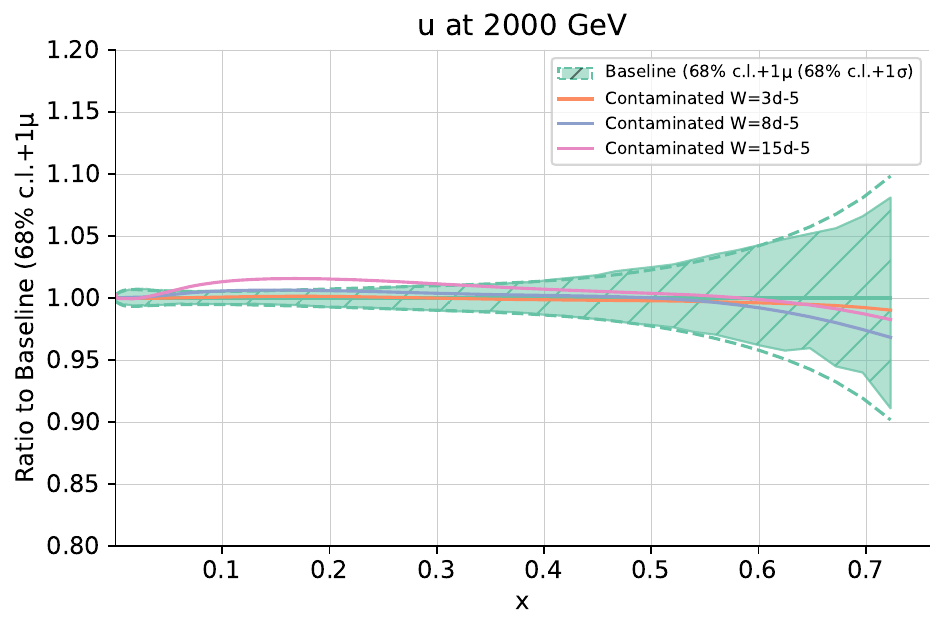}
  \includegraphics[width=0.49\linewidth]{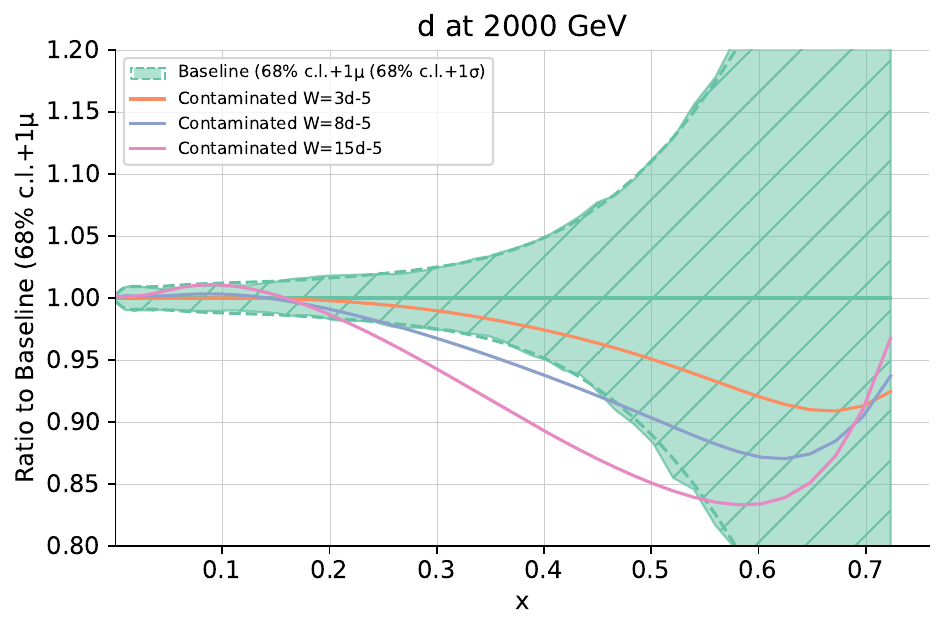}
  \caption{Same as Fig.~\ref{fig:Ypdfs} for $\hat{W}=3\cdot 10^{-5}$ (orange line),
    $\hat{W}=8\cdot 10^{-5}$ (blue line) and $\hat{W}=15\cdot
    10^{-5}$ (pink line).}
  \label{fig:Wpdfs}
      \end{figure}
      In Fig.~\ref{fig:Wpdfs}, we display the PDFs that are mostly affected
by the Scenario II new physics contamination, namely the up, down, anti-up and anti-down
distributions at $Q=$ 2 TeV in the large-$x$ region. We see that for
the critical value $\hat{W} = 8\cdot 10^{-5}$ the shift in the
antiquark PDFs is above the 2$\sigma$ level for all of the distributions
from $x\gtrsim 0.2$, apart from the up-quark PDF in which the shift is visible but below the 2$\sigma$ level.
\section{Random seed dependence}
\label{app:random}
As described in Eq.~\eqref{eq:observed_data}, the pseudodata used in this study is stochastic, fluctuated around the supposed law of Nature in order to simulate random experimental noise. This noise is generated in a reproducible manner using the NNPDF closure test code by selecting a particular \textit{seed} for the generation algorithm; different choices of seed lead to different choices of noise.

This has consequences for the resulting contaminated PDF fits, which in principle can depend on the seed used for the random noise. In certain parts of this work, in particular in the production of Figs.~\ref{fig:Y_chi2_nsigma_dist} and \ref{fig:W_chi2_nsigma_dist}, we have made the approximation that the contaminated PDFs do not depend significantly on the choice of random seed; rather, we hope that their behaviour is most importantly affected by whether or not new physics is present in the pseudodata or not. This is a useful approximation to make, since it avoids the requirement of running a large quantity of PDF fits, which is computationally expensive.

We justify this approximation in this brief appendix by comparing the PDF luminosities in various contaminated fits produced using different seeds for the random pseudodata. The luminosities are the relevant quantity to compare, since these are the quantities which enter the theoretical predictions for the hadronic data, in particular the Drell-Yan data, the focus of this study.

\begin{figure}[h!]
  \includegraphics[width=0.49\linewidth]{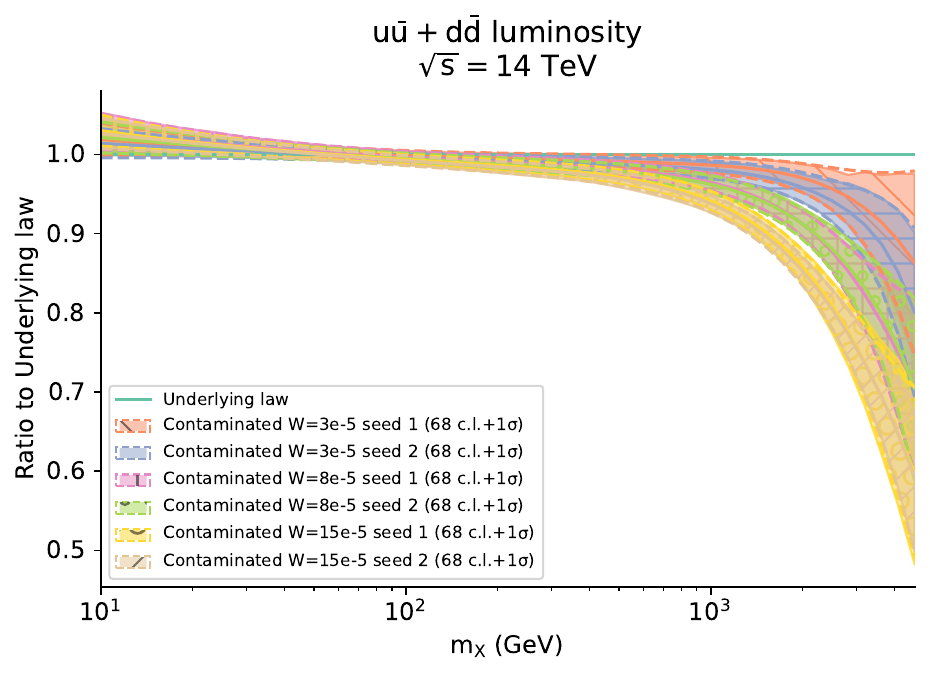}
  \includegraphics[width=0.49\linewidth]{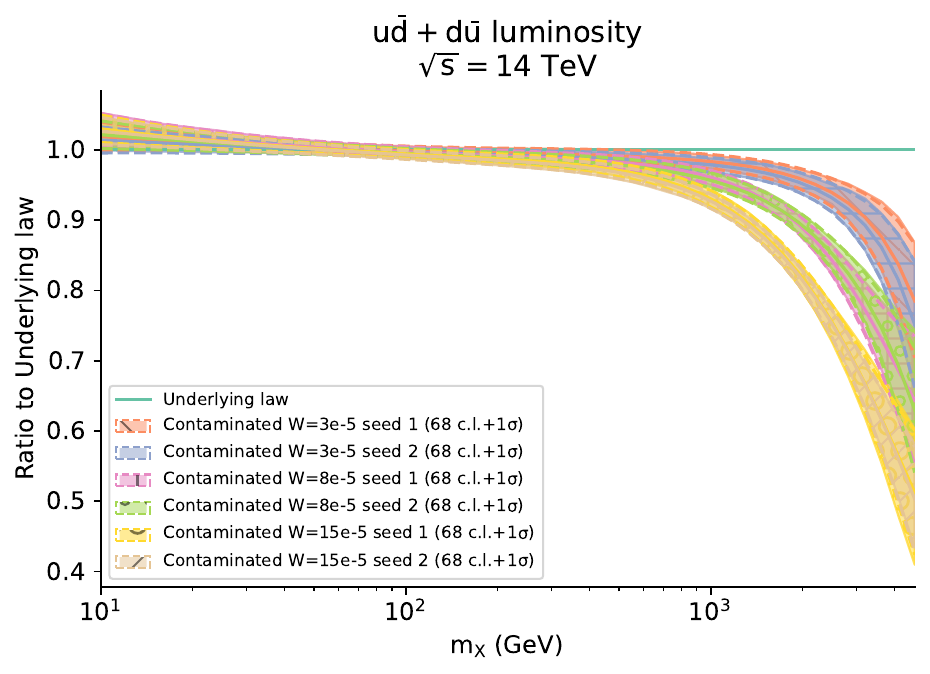}
  \caption{Comparison between luminosities obtained in contaminated fits using two different random seeds in the generation of pseudodata. In each case, we display six contaminated fits: two fits for each of the benchmark values $\hat{W} = 3 \times 10^{-5}, 8 \times 10^{-5}, 15 \times 10^{-5}$, trained on pseudodata generated with random seed $1$ and random seed $2$ respectively.}
\label{fig:lumi_random}
\end{figure}

In Fig.~\ref{fig:lumi_random}, we plot the luminosities obtained
from contaminated fits resulting from setting the $\hat{W}$ parameter to the benchmark values $\hat{W} = 3 \times 10^{-5}$, $\hat{W} = 8 \times 10^{-5}$ and $\hat{W} = 15 \times 10^{-5}$. We display the results for two separate contaminated fits for each of the benchmark values; in each case, one of the fits results from the use of a particular random seed (called \textit{seed $1$} in the plots), whilst the other results from the use of another random seed (called \textit{seed $2$} in the plots). 
We observe that the luminosities are completely statistically equivalent between the two seeds, but that across different benchmark values of $\hat{W}$, there is indeed a statistical difference between the luminosities. This justifies that the leading effect on the contaminated fits is the injection of new physics into the pseudodata, rather than the random noise added to the pseudodata. In particular, the approximation in Sect.~\ref{sec:dy} is fully justified. Similar conclusions hold for the $\hat{Y}$ parameter.

\FloatBarrier
\renewcommand{\em}{}
\bibliographystyle{UTPstyle}
\bibliography{references}

\end{document}